\documentclass{article}

\usepackage{cite}
\usepackage{amssymb,amsfonts}
\usepackage[ruled, vlined]{algorithm2e}
\usepackage{graphicx}
\usepackage{textcomp}
\usepackage{braket}
\usepackage{tabularx}
\usepackage{lscape} 
\usepackage{xcolor}
\usepackage{tikzit}
\usepackage{wrapfig}
\usepackage{mathtools, bbm} %
\usepackage{color, soul}
\usepackage{titlesec}
\def\BibTeX{{\rm B\kern-.05em{\sc i\kern-.025em b}\kern-.08em
    T\kern-.1667em\lower.7ex\hbox{E}\kern-.125emX}}
\usepackage[hidelinks]{hyperref}
\usepackage{float}
\usepackage{listings}
\usepackage[capitalise]{cleveref}
\usepackage{color}
\usepackage{subcaption}
\usepackage[section]{placeins}  %
\usepackage{rotating}  %
\usepackage[margin=1in]{geometry}
\usepackage[colorinlistoftodos,
            prependcaption,
            textsize=scriptsize,
            backgroundcolor=yellow,
            linecolor=lightgray,
            bordercolor=lightgray]{todonotes}
\usepackage{yfonts}
\usepackage{amsthm}

\newtheorem*{definition*}{Definition}
\newtheorem*{proposition}{Proposition}

\titleformat*{\section}{\Large\bfseries}
\titleformat*{\subsection}{\large\bfseries}
\titleformat*{\subsubsection}{\bfseries}

\tikzstyle{gate}=[shape=rectangle, text height=1.5ex, text depth=0.25ex, yshift=0.5mm, fill=white, draw=black, minimum height=3mm, yshift=-0.5mm, minimum width=3mm, font={\small}, tikzit category=circuit, inner sep=2pt]
\tikzstyle{big gate}=[shape=rectangle, text height=1.5ex, text depth=0.25ex, yshift=0.5mm, fill=white, draw=black, minimum height=10mm, yshift=-0.5mm, minimum width=5mm, font={\small}, tikzit category=circuit]
\tikzstyle{Z dot}=[inner sep=0mm, minimum size=2mm, shape=circle, draw=black, fill={rgb,255: red,221; green,255; blue,221}, tikzit category=zx]
\tikzstyle{Z phase dot}=[minimum size=5mm, font={\footnotesize\boldmath}, shape=rectangle, rounded corners=2mm, inner sep=0.2mm, outer sep=-2mm, scale=0.8, tikzit shape=circle, draw=black, fill={rgb,255: red,221; green,255; blue,221}, tikzit draw=blue, tikzit category=zx]
\tikzstyle{X dot}=[Z dot, shape=circle, draw=black, fill={rgb,255: red,255; green,136; blue,136}, tikzit category=zx]
\tikzstyle{X phase dot}=[Z phase dot, tikzit shape=circle, tikzit draw=blue, fill={rgb,255: red,255; green,136; blue,136}, font={\footnotesize\boldmath}, tikzit category=zx]
\tikzstyle{hadamard}=[fill=yellow, draw=black, shape=rectangle, inner sep=0.6mm, minimum height=1.5mm, minimum width=1.5mm, tikzit category=zx]
\tikzstyle{Q Z dot}=[fill={rgb,255: red,221; green,255; blue,221}, draw=black, shape=circle, thick, tikzit category=zx, tikzit draw=red, minimum size=2mm, inner sep=0mm]
\tikzstyle{Q X dot}=[fill={rgb,255: red,255; green,136; blue,136}, draw=black, shape=circle, tikzit category=zx, tikzit draw=red, thick, inner sep=0mm, minimum size=2mm]
\tikzstyle{Q hadamard}=[fill=yellow, draw=black, thick, shape=rectangle, inner sep=0.6mm, minimum height=1.5mm, minimum width=1.5mm, tikzit category=zx, tikzit draw=red]
\tikzstyle{paulibox}=[fill={rgb,255: red,221; green,221; blue,255}, draw=black, shape=rectangle, inner sep=0.6mm, minimum height=5mm, minimum width=5mm, font={\footnotesize}, text height=1.5ex, text depth=0.25ex, tikzit category=zx]
\tikzstyle{vertex}=[inner sep=0mm, minimum size=1mm, shape=circle, draw=black, fill=black, tikzit category=misc]
\tikzstyle{vertex set}=[inner sep=0mm, minimum size=1mm, shape=circle, draw=black, fill=white, font={\footnotesize\boldmath}, tikzit category=misc]
\tikzstyle{small black dot}=[fill=black, draw=black, shape=circle, inner sep=0pt, minimum width=1.2mm, tikzit category=circuit]
\tikzstyle{cnot ctrl}=[fill=black, draw=black, shape=circle, inner sep=0pt, minimum width=1.2mm, tikzit category=circuit]
\tikzstyle{cnot targ}=[fill=white, draw=white, shape=circle, tikzit category=circuit, label={center:$\oplus$}, inner sep=0pt, minimum width=2.1mm, tikzit fill={rgb,255: red,102; green,204; blue,255}, tikzit draw=black]
\tikzstyle{ket}=[fill=white, draw=black, shape=regular polygon, regular polygon sides=3, regular polygon rotate=-30, scale=0.7, inner sep=1pt, tikzit category=circuit, tikzit shape=rectangle, tikzit fill=green]
\tikzstyle{Q ket}=[fill=white, draw=black, thick, shape=regular polygon, regular polygon sides=3, regular polygon rotate=-30, scale=0.7, inner sep=1pt, tikzit category=circuit, tikzit shape=rectangle, tikzit fill=green, tikzit draw=red]
\tikzstyle{bra}=[fill=white, draw=black, shape=regular polygon, regular polygon sides=3, regular polygon rotate=30, scale=0.7, inner sep=1pt, tikzit category=circuit, tikzit shape=rectangle, tikzit fill=red]
\tikzstyle{scalar}=[shape=rectangle, text height=1.5ex, text depth=0.25ex, yshift=0.5mm, fill=white, draw=black, minimum height=5mm, yshift=-0.5mm, minimum width=5mm, font={\small}]
\tikzstyle{clabel}=[fill=white, draw=none, shape=rectangle, tikzit fill={rgb,255: red,56; green,255; blue,242}, font={\footnotesize}, inner sep=1pt, tikzit category=labels]
\tikzstyle{empty diagram}=[draw={gray!40!white}, dashed, shape=rectangle, minimum width=1cm, minimum height=1cm, tikzit category=misc]
\tikzstyle{amap}=[fill=white, draw=black, shape=NEbox, tikzit category=asymmetric, tikzit fill=yellow, tikzit shape=rectangle]
\tikzstyle{amap conj}=[fill=white, draw=black, shape=NWbox, tikzit category=asymmetric, tikzit fill=green, tikzit shape=rectangle]
\tikzstyle{amap adj}=[fill=white, draw=black, shape=SEbox, tikzit category=asymmetric, tikzit fill=red, tikzit shape=rectangle]
\tikzstyle{amap trans}=[fill=white, draw=black, shape=SWbox, tikzit category=asymmetric, tikzit fill=orange, tikzit shape=rectangle]
\tikzstyle{astate}=[fill=white, draw=black, shape=NEtriangle, tikzit category=asymmetric, tikzit shape=circle, tikzit fill=yellow]
\tikzstyle{astate conj}=[fill=white, draw=black, shape=NWtriangle, tikzit category=asymmetric, tikzit shape=circle, tikzit fill=green]
\tikzstyle{astate adj}=[fill=white, draw=black, shape=SEtriangle, tikzit category=asymmetric, tikzit shape=circle, tikzit fill=red]
\tikzstyle{astate trans}=[fill=white, draw=black, shape=SWtriangle, tikzit category=asymmetric, tikzit shape=circle, tikzit fill=orange]
\tikzstyle{white dot}=[inner sep=0mm, minimum size=2mm, shape=circle, draw=black, fill={rgb,255: red,250; green,250; blue,250}]
\tikzstyle{white phase dot}=[minimum size=5mm, font={\footnotesize\boldmath}, shape=rectangle, rounded corners=2mm, inner sep=0.2mm, outer sep=-2mm, scale=0.8, tikzit shape=circle, draw=black, fill={rgb,255: red,250; green,250; blue,250}, tikzit draw=blue]
\tikzstyle{hbox}=[shape=rectangle, text height=2mm, fill={rgb,255: red,255; green,235; blue,61}, draw=black, minimum height=2mm, minimum width=2mm, font={\small}, tikzit category=zh, inner sep=0pt, rounded corners=0.5mm]
\tikzstyle{Z dot (zh)}=[inner sep=0mm, minimum size=2mm, shape=circle, draw=black, fill={rgb,255: red,250; green,250; blue,250}, tikzit category=zh]
\tikzstyle{X dot (zh)}=[Z dot, shape=circle, draw=black, fill={rgb,255: red,193; green,193; blue,193}, tikzit category=zh]
\tikzstyle{triangle}=[fill=yellow, draw=black, shape=isosceles triangle, isosceles triangle apex angle=60, minimum size=2.5mm, inner sep=0mm]
\tikzstyle{labelled hbox}=[shape=rectangle, text height=1.75ex, text depth=0.5ex, fill={rgb,255: red,255; green,235; blue,61}, draw=black, minimum height=3mm, minimum width=4mm, font={\small}, tikzit category=zh, inner sep=1.3pt, rounded corners=0.5mm]
\tikzstyle{Z phase dot (zh)}=[Z phase dot, tikzit shape=circle, tikzit draw=blue, fill={rgb,255: red,250; green,250; blue,250}, font={\footnotesize\boldmath}, tikzit category=zh]
\tikzstyle{X phase dot (zh)}=[Z phase dot, tikzit shape=circle, tikzit draw=blue, fill={rgb,255: red,193; green,193; blue,193}, font={\footnotesize\boldmath}, tikzit category=zh]
\tikzstyle{W node}=[fill=black, draw=black, shape=regular polygon, regular polygon sides=3, minimum size=2mm]
\tikzstyle{Z dot (zw)}=[fill=white, draw=black, shape=circle, minimum width=1.2mm, inner sep=0pt]
\tikzstyle{W tri}=[fill=black, draw=black, shape=isosceles triangle, inner sep=0pt, minimum size=3.5mm, isosceles triangle apex angle=60, rotate=180]

\tikzstyle{hadamard edge}=[-, dashed, dash pattern=on 2pt off 0.5pt, thick, draw={rgb,255: red,68; green,136; blue,255}]
\tikzstyle{Q edge}=[-, thick, tikzit draw=red]
\tikzstyle{box edge}=[-, dashed, dash pattern=on 2pt off 0.5pt, thick, draw={rgb,255: red,203; green,192; blue,225}]
\tikzstyle{brace edge}=[-, tikzit draw=blue, decorate, decoration={brace,amplitude=1mm,raise=-1mm}]
\tikzstyle{diredge}=[->, thick]
\tikzstyle{double edge}=[-, double, shorten <=-1mm, shorten >=-1mm, double distance=2pt]
\tikzstyle{gray edge}=[-, {gray!60!white}]
\tikzstyle{pointer edge}=[->, very thick, gray]
\tikzstyle{boldedge}=[-, line width=1.6pt, shorten <=-0.17mm, shorten >=-0.17mm]
\tikzstyle{bidir edge}=[<->, very thick, draw={rgb,255: red,191; green,191; blue,191}]
\tikzstyle{purple edge}=[->, thick, draw={rgb,255: red,225; green,117; blue,216}]
\tikzstyle{green edge}=[->, thick, draw={rgb,255: red,167; green,231; blue,137}]
\tikzstyle{orange edge}=[->, thick, draw={rgb,255: red,245; green,170; blue,63}]
\tikzstyle{blue edge}=[->, thick, draw={rgb,255: red,68; green,136; blue,255}]
\tikzstyle{any edge}=[->, thick, draw=cyan]
\tikzstyle{red edge}=[->, thick, draw={rgb,255: red,255; green,136; blue,136}]
\tikzstyle{bidiredge}=[<->, thick]
\tikzstyle{dashed diredge}=[->, dashed, dash pattern=on 1pt off 0.5pt]
\tikzstyle{bidashed diredge}=[<->, dashed, dash pattern=on 1pt off 0.5pt]

\usepackage{multicol}

\definecolor{darkgreen}{RGB}{0, 153, 0}

\DeclareCaptionLabelFormat{adja-page}{\hrulefill\\#1 #2 \emph{(previous page)}}

\begin{document}

\title{Scalable and interpretable quantum natural language processing:\\
an implementation on trapped ions}

\author{Tiffany Duneau${}^{1,2}$, Saskia Bruhn${}^{1,}$\footnote{Equal contribution.} ,
Gabriel Matos${}^{1,*}$,
Tuomas Laakkonen${}^1$,\\
Katerina Saiti${}^3$,
Anna Pearson${}^{1,*}$,
Konstantinos Meichanetzidis${}^1$,
Bob Coecke${}^1$\\
~\\
${}^1$Quantinuum, ${}^2$University of Oxford, ${}^3$Leiden University}

\maketitle

\begin{abstract}
We present the first implementation of text-level quantum natural language processing, a field where quantum computing and AI have found a fruitful intersection.  We focus on the QDisCoCirc model, which is underpinned by a compositional approach to rendering AI interpretable: the behaviour of the whole can be understood in terms of the behaviour of parts, and the way they are put together. Interpretability is crucial for understanding the unwanted behaviours of AI.
By leveraging the compositional structure in the model's architecture, we introduce a novel setup which enables `compositional generalisation’: we classically train components which are then composed to generate larger test instances, the evaluation of which asymptotically requires a quantum computer. Another key advantage of our approach is that it bypasses the trainability challenges arising in quantum machine learning. 
The main task that we consider is the model-native task of question-answering, and we handcraft toy scale data that serves as a proving ground.  We demonstrate an experiment on Quantinuum's H1-1 trapped-ion quantum processor, which constitutes the first proof of concept implementation of scalable compositional QNLP.  We also provide resource estimates for classically simulating the model.    
The compositional structure allows us to inspect and interpret the word embeddings the model learns for each word, as well as the way in which they interact. This improves our understanding of how it tackles the question-answering task.  As an initial comparison with classical baselines, we considered transformer and LSTM models, as well as GPT-4, none of which succeeded at compositional generalisation.
\end{abstract}

\clearpage
\setcounter{tocdepth}{2} 
\tableofcontents
\clearpage

\section{Introduction}

Artificial intelligence permeates a wide range of activity, from academia to industrial real-world applications, with natural language processing (NLP) taking centre stage. In parallel, quantum computing has seen a recent surge in development, with the advent of \emph{noisy intermediate-scale quantum} (NISQ) processors~\cite{preskill2018quantum}. These are reaching scales where they become hard to simulate on classical computers within a reasonable resource budget~\cite{decross2024computational}. The merging of these two fields has given rise to \emph{quantum natural language processing} (QNLP).

One important feature that distinguishes our line of work~\cite{WillC, QNLP-foundations, GrammarAwareSentenceClassification, Lorenz_2023, kartsaklis2021lambeq} from other contributions to QNLP~\cite{widdows2023nearterm, widdows2024natural}, and the broad field of quantum machine learning~\cite{dunjko2017machine,Benedetti_2019,Schuld2014}, is that it provides a path towards \emph{explainability} and \emph{interpretability}.
Indeed, while the advancements of contemporary artificial intelligence are impressive, to achieve general applicability and high performance, these model architectures are set up as black boxes trained on large amounts of data: when things go wrong, one typically does not understand the reason why. 
In previous work, we have proposed \emph{DisCoCat}~\cite{CSC, teleling, FrobMeanI, GrefSadr, KartSadr}, a quantum-inspired model for language that aims to be inherently explainable and interpretable by making use of the principle of \emph{compositionality}~\cite{compinterp}.  
Here, we take compositionality to mean that the behaviour of the whole can be understood in terms of the behaviour of the parts, along with the way they are combined \cite{compositionality}.  In the case of language, this includes linguistic meaning, as well as linguistic structure, such as grammar. DisCoCat's origin in categorical quantum mechanics~\cite{AC1, CKbook} makes it an ideal framework to exploit this structure to design QNLP models. These models formed the foundation of our earlier work, and have already been implemented on NISQ quantum processors for the task of sentence classification \cite{GrammarAwareSentenceClassification, Lorenz_2023, kartsaklis2021lambeq}.  However, the particular nature of these models have certain shortcomings, which are discussed in~\cite{tkb}.

This work focuses on the novel \emph{DisCoCirc} framework \cite{CoeckeText, wang2023distilling}, where sentences are represented by circuits that can be further composed into \emph{text circuits}. This composition of sentences, which DisCoCat did not allow, is crucial for scalability. Here, we consider the applicability of DisCoCirc models within a quantum setup, which we refer to as QDisCoCirc.
We present experimental results for the task of question answering with QDisCoCirc.  This constitutes the first proof-of-concept implementation of \emph{scalable compositional QNLP}. Note that this work is part of a pair of papers on compositional quantum natural language processing. The article \cite{tkb}, which complements this one, focuses on the algorithms and complexity theory behind our work here, and provides some additional background.

We begin by constructing toy data in the form of simple stories with binary questions. Then, using parameterised quantum circuits to represent quantum word embeddings, we train the QDisCoCirc model in-task.  This provides us with word embeddings that can be used to compose texts of larger sizes compared to the texts used during training.  Our results in \cref{sec:CompGen} show, with statistical significance, that the accuracy of our models does not decay at test time when tested solely on instances larger than those used in training. This is what we call \emph{compositional generalisation}. 

Moreover, as highlighted earlier, the compositional structure allows for inspection of the trained model's internals, and we argue that such compositional setups are more interpretable than black box setups. Recent work~\cite{compinterp} has explored how compositionality can be used in constructing interpretable and explainable AI models, and defined a class of \emph{compositionally interpretable} models, which includes DisCoCirc. While the precise definition is given in terms of category theory, we summarise this informally here.

\begin{definition*}[{Informal, \cite{compinterp}}]
    \em A model is \emph{compositionally interpretable} if instances of the model are composed freely from generators, and the generators are equipped with \emph{human-friendly} interpretations.
\end{definition*}

In Section 4, we make this concrete by testing to what extent the axioms and relations that we would expect to hold are obeyed by the trained quantum word embeddings. However, \cite[Section 9]{compinterp} argues that DisCoCirc can also provide additional explainability benefits beyond compositional interpretability, due to the causal structure of the diagrams.

Importantly, our compositional approach provides a setup in which quantum circuit components may be pretrained via classical simulation. In this manner, we avoid the trainability challenges posed by barren plateaus in conventional quantum machine learning (QML) \cite{ragone2023unified}. This is reminiscent of the variational compilation of a component that is repeatably used in a quantum algorithm \cite{Mc_Keever_2023,Kikuchi_2023}.
A quantum computer is then necessary only at test time to evaluate larger instances, since the classical resources for simulating large text circuits are expected to grow exponentially asymptotically, as we show in our detailed resource estimation analysis in \cref{sec:unsimulable}.  To demonstrate compositional generalisation on a real device beyond instance sizes that we have simulated classically for one of our datasets, we use Quantinuum's H1-1 trapped-ion quantum processor, which features state-of-the-art two-qubit-gate fidelity \cite{h11_data_sheet}.
As a comparison with classical baselines, we trained both transformer and LSTM models, and also tested on GPT-4~\cite{openai2023gpt4}. These models exhibited no signs of compositional generalization, performing on par with random guessing on larger stories than those used in training. 

As discussed in \cite{tkb}, other DisCoCirc-style models could be proposed, for example, based on classical neural networks, which would not require a quantum computer, even at test time. However, we argue that a quantum model is most natural for this framework, since DisCoCirc, via DisCoCat, is ultimately derived from pregroup grammars \cite{LambekBook}, and in this case we have the following result:

\begin{proposition}[{Informal, \cite[Theorem 2.1]{coecke2018uniqueness} \& \cite[Theorem 5.49]{CKbook}}]
    \em If a model has the same abstract structure as pregroup grammars, composition of spaces must be analogous to the tensor product.
\end{proposition}

Indeed, it has been argued that this is necessary to fully capture the semantics of language \cite{GrefSadr,CSC}. This justifies using a quantum model such as QDisCoCirc, as it represents the most expressive setup that can capture tensor product structure while remaining efficient to evaluate at test time.

The structure of this paper is as follows: in \cref{sec:QA-DisCo}, we introduce the QDisCoCirc model, the task of question answering, the `following' datasets we generated for this task, and outline our training methodology. Our results are presented in \cref{sec:CompGen}, both from simulation and from runs on real hardware. Additionally, we present a comparison with classical baselines in this section. \Cref{sec:interpretability} looks at the interpretability of our model. Finally, \cref{sec:conclusion} concludes with a discussion and outlines directions for future work.

\section{Question answering with QDisCoCirc}
\label{sec:QA-DisCo}

\begin{figure}[ht]
    \centering
    \makebox[\textwidth][c]{$$\input{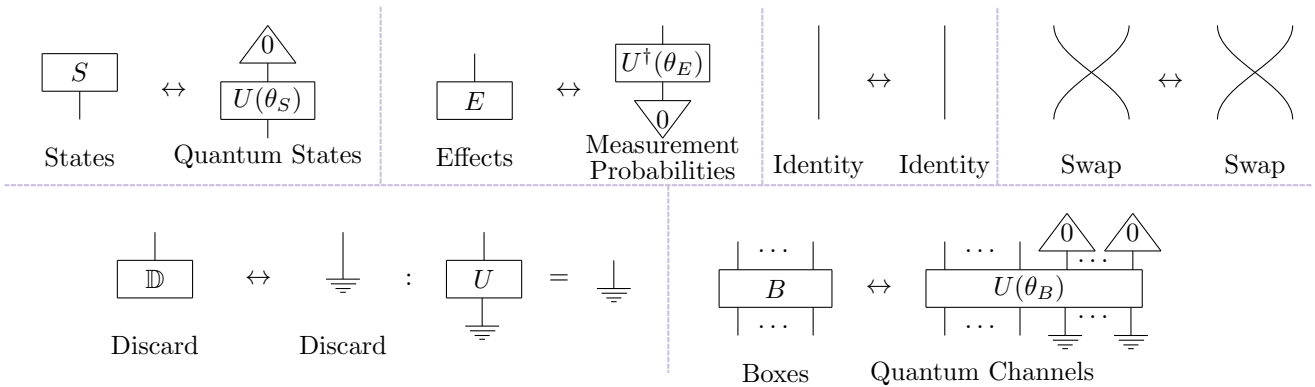}$$}
    \caption{Left-hand side of each subdivision: text generators, with which any text diagram is composed. There are three generators considered in this work: states, which encode inputs to a process, effects, which act as tests on states, and boxes, which transform states. We also consider special boxes, such as the identity, (the trivial operation) and the swap (the permutation of wires), as well as the special discard effect (the deletion of a wire). Right-hand side of each subdivision: the action of the semantic functor that instantiates the QDisCoCirc model used in this work. Every wire is assigned $n$ qubits. States are mapped to pure quantum states prepared by a unitary $U$ from $|0\rangle^{\otimes n}$. Effects are mapped to measurement probabilities of specific outcomes (state unpreparations, hence the dagger). The discard effect is interpreted as discarding the corresponding qubits (partial trace). Boxes, in general, are mapped to quantum channels, realised via a unitary, with potentially some ancillae that get discarded. The special boxes of identity and swap are mapped to the equivalent identity and swap unitary operations.}
    \label{fig:functor}
\end{figure}

\begin{figure}[ht]
    \centering
    $$\input{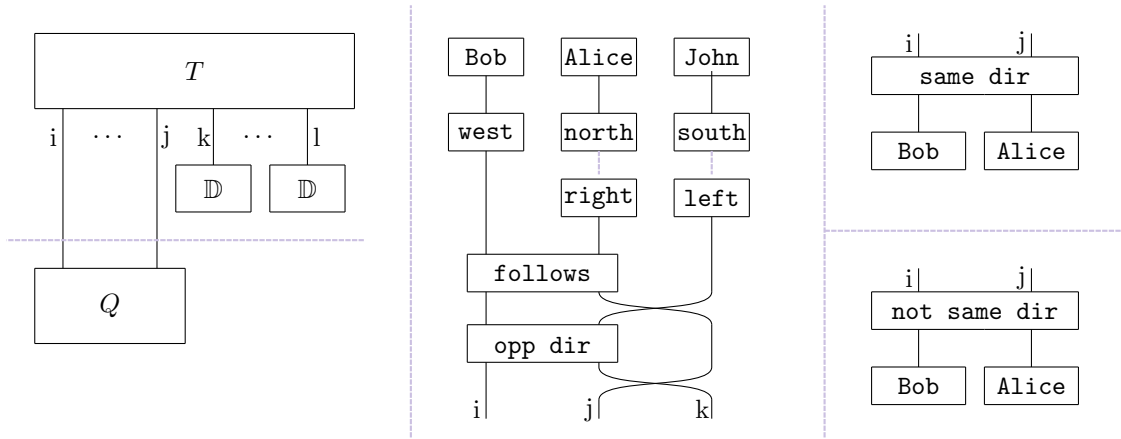}$$
    \caption{Left: Question answering as a process diagram in the DisCoCirc framework. A text-state $T$ prepares the state of affairs. A question-effect $Q$, which is posed as an affirmative statement, tests the extent to which the statement holds using the primitive of text-similarity (state overlap), as introduced in \cite{tkb}. Nouns irrelevant to the question are discarded. The horizontal dashed line is a guide to the eye and separates the text state from the question effect. Middle: An example text diagram from the `following' datasets that we have constructed for the story: \texttt{Bob walks west. Alice walks north. John walks south. Alice turns right. John turns left. Bob follows Alice. Bob goes in the opposite direction of John.} Here, noun-states (\texttt{Bob}, \texttt{Alice}, \texttt{John}) are transformed by boxes representing intransitive verbs (\texttt{walks west}, \texttt{walks north}, \texttt{walks south}, \texttt{turns right}, \texttt{turns left}) and transitive verbs (\texttt{follows}, \texttt{goes in the opposite direction of}). The diagram prepares the composite text state. Right: The question effects (\texttt{goes in the same direction as}, \texttt{does not go in the same direction as}) test the degree to which \texttt{Bob} and \texttt{Alice} satisfy those statements. 
    } 
    \label{fig:QAdiagram}
\end{figure}

\begin{figure}[ht]
    \centering
        $$\input{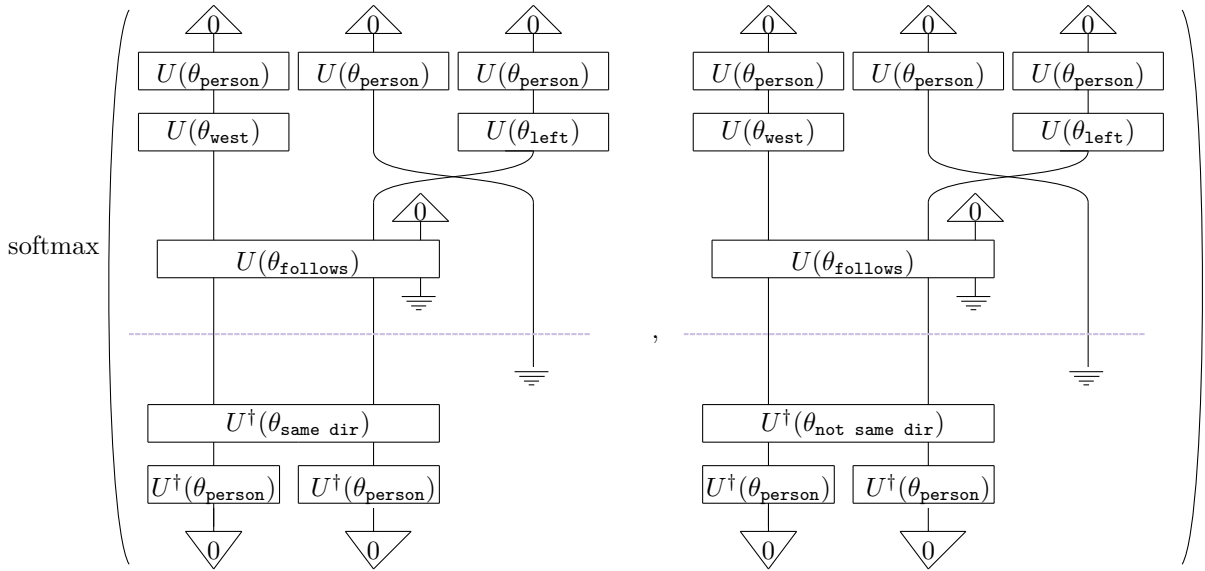}$$
    \caption{Our pipeline for question answering with QDisCoCirc. The effects for the positive and negative answer to the binary question are applied to the same quantum text state, using the primitive of text overlap (the dashed line separates the two texts whose overlap is being measured). Each diagram represents the probability of the all-zeros state being measured in the computational basis. Then, these two probabilities are passed to the softmax function to determine the answer to the question.}
    \label{fig:softmax_diags}
\end{figure}

\subsection{The QDisCoCirc model} \label{sec:qdiscocirc}

QDisCoCirc is a quantum model for text. It is constructed in the DisCoCirc framework by giving quantum semantics to text diagrams.
Text diagrams are read from top to bottom, and are composed of \emph{states}, \emph{effects}, and \emph{boxes},
which are to be understood as inputs, output \emph{tests}, and \emph{processes} that transform inputs into outputs \cite{coecke2020mathematics,wangmascianica2023distilling}.
We also consider special boxes, such as the `identity' and `swap' boxes, as well as the special `discard' effect (denoted by $\mathbb{D}$).
These constitute the \emph{generators} of text diagrams, and are displayed in \cref{fig:functor}.
A given sentence can be parsed into a text diagram, where each word is assigned a state or a box according to its part of speech \cite{liu2023pipeline}. Nouns are mapped to states, and verbs are mapped to boxes. Each state corresponds to a \emph{noun wire} (a line in the diagram connecting to the state). In this work, diagrams are constructed specifically for the chosen task. For simplicity, we represent expressions like e.g. \texttt{walks north, goes in the opposite direction of, goes in the same direction as}, etc. as single verb-like boxes; more general diagrams can be obtained using the DisCoCirc parser \cite{liu2023pipeline}. Sentences are composed sequentially according to the reading order of the text.

A QDisCoCirc model is instantiated via a \emph{semantic functor}, transforming text diagrams into quantum circuits.
This mapping is structure-preserving, as it is applied component-wise on the generators.
The specific functor chosen in this work maps
states to quantum states (which are prepared by unitary transformations from a fixed reference product state),
and effects to measurements (whose values are the probabilities of specific outcomes). %
The identity and swap boxes are mapped to the usual identity and swap gates in quantum circuits, and the discard effect is mapped to the discard of a quantum system, i.e. a partial trace.
Finally, boxes are mapped to quantum channels, which are realised by unitaries with ancilla qubits that get discarded.
The action of the semantic functor to the generators is shown in \cref{fig:functor}.

The construction of a QDisCoCirc model involves a number of hyperparameters, such as the number of qubits assigned to each wire, the parameterised quantum circuit ansatz used to implement the unitaries, and the ancillae that get discarded when implementing quantum channels.
For more details on the specific semantic functor we use in this work, and the specific ansatz we choose, see \cref{app:word-func}.

\subsection{The task of question answering}  \label{sec:question-answering}

Question answering can be represented natively in DisCoCirc via use of the primitive operation of \emph{text similarity}, as presented in \cite{tkb}. This works as follows: a text circuit states the facts, and the question is posed as an effect representing it as an affirmative statement.
The effect is applied to the noun wires relevant to the question, and the irrelevant nouns are discarded.
We show the structure of a text diagram for the question answering task in \cref{fig:QAdiagram}.

Using the semantic functor, we instantiate the question answering task as  quantum circuits. These are then evaluated and subjected to classical post-processing, yielding the answer to the question.
In particular, in this work, we focus on binary questions, which correspond to one effect for the positive answer and one effect for the negative answer.
Evaluating the two resulting circuits, we obtain two scalars. These are then passed to a classical nonlinearity, in our case a softmax function, to obtain the answer to the question, as shown in \cref{fig:softmax_diags}. We took this approach following the method set out in \cite{tkb}, but several alternatives exist. For instance, we could train an observable so that the measured value corresponds to the answer we wish to obtain, although this does not keep to the principle established in \cite{tkb} of modelling state overlaps as the overlaps in meaning between text circuits. By choosing our method instead, we obtain a more transparently interpretable model. We could also have measured only one effect (for instance, the affirmative statement), but we found that by using the maximum of two effects the Hilbert space can be split more evenly, making the task easier to learn.

It is important to note that, in practice, we approximate these effects using a finite number of measurement shots, and so it is possible that the measured outcome may not correspond to the `true' outcome (in the limit of infinitely many shots) -- indeed, in theory we would expect that exactly calculating which of the two effects has a higher probability is hard even for a quantum computer. In practice, we have found that only a few measurement shots are required.

\subsection{Generating the `following' datasets}\label{sec:datasets}
We generated two custom small-scale toy datasets, the \emph{two-directional} and the \emph{four-directional} datasets,
in order to perform our proof-of-concept experiments in a controlled setup, in the spirit of the bAbI datasets introduced in Ref.~\cite{bAbI}.
In one dataset, the texts describe \emph{actors} who initially walk in one of two cardinal directions; in the other, the actors walk in one of four cardinal directions.
Throughout the text, each actor can perform actions, which are described by intransitive verbs and transitive verbs that result in their directions changing. 
From this basic set of possible actions, we generated texts with varying numbers of actors and sentences. We refer to the number of actors in a text by \emph{text width} (or \emph{number of actors}), and the number of sentences in a text by \emph{text depth}.
For each text, we ask the binary question of whether two actors are going in the same direction.

\paragraph{Two-directional dataset:}
The actors can walk in two directions, north and south, and turn by $180^{\circ}$. The set of possible actions, i.e. verbs, in this dataset, is:
\begin{equation}\nonumber
        \{ \texttt{walks~north}, \hspace{6pt} \texttt{walks~south}, \hspace{6pt} \texttt{turns~around}, \hspace{6pt} \texttt{follows}, \hspace{6pt} \texttt{goes~in~the~opposite~direction~of}\}.
\end{equation}
\paragraph{Four-directional dataset:}
The actors can walk in all four cardinal directions, turn by $90^{\circ}$ and $180^{\circ}$. The set of possible actions in this dataset is:
\begin{equation}\nonumber
    \begin{aligned}
        \{ & \texttt{walks~north}, \hspace{6pt} \texttt{walks~south}, \hspace{6pt} \texttt{walks~east}, \hspace{6pt} \texttt{walks~west},\hspace{6pt}  \texttt{turns~right},\\ 
        &\texttt{turns~left}, \hspace{6pt} \texttt{turns~around},\hspace{6pt}  \texttt{follows},\hspace{6pt}  \texttt{goes~in~the~opposite~direction~of} \}.
    \end{aligned}
\end{equation}
The set of questions we use for both datasets is:
\begin{equation}\nonumber
    \begin{aligned}
        \{ & \texttt{goes in the same direction as}, \hspace{6pt} \texttt{does not go in the same direction as} \}.
    \end{aligned}
\end{equation}

\Cref{fig:QAdiagram} shows an example of a diagram from the four-directional dataset for an example story.
For both these datasets, we generated sub-datasets with different story densities.
The density of a story is related to the entanglement generated by the story, as interactions between actors (transitive verbs) are instantiated as entangling operations. We define the density of a story to be the number of two-actor interactions within the story divided by the number of sentences in the story and create the following five different subsets according to this definition.

\paragraph{simple, deeper:} These datasets are generated by randomly applying actions to actors until a chosen number of sentences is reached.  The simple dataset contains stories from 2 to 30 actors, and the deeper dataset contains stories from 6 to 30 actors. The stories in the deeper dataset also have more sentences than the stories in the simple dataset.

\paragraph{less dense, dense, superdense:} To ensure high connectivity of the nouns in these datasets, we first generate fully connected stories, where each actor interacts with every other actor in the story exactly once. Then, we add some single-actor actions, shuffle the sentences, and cut the story off after a chosen number of sentences. The proportion of single-actor actions decreases going from less-dense to superdense.
Like the deeper dataset, these datasets contain stories from 6 to 30 actors, and have the same number of sentences as the stories in the deeper dataset.

Further information, such as dataset sizes (\Cref{tab:dataset_sizes}), the average story densities (\Cref{tab:story_densities}), and data characterisation plots (\Cref{fig:datasets/2q-depth}), can be found in \cref{app:datasets}. To obtain diagrams of the form of \cref{fig:QAdiagram} from our text-level stories we implemented a parser, based on the one in \cite{liu2023pipeline}, using the \texttt{DisCoPy}~\cite{deFelice2021} package. Our parser maps the individual words to their respective boxes and arranges them in the structure set by the DisCoCirc model. We then use the \texttt{lambeq}~\cite{kartsaklis2021lambeq} package to apply the semantic functor explained in \cref{app:word-func} to the diagrams to generate the quantum circuits used in training.

\subsection{Training methodology}

The resulting quantum circuits are converted into tensor networks,
which are then evaluated exactly via tensor contraction using \texttt{Tensornetwork}~\cite{roberts2019tensornetwork}
on Quantinuum's \textsl{duvel4} server (see \cref{app:classical-resources} for specifications).
We use \texttt{PyTorch}~\cite{Paszke_PyTorch_An_Imperative_2019} to track the parameters and
the Adam optimizer~\cite{adam_kingma_2015} to train them. The hyper-parameters are tuned using \texttt{Ax}~\cite{bakshy2018, AxTuningRepo}.

For each training epoch, the model is evaluated on the entire training dataset in batches. We save the parameters obtained at the end of each epoch and record the loss and validation accuracy. Every three epochs, starting from the first, we evaluate and record the accuracy of the model on the training dataset\footnote{We do not do this every epoch in order to keep model training times down.}. For each training run, we select the model from the epoch with the best validation accuracy, tie-breaking if necessary with the closest logged training accuracy, then loss.

The hyperparameters we considered and their final tuned values, along with further details on the training methodology, are summarised in \cref{tab:hyperparams} in \cref{app:training}. In each case, we optimised for the best validation accuracy.
The tuning was conducted using the four-directional dataset. We did not do any further tuning on the two-directional dataset, as the model already achieved a high accuracy with the first hyperparameters we chose (shown in \cref{app:training}). 
\Cref{app:training} shows the cross-validation performed on the two-directional dataset.

\subsubsection{Train-Validation-Test split}
\label{sec:train-val-test-split}

\paragraph{Two-directional dataset:}
The model is trained on the simple dataset stories with up to 8 actors. $20\%$ of the simple dataset with up to 8 actors is used as validation data; we refer to this set as \textit{Valid A}. The stories from all other (deeper to superdense) datasets with up to 8 actors, and the stories from all datasets (simple to superdense) with 9 to 20 actors, are used as a second compositionality validation dataset, referred to as \textit{Valid Comp}. This set is used to pick the model with the best generalisation performance. Stories from all datasets with 21 to 30 actors, which compile down to 20 qubits or less with qubit reuse, form the test set. We simulate up to 20 actors and send a selection of circuits between 9 and 30 actors to Quantinuum's ion-trap quantum computer H1-1 (referred to as \textit{Valid Comp (H1-1)} and \textit{Test (H1-1)} depending on which dataset the instances were sampled from). See \cref{sec:H1-imp} for more details of the machine.

\paragraph{Four-directional dataset:}
The model is trained on all (simple to superdense) stories with up to 8 actors. $20\%$ of the dataset with up to 8 actors is used as validation data. The stories from all datasets (simple to superdense) with 9 to 20 actors are used as a second compositionality validation dataset (\emph{Valid Comp}), which is used to pick the model with the best generalisation performance.  Stories from all datasets with 21 to 30 actors, which compile down to 20 qubits or less with qubit reuse, form the test set. We select a subset of these circuits to evaluate using shot-based noisy simulation (referred to as \emph{Test (qujax)}).  \\

\Cref{tab:data_splits} shows a detailed overview of the splits used in our experiments.

\begin{table}[ht]
    \centering
    \begin{tabular}{| l | lll | lll | }
        \hline
        \rule[-0.5em]{0pt}{1.5em} \bf{Dataset} & \multicolumn{3}{|c|}{\bf{Two-directional}} & \multicolumn{3}{|c|}{\bf{Four-directional}} \\
         & Samples & Actors & Density & Samples & Actors & Density \\ 
        \hline
        \rule[-0.5em]{0pt}{1.5em} \bf{Train} 
            & 80\% & 2-8 & simple & 80\% & 2-8 & all \\ 
        \hline
        \rule[-0.5em]{0pt}{1.5em} \bf{Valid A}
            & 20\% & 2-8 & simple & 20\% & 2-8 & all \\  
        \hline
        \rule[-0.5em]{0pt}{1.5em} \bf{Valid Comp}
            & all & 9-20 & simple            & all & 9-20 & all \\
        \rule[-0.5em]{0pt}{1.5em} 
            & all & 6-20 & deeper-superdense & & & \\
        \hline
        \rule[-0.5em]{0pt}{1.5em} \bf{Valid Comp (H1-1)} 
            & 120 & 9-20 & dense      & & - &\\
        \rule[-0.5em]{0pt}{1.5em} 
            & 120 & 9-20 & superdense & & - &\\
        \hline
        \rule[-0.5em]{0pt}{1.5em} \bf{Test}
            & all & 21-30 & all & all & 21 - 30 & all \\
        \hline 
        \rule[-0.5em]{0pt}{1.5em} \bf{Test (H1-1)} 
            & 82 & 21-30 & dense      & & - & \\ %
        \rule[-0.5em]{0pt}{1.5em}  & 47 & 21-30 & superdense & & - & \\ %
        \hline
        \rule[-0.5em]{0pt}{1.5em} \bf{Test (qujax)
        } 
        &  & - &      & 86 & 21-30 & dense \\
        \rule[-0.5em]{0pt}{1.5em} &  & - &  & 45 & 21-30 & superdense 
        \\
        \hline
    \end{tabular}
    \caption{Detailed splits used in our experiments. A visual representation is provided in \cref{app:datasets} \cref{fig:datasets/shapes,fig:datasets/2q-depth-H1,fig:datasets/4dir-emulated-estimations}. }
    \label{tab:data_splits}    
\end{table}

\subsubsection{Semantic rewrites} 

\begin{figure}[ht]
    \centering
    $$\input{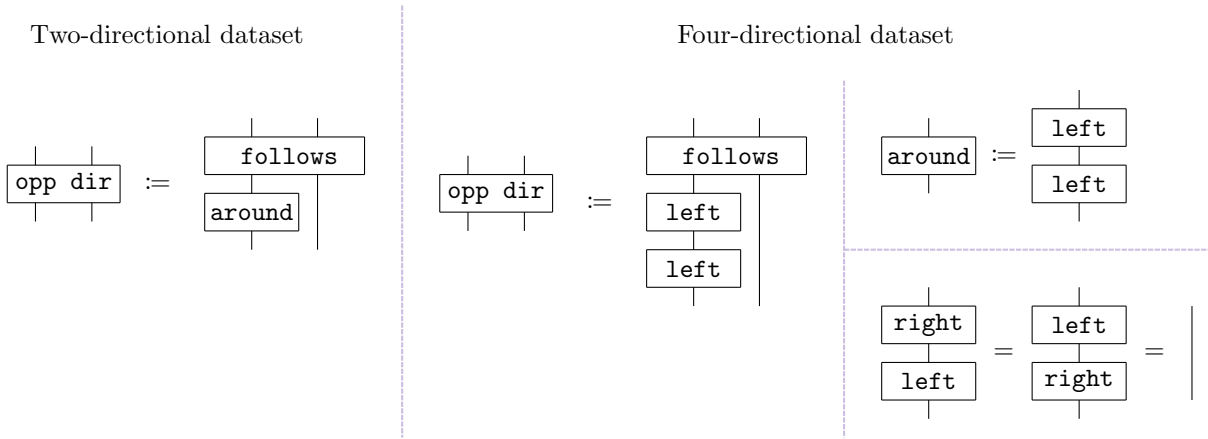}$$
    \caption{Hardcoded semantic rewrites for our model. For the two-directional dataset \texttt{goes in the opposite direction of} is implemented as a consecutive application of \texttt{follows} and \texttt{turns around} on the first actor. In the four-directional dataset we additionally replace \texttt{turns around} with two consecutive applications of \texttt{turns left}. Further, we enforce the consecutive application of \texttt{turns right} and \texttt{turns left} to be the identity.
      }
    \label{fig:axioms}
\end{figure}

To help the model learn a compositional solution, and reduce the number of parameters needed, we implemented \emph{semantic rewrites} into the diagrams. These rewrites are depicted in \cref{fig:axioms}, for both the two- and four-directional datasets.
The semantic functor defines the word circuits $U(\theta_\texttt{opp dir})$ and $U(\theta_\texttt{around})$ in terms of $U(\theta_\texttt{follows})$ and $U(\theta_\texttt{left})$.
Further, we impose \texttt{turns right} and \texttt{turns left} being inverse of each other by defining one circuit with respect to the other as $U(\theta_{\texttt{right}}) = U^\dagger(\theta_{\texttt{left}})$. 

\section{Compositional generalisation}
\label{sec:CompGen}

Having trained QDisCoCirc models for the two- and four-directional question answering datasets,
we can measure the degree to which the compositional structure of the model allows it to generalise to instances outside the training data distribution.
Specifically, we can quantify \emph{productivity}, i.e. whether a model trained on small examples can use the learned rules to correctly classify larger problem instances \cite{hupkes2020compositionality}.
In \Cref{app:compositionality} we explain how the other aspects of compositionality are manifest in the DisCoCirc formalism, but leave their exploration for future work.
We have chosen the number of actors
involved in the text as our metric of size, as this allowed us to easily quantify the worst-case resources required (see \cref{fig:resources/2-dir-summary}, and \cref{app:classical-resources} \cref{fig:resources/4-dir-summary}).
This is, on average, positively correlated with other size metrics, such as the text depth, and the circuit depth of the quantum circuits that compose the QDisCoCirc model. In \Cref{app:datasets}, we quantify this correlation.

Recall that the texts present in the \textit{Train} set include up to $8$ nouns; testing productivity entails observing whether the test accuracy for larger instances is correlated with the size of said instances. During the in-task training of the QDisCoCirc model, we used only exact tensor contractions.
We also used exact tensor contractions to obtain the \textit{Valid Comp} accuracy for instances featuring up to 20 nouns.
Pushing beyond 20 nouns, we use Quantinuum's H1-1 to evaluate accuracy on \textit{Test (H1-1)} for instances up to 30 nouns from the two-directional dataset, and used shot-based noisy simulation (as described in \cref{sec:noise_details}) to obtain the results for the four-directional \textit{Test (qujax)} dataset.

\begin{figure}[ht!]
    \centering

    \begin{tabular}{c}
        \includegraphics[width=\textwidth]{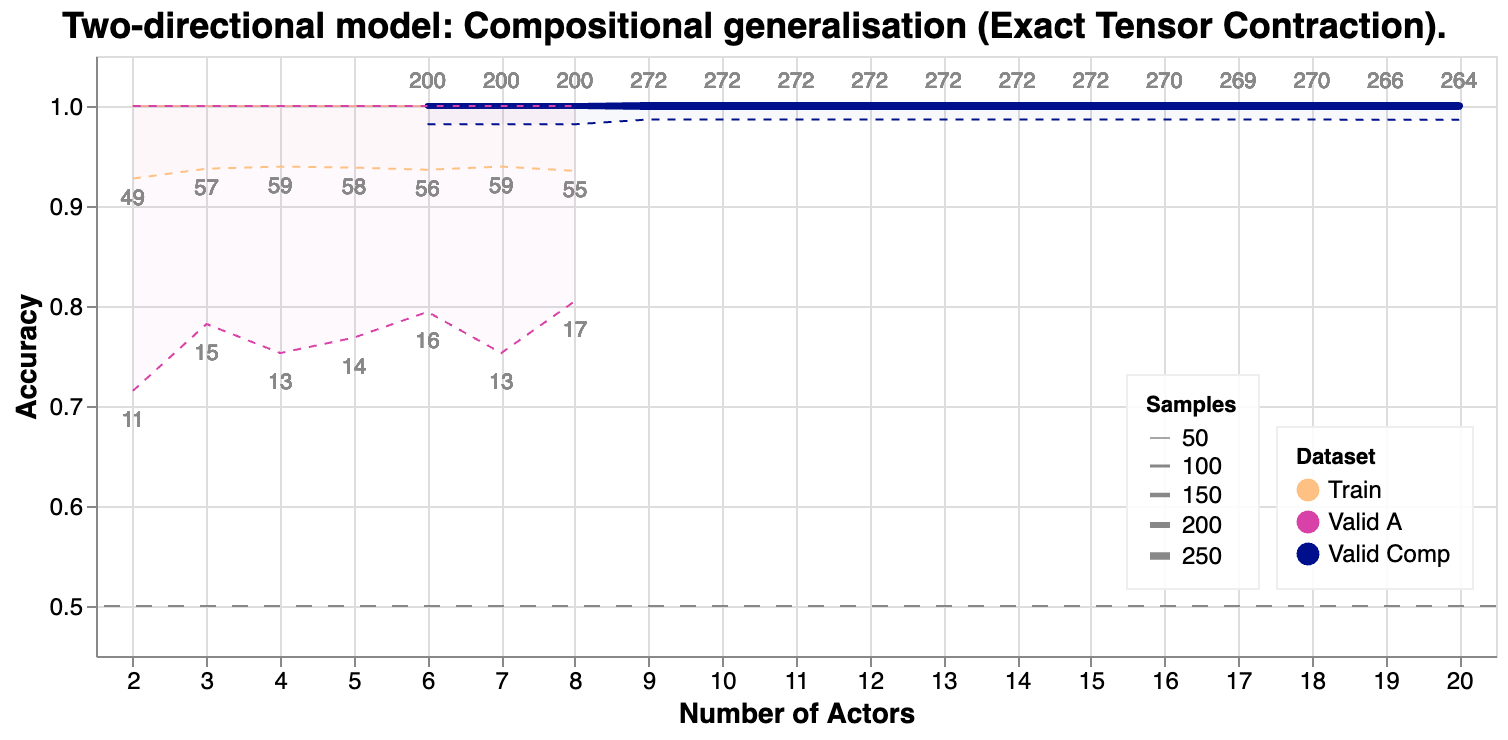}\\
        (a)\\
        \\
        \includegraphics[width=\textwidth]{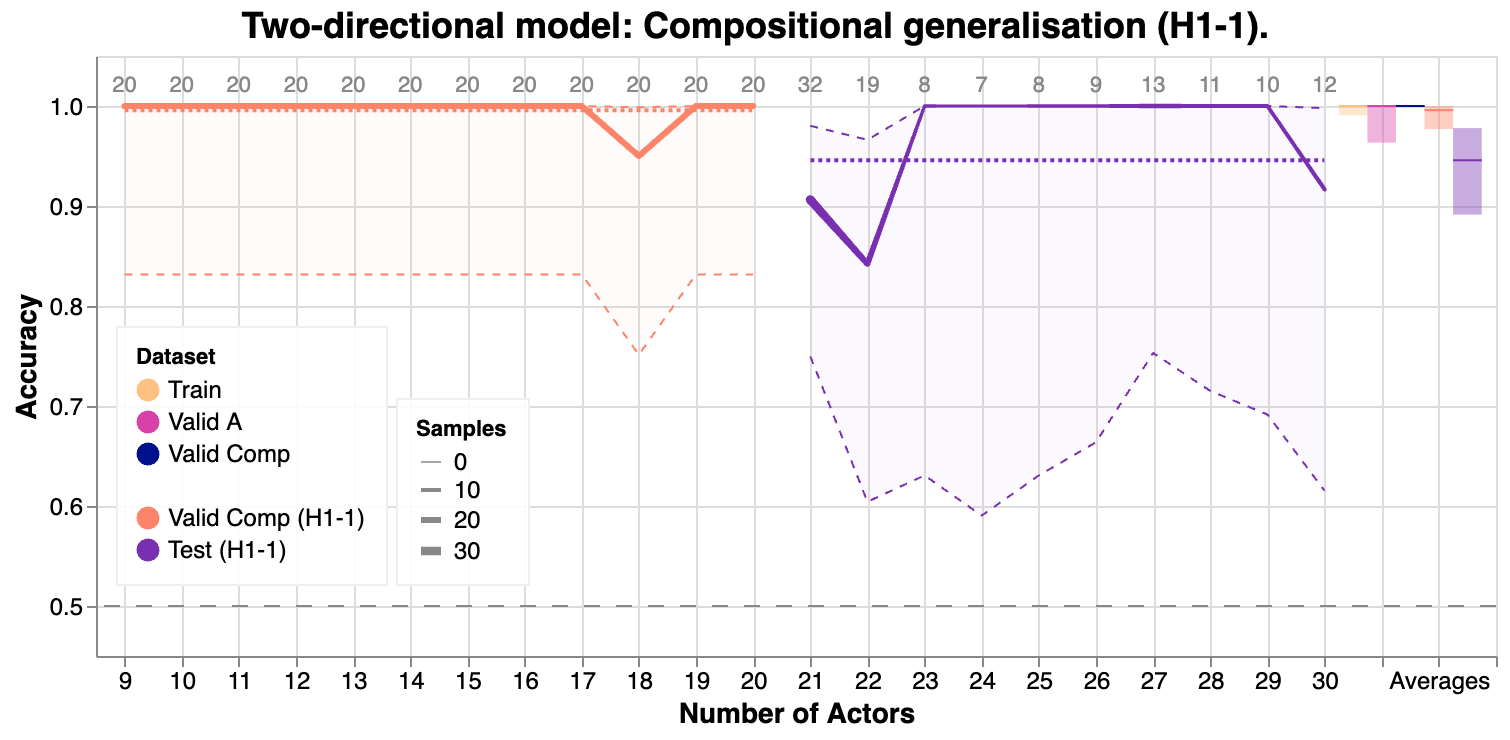}\\
        (b)
    \end{tabular}
    
    \caption{Compositional generalisation for the two-directional model. We label the total number of data points evaluated for each number of actors to contextualise the error bars. The error bars capture the 95\% Clopper-Pearson confidence interval \cite{clopper_pearson_1934} for the mean, taking into account the number of samples available. Results are plotted as lines with shaded error regions: a. Simulated datapoints, up to 20 actors. b. Datapoints evaluated on H1-1 (using $50$ shots per circuit). As there were only a few samples available for each number of actors, this leads to wider error margins. The dashed grey line represents the baseline obtained by random guessing. The average performance per dataset is shown as a dotted line for each split. The rightmost columns display the average performance for each dataset with shaded 95\% confidence intervals.
    }   
    \label{fig:compogen-2-dir}
\end{figure}

\begin{figure}[ht!]
    \centering
    \begin{tabular}{c}
    \includegraphics[width=\textwidth]{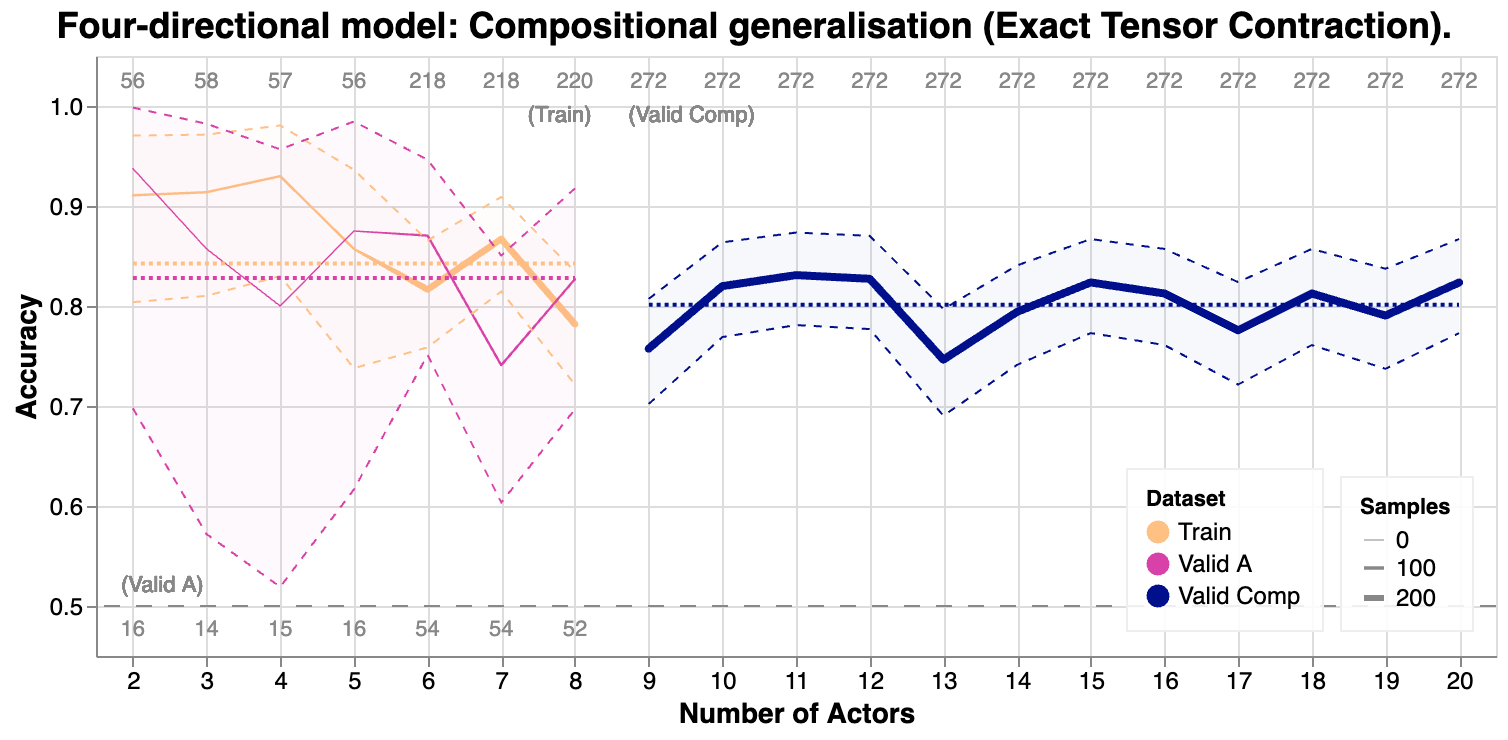}
    \\
    (a)
    \\
    \includegraphics[width=\textwidth]{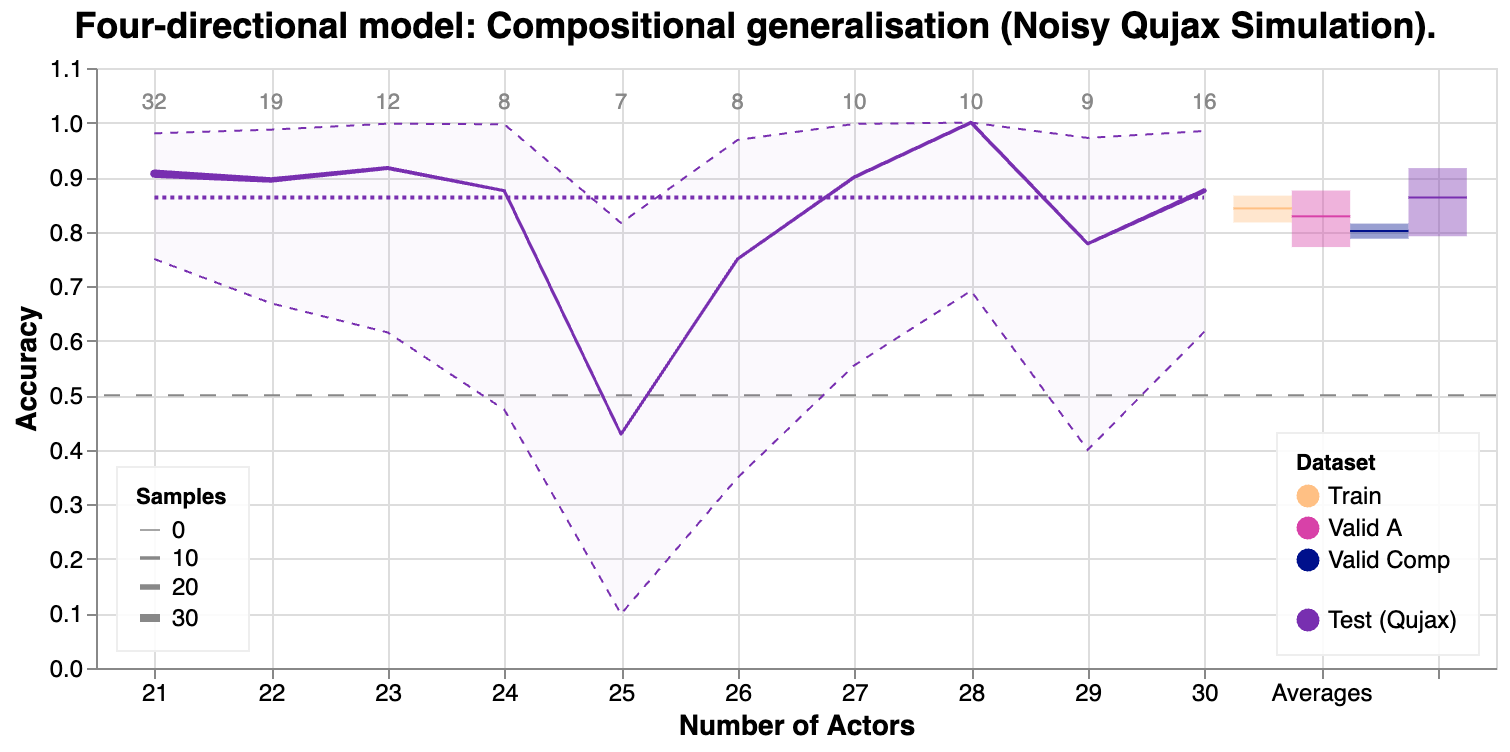}
    \\
    (b)
    \end{tabular}
    
    \caption{Compositional generalisation for the four-directional model. See \Cref{app:compogen}, \Cref{fig:compogen-4-dir-densities} for a breakdown of the generalisation curves per density. The shaded region represents the 95\% confidence interval for the mean, calculated using the Clopper-Pearson method. Note that for up to 6 actors, data points are from the simple dataset only, hence the reduced number of samples. We also plot the average accuracy across all actor widths for each dataset as a straight dotted line, with corresponding 95\% CI shaded on the rightmost column. a. Training and validation performance of the model. All datapoints shown were simulated exactly. b. Test performance of the model, simulated with noise using \texttt{qujax} with 1000 shots per circuit. The number of datapoints sampled for each number of nouns is recorded above the line.}
    \label{fig:compogen-4-dir}
\end{figure}

With statistical significance, the accuracy of the model on the \textit{Valid Comp} set does not decay for the two- and four-directional datasets in \cref{fig:compogen-2-dir} and \cref{fig:compogen-4-dir}. This also happens for the two-directional \textit{Test (H1-1)} sub-dataset in \cref{fig:compogen-2-dir}. For the four-directional \textit{Test (qujax)} sub-dataset in \cref{fig:compogen-4-dir}, we note that, although the accuracy and confidence intervals dip below the random guessing baseline for some noun widths, this can be attributed to the low number of sampled data points available for each width. Additionally, for the drop at 25 actors, approximately half of the samples selected were amongst those we expect the model to solve incorrectly a priori (see \Cref{sec:interpretability} for a discussion of this point), which is a much higher proportion than in the rest of the dataset.
We argue that this is experimental evidence that compositionality enables generalisation, as to perform inference we need only compose pre-trained components. In other words, we have provided experimental evidence that the quantum word circuits have learned effective compositional representations.

For comparative analysis, we trained both Transformer and LSTM models on both the two- and four-directional datasets to explore their capabilities in compositional generalisation. We maintained identical configurations for training, validation (\textit{Valid A}), and compositional validation (\textit{Valid Comp}) across all experiments, as detailed in \Cref{tab:data_splits}. Additionally, we included test evaluations of GPT-4 using both the two- and four-directional datasets. These models exhibited no signs of compositional generalization, performing on par with random guesses in stories with increased noun widths. This poor performance may be attributed to the constraints inherent in drawing direct comparisons with QDisCoCirc. Detailed methodologies, data analyses, and discussions of these results, as well as suggestions for future research, are documented comprehensively in \Cref{app:baselines}.

\subsection{Implementation on H1-1}
\label{sec:H1-imp}

We now present in more detail the experimental methodology and results from the execution of the question-answering task for the two-directional dataset on Quantinuum's H1-1 quantum computer~\cite{h11_experiment, h11_data_sheet}.
H1-1 is a state-of-the-art ion-trap quantum computer supporting $20$ physical qubits, reaching $0.998$ two-qubit gate fidelity and featuring any-to-any gate connectivity. This connectivity allows for any text circuit, which will in general be non-local, to be compiled without additional overhead. Further, the circuit ansatz we have chosen in Section~\ref{sec:qdiscocirc} is hardware efficient, in the sense that its entangling gates can be efficiently expressed in terms of the native gate set of H1-1 (see \cref{sec:ansatz_to_hardware}).

As discussed in \cref{sec:train-val-test-split}, the \textit{Valid Comp (H1-1)} dataset (a randomly sampled balanced subset from the \textit{Valid Comp} dataset) was selected to be evaluated on the H1-1 device. The \textit{Test (H1-1)} set is formed from a further $82$ data points sampled from the `dense' \textit{Test} set and $47$ data points from the `superdense' \textit{Test} set -- these circuits were those that compiled down to 20 or fewer qubits with qubit reuse, and hence would fit on the device. Each circuit was repeated for a total of $50$ shots. Note that each data point corresponds to two circuits: one for the positive question and another for the negative question. The circuits were first converted to \texttt{TKET}~\cite{tket}, before undergoing a qubit reuse compilation pass using the local greedy algorithm described in~\cite{qubit_reuse}. This pass attempts to reduce the number of qubits used in the circuit by identifying those which do not need to be kept for the full computation, and then reusing them where new qubits would need to be introduced. The result is a reduction in the number of qubits necessary to execute the circuit, in exchange for a higher number of measurement errors (which occur when resetting the qubits to be reused) and a deeper circuit. 
The increase in the circuit depth resulting from qubit reuse is shown in \cref{fig:datasets/2q-depth} (all datasets) and \cref{fig:datasets/2q-depth-H1} (subset sent to H1-1) of \cref{app:datasets}, while the resulting reduction in the number of qubits needed to execute the circuits is shown in \cref{fig:datasets/qubit-reuse} (all datasets) and Fig.~\ref{fig:datasets/qubit-reuse-H1} (subset sent to H1-1) of \cref{app:datasets}.

The results of evaluating the circuits on the H1-1 device are displayed in Fig.~\ref{fig:compogen-2-dir}. As a result of the qubit reuse described above, we were able to compile circuits exceeding the $20$ qubit limit of the H1-1 device down to this threshold, including some circuits with up to 30 actors. The largest of these circuits originally had 108 qubits. This large number of qubits arises mainly from the ancillas used for certain words, as each actor only takes one qubit in our model. For a discussion on the use of ancilla qubits see \cref{app:word-func}. As discussed in \cref{sec:CompGen}, we see that there is no significant drop in the model accuracy when the number of actors increases beyond what was used during training, confirming that the model successfully generalises to longer texts. This also shows that the H1-1 device noise levels do not corrupt this compositional generalisation for the text lengths considered. This agrees with the results in \cref{sec:noise}, where we numerically simulate an evaluation of the model on the \textit{Valid Comp} set using a noise model that emulates the native noise of the H1-1 device.

\subsection{Towards classically unsimulable instances}
\label{sec:unsimulable}

Finally, since we have demonstrated that productivity provides a scaffold for constructing circuits that solve the question-answering task for the two datasets we have created, it is natural to investigate at which instance size a quantum computer is necessary for executing the circuits involved.
This requires the quantification of the classical resources necessary to simulate the quantum circuits, as well as the quantification of the effect of noise in the circuits to the output of the model.
This is because the presence of noise can be used to reduce the cost of classical simulation using approximate simulation methods \cite{Aharonov_2023,fontana2023classical}.

Here, we present a thorough study of the classical resources necessary for simulating the circuits involved in the question-answering task for the dataset we have introduced in this work.
We consider the cost of exact tensor contraction for establishing a reasonable upper bound on the simulation cost that is less naive than state vector simulation.
Further, we perform noisy simulations of the circuits using the simple noise model of a depolarising channel to quantify the effect of noise on the accuracy of the task at hand.

\subsubsection{Tensor contraction cost estimates}

\begin{figure}[ht]
    \centering
    \includegraphics[width=0.9\textwidth]{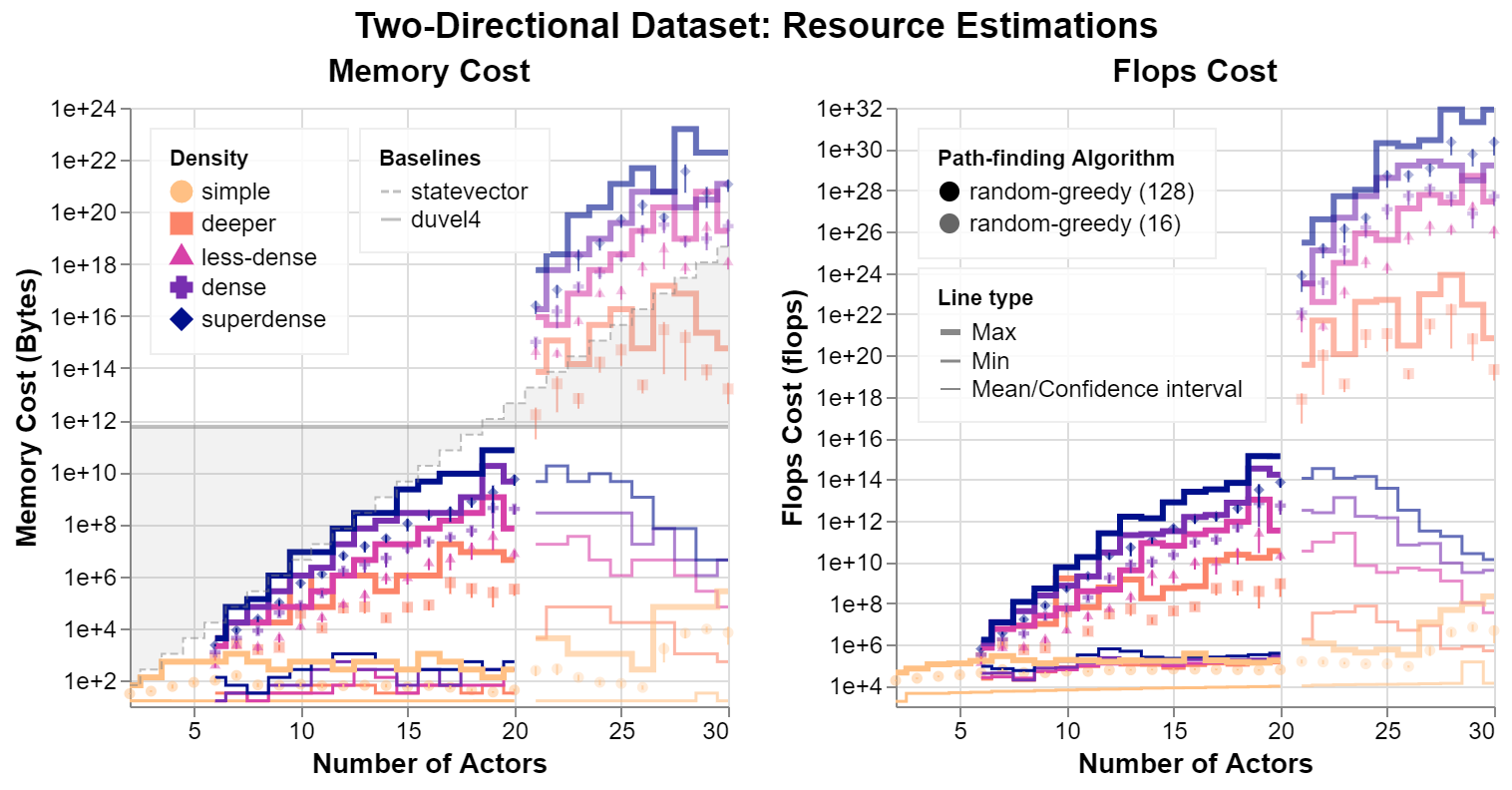}
    \caption{Resource estimates for the two-directional dataset. For each density, we plot the required memory and FLOPS (floating-point operations per second) for the exact evaluation of each data point as a full tensor contraction. In each case, the thickest line shows the most expensive contraction, while the thin line gives the lower bound. The mean for each number of actors is plotted as a point, with a vertical line showing the 95\% confidence interval for the mean. Fewer iterations were used to compute the paths for data points with more than 20 actors, to keep the time spent reasonable (see App.~\ref{app:classical-resources} for further details). We show two baselines for the memory cost to contextualise this data. To train and evaluate the models we used Quantinuum's \textsl{duvel4} server, which has 512GB of memory. The second baseline reflects performing a statevector-style simulation of the circuit. We assume the statevector requires at most $n + 1$ qubits, that is: one qubit per noun, $n$ nouns, and a single reused ancilla qubit. The circuit must be evaluated as a doubled state (or density matrix) due to the presence of discards. The total memory cost is then $2^{2(n + 1)}$. 
    }
    \label{fig:resources/2-dir-summary}
\end{figure}

We provide resource estimates for the exact simulation of our circuits using tensor networks, which are one of the most competitive methods for simulating quantum systems exactly~\cite{Or_s_2019,Gray_2021}. The tensor contraction paths are computed using \texttt{Opt-Einsum}~\cite{Smith2018}. We use the randomised greedy optimiser, picking the best path from a fixed number of repeats. In App.~\ref{app:classical-resources} we provide a detailed study of the diminishing returns in path quality versus the number of path samples chosen, which justifies the choice of this path finding heuristic.
Instances that we simulated were evaluated using \texttt{Tensornetwork}~\cite{roberts2019tensornetwork} using the saved contraction paths.
\cref{fig:resources/2-dir-summary} displays the exact tensor contraction costs for the two-directional dataset. It shows an exponential scaling of the space and time costs for the exact classical simulation of the text circuits involved in our question-answering task. The same data for the four-directional dataset can be found in \cref{fig:resources/4-dir-summary} of \cref{app:classical-resources}.

\subsubsection{The effects of noise} \label{sec:noise}
\begin{figure}[ht]
    \centering
    \includegraphics[width=0.9\textwidth]{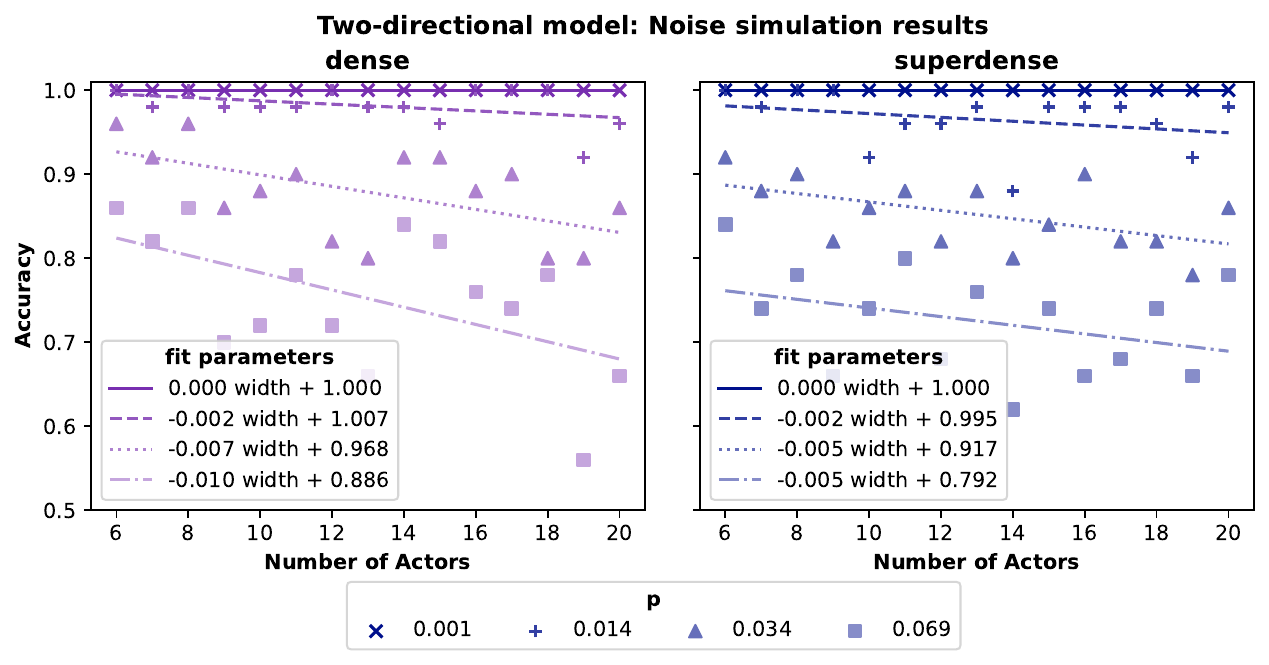}
    \caption{\emph{Valid Comp} accuracy for the two directional dataset, stratified by number of actors, evaluated numerically using the depolarising noise levels listed in the figure legend, where $p$ is as in \cref{eq:noise_strength}. We see that when the noise level models that of the H1-1 device ($p \approx 0.001$), no drop in accuracy occurs. As the noise increases, there is a decrease in the accuracy with the number of actors, reflecting the fact that more actors are correlated with deeper circuits. This shows a direct correspondence between noise present in the device and the maximum admissible text depth, which indicates that classical simulation techniques that leverage robustness to noise do not scale as the length of the text increases. Simulations were run using a Monte-Carlo sampling approach as described in \cref{sec:noise_details}.}
    \label{fig:noise_simulations}
\end{figure}

Finally, we quantify the effect of noise on the \textit{Valid Comp} accuracy of the model, and examine how its influence on the model's predictions depends on the number of actors that the texts feature.
To this end, we employ a noise model based on the emulator~\cite{h11_emulator} for Quantinuum's H1-1 ion-trap quantum computer, which is applied after the circuits have been compiled down to the device's native gate set. We focus our noise investigations on the two-directional dataset, since this is the one we ran on the H1-1 device.

For simplicity, we only consider a symmetric depolarising 2-qubit gate noise, which affects the device's $R_{ZZ} = e^{-i\theta Z_jZ_{k}}$ ($j\neq k$) gates with probability
\begin{align} \label{eq:noise_strength}
    P_{2q}(\theta) = \left(a \frac{|\theta|}{\pi}^c + b\right) p, \quad 
\end{align}
where $
    a = 1.651,
    b = 0.175,
    c = 1$, $
    p = s p_0$,
    $p_0 = 1.38 \times 10^{-3}$, and $s$ is a scaling factor parameter that we vary to control the noise level. We noramalise the angles to be in the interval $[-\pi, \pi)$.
Note that $P_{2q}$ depends on the angle of rotation used, and that $p=p_0, s=1$ corresponds to the noise level of the H1-1 device. The simulations were run using \texttt{qujax}'s~\cite{qujax} statevector simulator by following a Monte-Carlo sampling approach (see App.~\ref{sec:noise_details} for more details.)

In \cref{fig:noise_simulations}, we plot how the accuracy varies with increasing circuit depth for the different noise scaling factors, where we consider $s=1,10,25,50$. As the noise probability increases, we observe a decrease in the model's accuracy with the number of actors. This is because the number of actors in a text and the depth of the corresponding circuit are correlated (\cref{app:datasets} \cref{fig:datasets/shapes}), and the effect of noise on deeper circuits is more pronounced. For larger noise levels, there is an approximately linear decay of the accuracy; this decay is expected to plateau at $0.5$, indicating the model's performance has degraded to effectively random guessing.
Note that the two-qubit gate noise present in H1-1 is not enough to cause the model's accuracy to decrease for the number of actors considered in this noise study, i.e. up to 20 actors.

It is interesting to investigate the noise threshold below which the model's accuracy is not correlated with text size for a given task and dataset.
Classical simulation techniques that leverage the presence of noise (such as those in e.g.~\cite{Noh2020}) would then aim to inject as much noise as possible to reduce the classical resources for simulating the circuits,
while not exceeding the threshold.
It is beyond the scope of this work to determine for which text sizes the classical resources outweigh the quantum resources for achieving the same test accuracy.
Note that resources can be measured either in the number of primitive operations, or even time-to-solution, as the time per operation differs vastly between quantum and classical computers as they stand today.

\section{Compositional interpretability}
\label{sec:interpretability}

The built-in compositional structure of QDisCoCirc allows us to interpret its behaviour.
We begin by studying how each component in the quantum representation of the texts behave. Then, we use its inherent compositional structure to piece together how the model performs on our datasets. We confirm our analysis by examining how well the axioms we expect to hold are respected. In \cref{sec:comp_interpret_2_dir}, we observe that the two-directional model performs well due to behaving similarly to an ideal model we construct by hand in \cref{app:interpretability}. In \cref{sec:comp_interpret_4_dir}, we explain how the four-directional model does not perfectly correspond to this ideal model, but rather approximates it, performing well nonetheless. For it to correspond to an ideal model, the four-directional model would require two qubits per wire instead of our choice of one qubit per wire (made due to hardware constraints).

Throughout this section, we make use of the following representations:
\begin{itemize}
    \item Single-qubit states are visualised on a Bloch sphere in the standard way, where pure states lie on the surface of the sphere, and mixed states lie inside the sphere.
    \item Following~\cite{Altepeter_2009}, two-qubit states $\rho_{1, 2}$ are  visualised as coloured ellipsoid surfaces inside a pair of Bloch spheres in terms of partial projections.
In this representation, each sphere corresponds to one of the qubits.
The first Bloch sphere displays a surface defined by $\rho_{1,\psi} = (\mathbb{I} \otimes \langle\psi|)\rho_{1,2}(\mathbb{I} \otimes |\psi\rangle)$ for all pure states $|\psi\rangle$ that the second qubit can be in.
Each state $\rho_{1,\psi}$ is coloured as per the reference colour of $|\psi\rangle$, shown in Fig.~\ref{fig:interpret/2dir/directions-states}a.
The second Bloch sphere similarly displays a surface describing the state $\rho_{\psi, 2}$ of the second qubit given that the first qubit is in state $|\psi\rangle$.
\item We visualise single-qubit gates as rotations on the Bloch sphere, plotting the trajectory of reference points $|\phi\rangle \in \{\ket{0}$, $\ket{1}$, $\ket{+}$, $\ket{-}$, $\ket{i}$, $\ket{-i}\}$, under the action of the rotation $U$. The final state $U|\phi\rangle$ is coloured according to the colour of the initial state $\ket\phi$, as given in Fig.~\ref{fig:interpret/2dir/directions-states}a.
\end{itemize}

\subsection{Two-directional dataset}
\label{sec:comp_interpret_2_dir}
\begin{figure}[ht]
    \centering
    \includegraphics[width=\textwidth * 2 / 4]{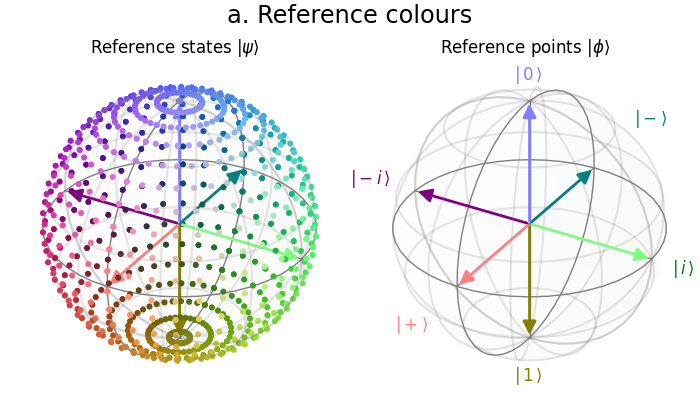}
    \\
    \includegraphics[width=\textwidth * 3 / 4]{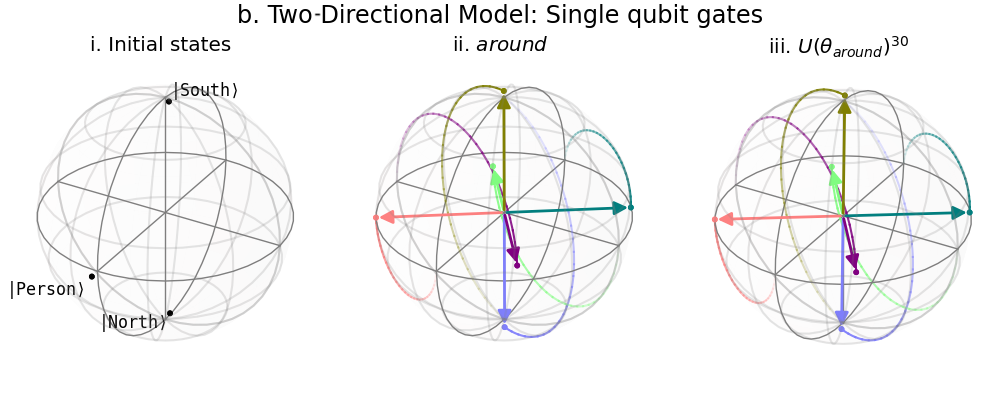}

    \caption{
    a. Colour mapping used for paired Bloch sphere pure states $\ket\psi$. The reference points $\ket{0}, \ket{1}, \ket{+}, \ket{-}, \ket{i}, \ket{-i}$ are highlighted as vectors.
    b. Single qubit gate analysis for the two directional model:
    i. Single qubit initialisations, viewed as states. The two directions are aligned with the computational basis.
    ii. The effect of the single qubit rotation \texttt{turns around} on the reference points.
    iii. Applying \texttt{turns around} 30 times resembles applying it just once. 
    }
    \label{fig:interpret/2dir/directions-states}
\end{figure}

\paragraph{Initial States}
Since these only ever occur at the start of a text, they can be represented as states (rather than rotations). Note that the state labelled $\ket{\texttt{North}}$ represents the state obtained from the entire sentence \texttt{Person walks north}, and similarly for $\ket{\texttt{South}}$, as this is how it is always encountered in the text. These are depicted in Fig.~\ref{fig:interpret/2dir/directions-states}b.i.

\paragraph{Single-qubit gates}
In Fig.\ref{fig:interpret/2dir/directions-states}b.ii, we plot the single-qubit rotation \texttt{turns around}. The gate is approximately a $\pi$ rotation about the x-axis (up to a small phase). Since the directions $\ket{\texttt{North}}$ and $\ket{\texttt{South}}$ are approximately on the poles of the sphere, the expected axioms also hold approximately (see App.~\ref{app:interpretability} Fig.~\ref{fig:interpret/2dir-1q-axioms}). Since the rotation is only \textit{approximately} $\pi$, there is a small offset that can accumulate if \texttt{turns around} is applied successively - Fig.~\ref{fig:interpret/2dir/directions-states}b.iii. shows the effect of applying \texttt{turns around} thirty times in a row, demonstrating that an extra $\pi$ rotation has been accumulated: thirty applications resemble a single application, whilst we would expect that turning around an even number of times should be equivalent to not turning around at all.
A perfect model (as described in App.~\ref{app:interpretability} Fig.~\ref{fig:interpretability/clifford-2dir}) would not exhibit such behaviour, however the model has learnt to accommodate this error, as will be shown in the next section.

\paragraph{Two-qubit gates}
\begin{figure}[ht!]
    \centering
    \begin{tabular}{c}
        \includegraphics[scale=0.5]{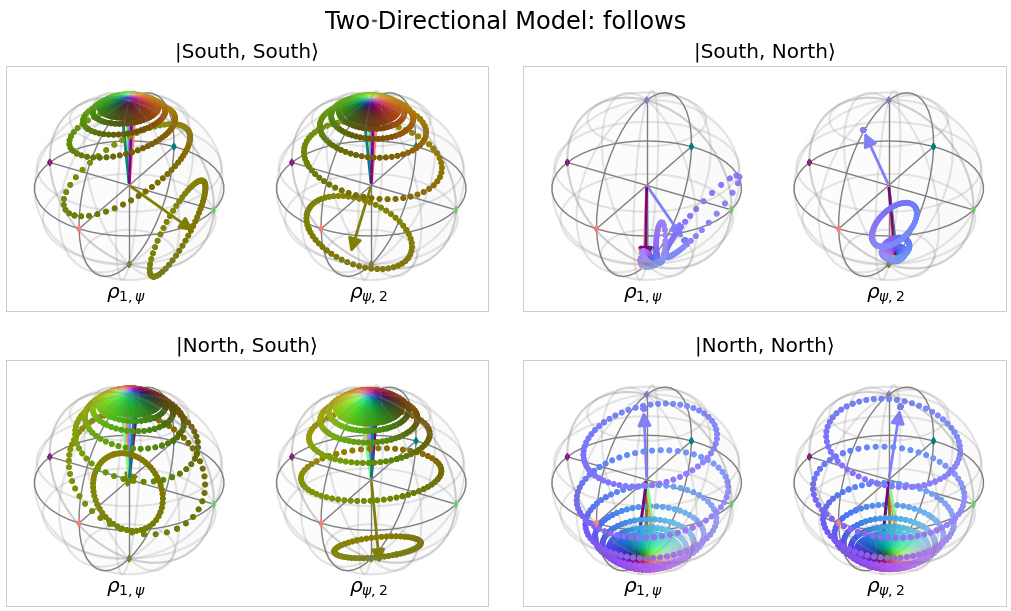} 
        \\ (a) \\
        \\
        \includegraphics[scale=0.5]{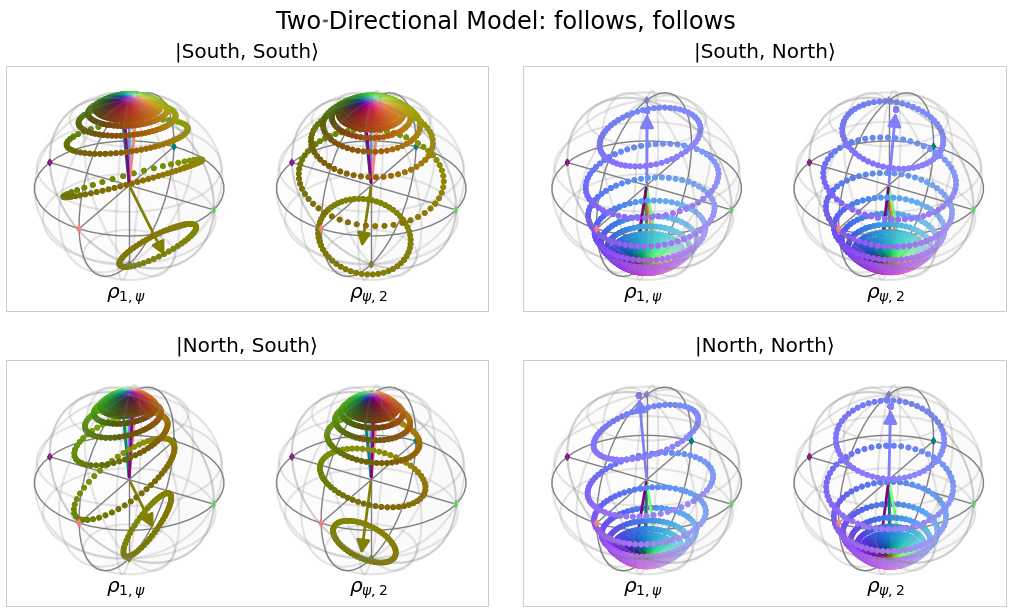}
        \\ (b)
    \end{tabular}
    $$\begin{tikzpicture}
	\begin{pgfonlayer}{nodelayer}
		\node [style=none] (194) at (22.5, -2.5) {$=$};
		\node [style=none] (195) at (19.75, -2.5) {\texttt{follows}};
		\node [style=none] (196) at (18.25, -2) {};
		\node [style=none] (197) at (21.25, -2) {};
		\node [style=none] (198) at (21.25, -3) {};
		\node [style=none] (199) at (18.25, -3) {};
		\node [style=none] (260) at (19, -2) {};
		\node [style=none] (261) at (19, -1.5) {};
		\node [style=none] (262) at (20.5, -2) {};
		\node [style=none] (263) at (20.5, -1.5) {};
		\node [style=none] (264) at (20.5, -3) {};
		\node [style=none] (265) at (20.5, -3.5) {};
		\node [style=none] (266) at (19, -3) {};
		\node [style=none] (267) at (19, -3.5) {};
		\node [style=none] (268) at (25, -1.75) {\texttt{follows}};
		\node [style=none] (269) at (23.5, -1.25) {};
		\node [style=none] (270) at (26.5, -1.25) {};
		\node [style=none] (271) at (26.5, -2.25) {};
		\node [style=none] (272) at (23.5, -2.25) {};
		\node [style=none] (273) at (24.25, -1.25) {};
		\node [style=none] (274) at (24.25, -0.75) {};
		\node [style=none] (275) at (25.75, -1.25) {};
		\node [style=none] (276) at (25.75, -0.75) {};
		\node [style=none] (277) at (25.75, -2.25) {};
		\node [style=none] (278) at (25.75, -2.75) {};
		\node [style=none] (279) at (24.25, -2.25) {};
		\node [style=none] (280) at (24.25, -2.75) {};
		\node [style=none] (281) at (25, -3.25) {\texttt{follows}};
		\node [style=none] (282) at (23.5, -2.75) {};
		\node [style=none] (283) at (26.5, -2.75) {};
		\node [style=none] (284) at (26.5, -3.75) {};
		\node [style=none] (285) at (23.5, -3.75) {};
		\node [style=none] (290) at (24.25, -4.25) {};
		\node [style=none] (291) at (24.25, -3.75) {};
		\node [style=none] (292) at (25.75, -4.25) {};
		\node [style=none] (293) at (25.75, -3.75) {};
	\end{pgfonlayer}
	\begin{pgfonlayer}{edgelayer}
		\draw (199.center) to (198.center);
		\draw (198.center) to (197.center);
		\draw (197.center) to (196.center);
		\draw (196.center) to (199.center);
		\draw (261.center) to (260.center);
		\draw (263.center) to (262.center);
		\draw (264.center) to (265.center);
		\draw (266.center) to (267.center);
		\draw (272.center) to (271.center);
		\draw (271.center) to (270.center);
		\draw (270.center) to (269.center);
		\draw (269.center) to (272.center);
		\draw (274.center) to (273.center);
		\draw (276.center) to (275.center);
		\draw (277.center) to (278.center);
		\draw (279.center) to (280.center);
		\draw (285.center) to (284.center);
		\draw (284.center) to (283.center);
		\draw (283.center) to (282.center);
		\draw (282.center) to (285.center);
		\draw (291.center) to (290.center);
		\draw (293.center) to (292.center);
	\end{pgfonlayer}
\end{tikzpicture}$$

    \caption{Visualising the two-qubit \texttt{follows} gate for the two-directional model to confirm axiom \ref{eq:follows-twice-axiom}. The label $\ket{x, y}$ identifies the state $U(\theta_{\texttt{follows}})\ket{x, y}$. The expected joint final state is $\ket{y, y}$. Note that we mark the reference points with a coloured dot on the paired Bloch spheres as a visual aid; these are not to be considered part of the plotted ellipsoids.
    a. Single application of the \texttt{follows} gate. b. \texttt{follows} applied twice.
    }
    \label{fig:interpret/2dir/following}
\end{figure}
\begin{figure}[ht!]
    \captionsetup{labelformat=adja-page}
    \ContinuedFloat
    \caption{
    Reading the plot: 
    a. Consider the paired Bloch sphere for $\ket{\texttt{South, South}}$ for a single application of \texttt{follows}. The left-most sphere shows that $\rho_{1, \psi} \approx \ket{0} \approx \ket{\texttt{South}}$ for $\ket\psi \not\approx \ket{1} \approx \ket{\texttt{North}}$, and vice-versa for $\rho_{\psi, 2}$. This suggests that the pair of qubits is `mostly in the $\ket{00} \approx \ket{\texttt{South, South}}$ state'.
    Although somewhat mixed, many of the points are close to the surface of the sphere, suggesting that the qubits are entangled (though not maximally so). 
    This aligns with the interpretation that \texttt{follows} acts as an approximate projection of the first qubit onto the state of the second (as measured in the $\{\ket{\texttt{North}},\ket{\texttt{South}}\}$ basis), with an additional entangling component that allows for $\rho_{1, \texttt{North}} \approx \ket{\texttt{North}}$ (whereas we would expect a perfect projection to give $\rho'_{1, \texttt{North}} \approx \ket{\texttt{South}}$).
    The paired state for $\ket{\texttt{North, South}}$ is very similar, whilst $\ket{\texttt{North, North}}$ is instead concentrated towards the $\ket{11} \approx \ket{\texttt{North, North}}$ state. For $\ket{\texttt{South, North}}$, we can see the state as being even more concentrated towards $\ket{\texttt{North, North}}$, to the point where the state is almost separable.
    b. The paired Bloch spheres describing the application of \texttt{follows} twice are visually quite similar to those describing \texttt{follows} once, however the colours are rotated slightly along the vertical axis, representing an induced phase shift. In the $\ket{\texttt{South, North}}$ case, there is a more significant difference, as the second application of \texttt{follows} results in a more entangled state, though in both cases the final state remains approximately $\ket{\texttt{North, North}}$. We thus conclude that axiom \ref{eq:follows-twice-axiom} holds at least approximately.
    }
    \label{fig:interpret/2dir/following}
\end{figure}

As our ansatz contains some built-in semantic rewrites, there is only one two-qubit gate: \texttt{follows}. We consider its action on a set of states defined by the possible directions the two people involved might be initially facing. Since $\ket{\texttt{North}}$ and $\ket{\texttt{South}}$ are approximately orthogonal, this acts as a basis, and so investigating the action of \texttt{follows} on each of these initial states fully characterises the gate.
Depicted in Fig.~\ref{fig:interpret/2dir/following}, the states considered encode texts of the form:
\begin{equation}
    \texttt{Person}_1 \texttt{ walks }x\texttt{. Person}_2\texttt{ walks }y\texttt{. Person}_1\texttt{ follows Person}_2\texttt{.}
\end{equation}
for $x, y \in \{\texttt{North}, \texttt{South}\}$.
The gate behaves largely as a joint projection of the qubits onto the $\ket{\texttt{North}}$ - $\ket{\texttt{South}}$ axis, controlled by the state of qubit 2. The joint aspect of this projection entangles the qubits in such a way that some certainty as to the original directions faced is lost, however since the state of $\texttt{Person}_1$ remains correlated with the new location of $\texttt{Person}_2$, and largely independent of the original direction $\texttt{Person}_1$ was facing, the \texttt{follows} gate seems to have correctly captured the notion of \texttt{following}.

We also visually confirm that the axiom 
\begin{equation}\label{eq:follows-twice-axiom}
    \texttt{Person}_1 \texttt{ follows Person}_2\texttt{. Person}_1\texttt{ follows Person}_2\texttt{.} = \texttt{Person}_1\texttt{ follows Person}_2\texttt{.}
\end{equation}
approximately holds, by comparing the visualisations for both gates. Applying following a second time induces a slight phase shift, but the state remains at the relevant pole with high probability.

We also note that the projective aspect of \texttt{follows} can correct for some accumulated errors from \texttt{turns around}. Although we would expect even powers of \texttt{turns around} to be approximately the identity, we find that $U^{30}(\theta_{\texttt{around}}) \approx U(\theta_{\texttt{around}})$, as demonstrated in Fig.~\ref{fig:interpret/2dir/directions-states}b.iii. Provided \texttt{follows} is interspersed between \texttt{turns around} sufficiently frequently, the states will be projected back towards the pole they are closest to, thus clearing the error accumulated so far. Sufficiently frequently in this case means at most 15 \texttt{turns around} gates have been applied in succession, as from this point the probability of projecting to the correct state will dip below 50\%.

\paragraph{Questions}
Both question effects are maximally entangled, approximately corresponding to two states from the Bell basis. Written in terms of our initial states, we have
$$|\texttt{same dir}\rangle \approx \frac{1}{\sqrt{2}} (\ket{\texttt{North}} \otimes \ket{\texttt{North}} + \ket{\texttt{South}} \otimes \ket{\texttt{South}})$$
$$|\texttt{opp dir}\rangle \approx \frac{1}{\sqrt{2}} (\ket{\texttt{South}} \otimes \ket{\texttt{North}} + \ket{\texttt{North}} \otimes \ket{\texttt{South}})$$

which intuitively captures the meaning of these questions (see Fig.~\ref{fig:interpret/questions} in App.~\ref{app:interpretability} for visualisations).

\subsection{Four-directional dataset}
\label{sec:comp_interpret_4_dir}
\begin{figure}[ht]
    \centering
    \includegraphics[width=\textwidth]{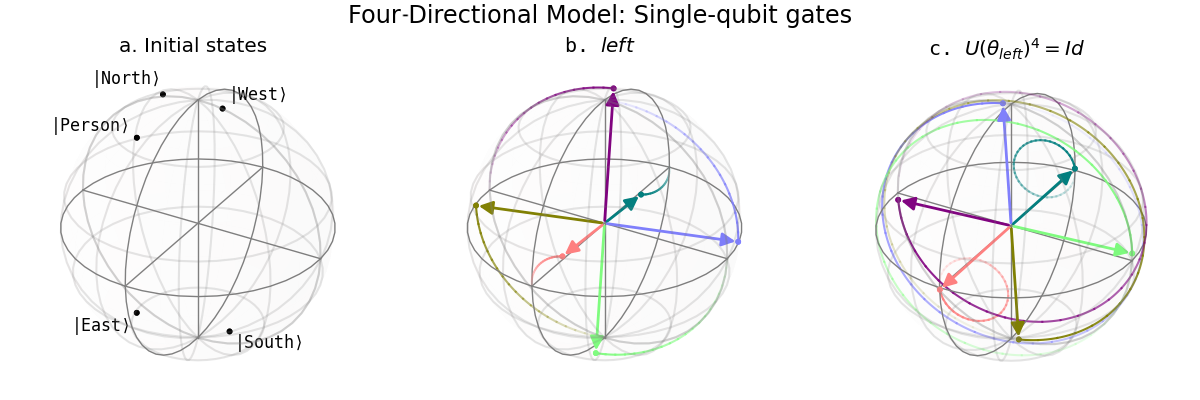}
    $\begin{tikzpicture}
	\begin{pgfonlayer}{nodelayer}
		\node [style=none] (194) at (22, -4) {$=$};
		\node [style=none] (262) at (20.5, -6.25) {};
		\node [style=none] (263) at (20.5, -1.5) {};
		\node [style=none] (268) at (25, -1.75) {\texttt{left}};
		\node [style=none] (269) at (23.5, -1.25) {};
		\node [style=none] (270) at (26.5, -1.25) {};
		\node [style=none] (271) at (26.5, -2.25) {};
		\node [style=none] (272) at (23.5, -2.25) {};
		\node [style=none] (273) at (25, -1.25) {};
		\node [style=none] (274) at (25, -0.75) {};
		\node [style=none] (277) at (25, -2.25) {};
		\node [style=none] (278) at (25, -2.75) {};
		\node [style=none] (281) at (25, -3.25) {\texttt{left}};
		\node [style=none] (282) at (23.5, -2.75) {};
		\node [style=none] (283) at (26.5, -2.75) {};
		\node [style=none] (284) at (26.5, -3.75) {};
		\node [style=none] (285) at (23.5, -3.75) {};
		\node [style=none] (292) at (25, -4.25) {};
		\node [style=none] (293) at (25, -3.75) {};
		\node [style=none] (294) at (25, -4.75) {\texttt{left}};
		\node [style=none] (295) at (23.5, -4.25) {};
		\node [style=none] (296) at (26.5, -4.25) {};
		\node [style=none] (297) at (26.5, -5.25) {};
		\node [style=none] (298) at (23.5, -5.25) {};
		\node [style=none] (301) at (25, -5.25) {};
		\node [style=none] (302) at (25, -5.75) {};
		\node [style=none] (303) at (25, -6.25) {\texttt{left}};
		\node [style=none] (304) at (23.5, -5.75) {};
		\node [style=none] (305) at (26.5, -5.75) {};
		\node [style=none] (306) at (26.5, -6.75) {};
		\node [style=none] (307) at (23.5, -6.75) {};
		\node [style=none] (308) at (25, -7.25) {};
		\node [style=none] (309) at (25, -6.75) {};
	\end{pgfonlayer}
	\begin{pgfonlayer}{edgelayer}
		\draw (263.center) to (262.center);
		\draw (272.center) to (271.center);
		\draw (271.center) to (270.center);
		\draw (270.center) to (269.center);
		\draw (269.center) to (272.center);
		\draw (274.center) to (273.center);
		\draw (277.center) to (278.center);
		\draw (285.center) to (284.center);
		\draw (284.center) to (283.center);
		\draw (283.center) to (282.center);
		\draw (282.center) to (285.center);
		\draw (293.center) to (292.center);
		\draw (298.center) to (297.center);
		\draw (297.center) to (296.center);
		\draw (296.center) to (295.center);
		\draw (295.center) to (298.center);
		\draw (301.center) to (302.center);
		\draw (307.center) to (306.center);
		\draw (306.center) to (305.center);
		\draw (305.center) to (304.center);
		\draw (304.center) to (307.center);
		\draw (309.center) to (308.center);
	\end{pgfonlayer}
\end{tikzpicture}$
    \caption{Visualising single qubit gates for the four-directional model: a. Single qubit initialisations. b. Single qubit rotation \texttt{turns left} - the other rotations are defined in terms of \texttt{turns left}, see App.~\ref{app:interpretability} for an explicit visualisation. c. \texttt{turns left} four times is (approximately) the identity.
    }
    \label{fig:interpret/4dir/directions}
\end{figure}

\paragraph{Initial states}
For four directions, we have two additional inital states, $\ket{\texttt{East}}$ and $\ket{\texttt{West}}$, shown in Fig.~\ref{fig:interpret/4dir/directions}a. Like the two-directional model, this model places $\ket{\texttt{North}}$ and $\ket{\texttt{South}}$ at approximately orthogonal locations. $\ket{\texttt{East}}$ and $\ket{\texttt{West}}$ are similarly orthogonal, however these two `bases' are not maximally distinct from each other: $\ket{\texttt{North}}$ is somewhat close to $\ket{\texttt{West}}$, and $\ket{\texttt{South}}$ and $\ket{\texttt{East}}$ are similar.

\paragraph{Single-qubit gates}
Though there are three single qubit gates in the text, the semantic rewrties built into the ansatz express each of these in terms of \texttt{turns left} only. \texttt{turns left} is approximately a $\pi / 2$ rotation about the x-axis as shown in in Fig.~\ref{fig:interpret/4dir/directions}(b), so we can derive that \texttt{turns around} is approximately a $\pi$ rotation, and \texttt{turns right} a $3\pi/2$ rotation about the same axis. The derived turns are depicted in App.~\ref{app:interpretability} Fig.~\ref{fig:interpret/4dir/1q-rotations}.

The single-qubit axioms hold less well in the four directions model, due to the fact that the directions are clustered into pairs. Fig.~\ref{fig:interpret/4dir/directions}(c) shows that \texttt{turns left}  has the expected period of (approximately) 4. The axioms concerning \texttt{turns around} mostly hold, but not the axiom establishing the quarter-turns between the initial directions, as per App.~\ref{app:interpretability} Fig.~\ref{fig:interpret/4dir-1q-axioms}. The favouring of \texttt{turns around} can be explained by its featuring in the decomposition of \texttt{goes in the opposite direction}, as this will bias the dataset towards the occurrence of \texttt{turns around} compared to \texttt{turns left} or \texttt{turns right}. As such, the optimisation will tend to favour parameters where \texttt{turns around} interacts correctly since this will have a greater impact on the accuracy of the model.

\paragraph{Two-qubit gates}
We have only one two qubit gate to consider: \texttt{follows}. We additionally consider initial states including the two new directions.
Like the two directions model, the four directions model projects the state of $\texttt{Person}_1$ onto a state matching $\texttt{Person}_2$ independent of the initial state of $\texttt{Person}_1$, this time with significantly less entanglement. Since such a projection can only occur along a single axis, we see an even clearer pairing of the directions, such that the model is effectively collapsing into a two-directions model over states we shall call \texttt{North-West} ($\ket{\texttt{NW}}$) and \texttt{South-East}  ($\ket{\texttt{SE}}$).

\begin{figure}[H]
    \centering
    \begin{tabular}{c}
         \includegraphics[scale=0.47]{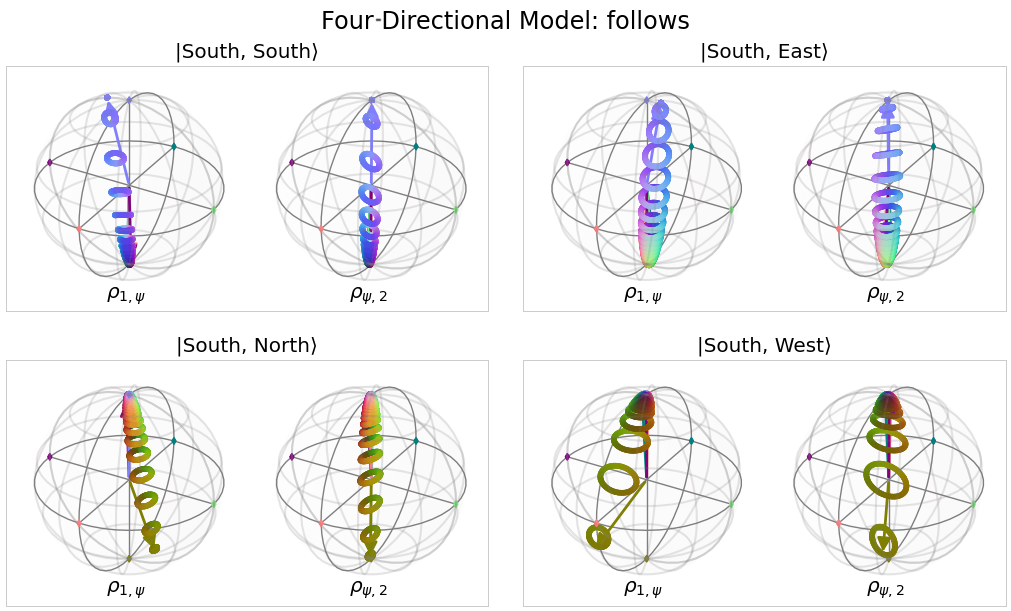} \\
         (a) \\
         \includegraphics[scale=0.47]{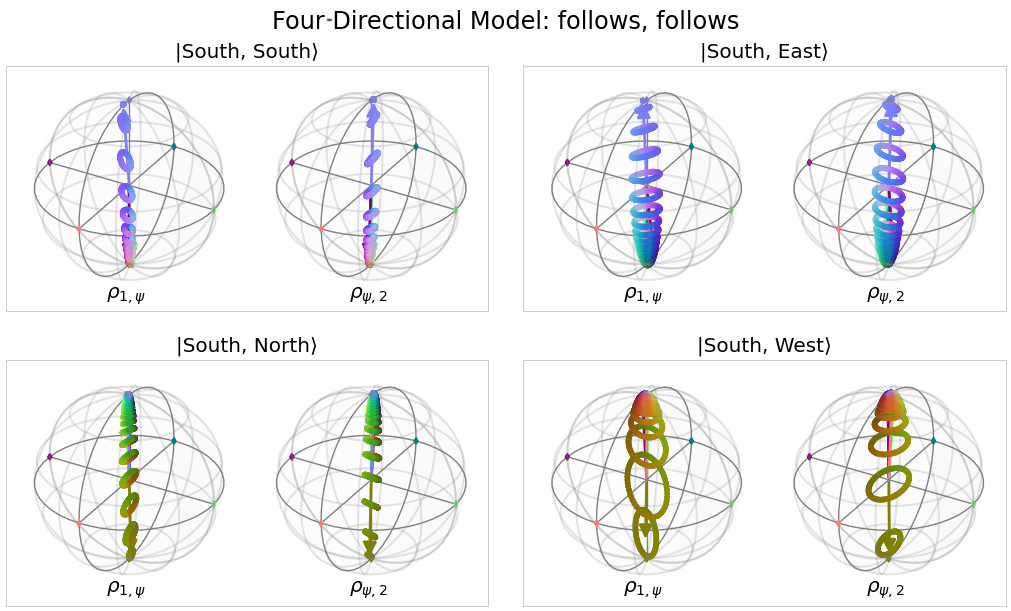}
         \\ (b)
    \end{tabular}
    $$\begin{tikzpicture}
	\begin{pgfonlayer}{nodelayer}
		\node [style=none] (194) at (22.5, -2.5) {$=$};
		\node [style=none] (195) at (19.75, -2.5) {\texttt{follows}};
		\node [style=none] (196) at (18.25, -2) {};
		\node [style=none] (197) at (21.25, -2) {};
		\node [style=none] (198) at (21.25, -3) {};
		\node [style=none] (199) at (18.25, -3) {};
		\node [style=none] (260) at (19, -2) {};
		\node [style=none] (261) at (19, -1.5) {};
		\node [style=none] (262) at (20.5, -2) {};
		\node [style=none] (263) at (20.5, -1.5) {};
		\node [style=none] (264) at (20.5, -3) {};
		\node [style=none] (265) at (20.5, -3.5) {};
		\node [style=none] (266) at (19, -3) {};
		\node [style=none] (267) at (19, -3.5) {};
		\node [style=none] (268) at (25, -1.75) {\texttt{follows}};
		\node [style=none] (269) at (23.5, -1.25) {};
		\node [style=none] (270) at (26.5, -1.25) {};
		\node [style=none] (271) at (26.5, -2.25) {};
		\node [style=none] (272) at (23.5, -2.25) {};
		\node [style=none] (273) at (24.25, -1.25) {};
		\node [style=none] (274) at (24.25, -0.75) {};
		\node [style=none] (275) at (25.75, -1.25) {};
		\node [style=none] (276) at (25.75, -0.75) {};
		\node [style=none] (277) at (25.75, -2.25) {};
		\node [style=none] (278) at (25.75, -2.75) {};
		\node [style=none] (279) at (24.25, -2.25) {};
		\node [style=none] (280) at (24.25, -2.75) {};
		\node [style=none] (281) at (25, -3.25) {\texttt{follows}};
		\node [style=none] (282) at (23.5, -2.75) {};
		\node [style=none] (283) at (26.5, -2.75) {};
		\node [style=none] (284) at (26.5, -3.75) {};
		\node [style=none] (285) at (23.5, -3.75) {};
		\node [style=none] (290) at (24.25, -4.25) {};
		\node [style=none] (291) at (24.25, -3.75) {};
		\node [style=none] (292) at (25.75, -4.25) {};
		\node [style=none] (293) at (25.75, -3.75) {};
	\end{pgfonlayer}
	\begin{pgfonlayer}{edgelayer}
		\draw (199.center) to (198.center);
		\draw (198.center) to (197.center);
		\draw (197.center) to (196.center);
		\draw (196.center) to (199.center);
		\draw (261.center) to (260.center);
		\draw (263.center) to (262.center);
		\draw (264.center) to (265.center);
		\draw (266.center) to (267.center);
		\draw (272.center) to (271.center);
		\draw (271.center) to (270.center);
		\draw (270.center) to (269.center);
		\draw (269.center) to (272.center);
		\draw (274.center) to (273.center);
		\draw (276.center) to (275.center);
		\draw (277.center) to (278.center);
		\draw (279.center) to (280.center);
		\draw (285.center) to (284.center);
		\draw (284.center) to (283.center);
		\draw (283.center) to (282.center);
		\draw (282.center) to (285.center);
		\draw (291.center) to (290.center);
		\draw (293.center) to (292.center);
	\end{pgfonlayer}
\end{tikzpicture}$$
    
    \caption{Visualising the effect of the two qubit \texttt{follows} gate of the four-directional model. For brevity, we consider its application to each of the possible initialisations for the second actor, fixing the first actor to always walk south. See Fig.~\ref{fig:interpret/4dir/following-full} in App.~\ref{app:interpretability} for the full plot. 
    The label $\ket{x, y}$ identifies the state $U(\theta_{\texttt{follows}})\ket{x, y}$. The expected joint final state is $\ket{y, y}$.
    The paired Bloch spheres for the four directional model are very similar to the two directional model, except that the surfaces are much thinner. This suggests that the \texttt{follows} gate is acting much more like a projection, keeping the final states relatively un-entangled (though they remain classically correlated).
    We verify the (approximate) idempotence of the two qubit \texttt{follows} gate. Comparing the result of applying \texttt{follows} twice (b), to applying \texttt{follows} only once (a), we see that a phase shift is applied, though the shape of the states are very similar.
    }
    \label{fig:interpret/4dir/following}
\end{figure}

\begin{figure}[ht]
    \centering
    \includegraphics[width=\textwidth]{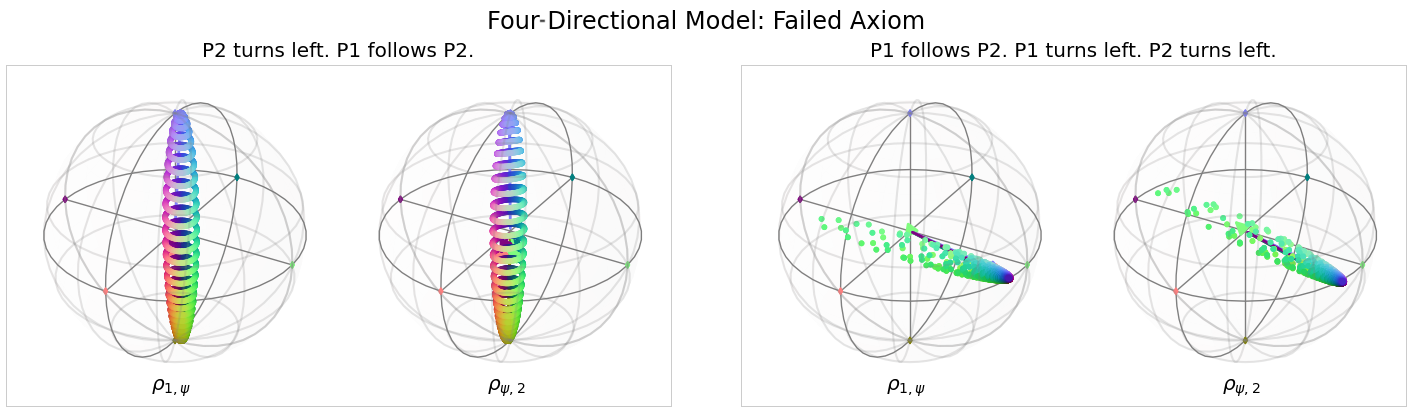}
    $$\begin{tikzpicture}
	\begin{pgfonlayer}{nodelayer}
		\node [style=none] (194) at (29.25, -2.25) {$=$};
		\node [style=none] (249) at (26.25, -3) {\texttt{follows}};
		\node [style=none] (250) at (24, -2.5) {};
		\node [style=none] (251) at (28.25, -2.5) {};
		\node [style=none] (252) at (28.25, -3.5) {};
		\node [style=none] (253) at (24, -3.5) {};
		\node [style=none] (254) at (25, -3.5) {};
		\node [style=none] (258) at (27.25, -3.5) {};
		\node [style=none] (259) at (27.25, -4) {};
		\node [style=none] (268) at (25, -2.5) {};
		\node [style=none] (269) at (25, -0.5) {};
		\node [style=none] (270) at (27.25, -2.5) {};
		\node [style=none] (271) at (27.25, -2) {};
		\node [style=none] (323) at (27.25, -1.5) {\texttt{left}};
		\node [style=none] (324) at (26.25, -1) {};
		\node [style=none] (325) at (28.25, -1) {};
		\node [style=none] (326) at (28.25, -2) {};
		\node [style=none] (327) at (26.25, -2) {};
		\node [style=none] (328) at (27.25, -0.5) {};
		\node [style=none] (329) at (27.25, -1) {};
		\node [style=none] (330) at (25, -4) {};
		\node [style=none] (331) at (32.5, -1.5) {\texttt{follows}};
		\node [style=none] (332) at (30, -1) {};
		\node [style=none] (333) at (34.75, -1) {};
		\node [style=none] (334) at (34.75, -2) {};
		\node [style=none] (335) at (30, -2) {};
		\node [style=none] (336) at (31, -2) {};
		\node [style=none] (337) at (33.75, -2) {};
		\node [style=none] (339) at (31, -1) {};
		\node [style=none] (340) at (31, -0.5) {};
		\node [style=none] (342) at (33.75, -3.5) {};
		\node [style=none] (343) at (33.75, -3) {\texttt{left}};
		\node [style=none] (344) at (32.75, -2.5) {};
		\node [style=none] (345) at (34.75, -2.5) {};
		\node [style=none] (346) at (34.75, -3.5) {};
		\node [style=none] (347) at (32.75, -3.5) {};
		\node [style=none] (349) at (33.75, -2.5) {};
		\node [style=none] (351) at (31, -3.5) {};
		\node [style=none] (352) at (31, -3) {\texttt{left}};
		\node [style=none] (353) at (30, -2.5) {};
		\node [style=none] (354) at (32, -2.5) {};
		\node [style=none] (355) at (32, -3.5) {};
		\node [style=none] (356) at (30, -3.5) {};
		\node [style=none] (357) at (31, -2.5) {};
		\node [style=none] (358) at (33.75, -1) {};
		\node [style=none] (359) at (33.75, -0.5) {};
		\node [style=none] (360) at (31, -4) {};
		\node [style=none] (361) at (33.75, -4) {};
	\end{pgfonlayer}
	\begin{pgfonlayer}{edgelayer}
		\draw (253.center) to (252.center);
		\draw (252.center) to (251.center);
		\draw (251.center) to (250.center);
		\draw (250.center) to (253.center);
		\draw (258.center) to (259.center);
		\draw (269.center) to (268.center);
		\draw (271.center) to (270.center);
		\draw (327.center) to (326.center);
		\draw (326.center) to (325.center);
		\draw (325.center) to (324.center);
		\draw (324.center) to (327.center);
		\draw (328.center) to (329.center);
		\draw (254.center) to (330.center);
		\draw (335.center) to (334.center);
		\draw (334.center) to (333.center);
		\draw (333.center) to (332.center);
		\draw (332.center) to (335.center);
		\draw (340.center) to (339.center);
		\draw (347.center) to (346.center);
		\draw (346.center) to (345.center);
		\draw (345.center) to (344.center);
		\draw (344.center) to (347.center);
		\draw (356.center) to (355.center);
		\draw (355.center) to (354.center);
		\draw (354.center) to (353.center);
		\draw (353.center) to (356.center);
		\draw (359.center) to (358.center);
		\draw (336.center) to (357.center);
		\draw (337.center) to (349.center);
		\draw (342.center) to (361.center);
		\draw (351.center) to (360.center);
	\end{pgfonlayer}
\end{tikzpicture}$$
    
    \caption{Demonstration that axiom \ref{eq:left-follows-axiom} does not hold for this model. In both cases shown, the initial directions were $\ket{\texttt{South, South}}$. The resulting states are respectively approximately a classical ensemble $\frac{1}{2} (\ket{00}\bra{00} + \ket{11}\bra{11})$, and the pure, separable state $\ket{ii}\bra{ii}$. This reinforces the finding that \texttt{follows} involves a projection, and highlights the inherent difficulty faced by the four directional model in encoding two bits of information into a single qubit.
    }
    \label{fig:interpret/4dir/left-following}
\end{figure}

Fig.~\ref{fig:interpret/4dir/following}a. visualises the following gate on a subset of the 16 initial states.
We visually confirm the approximate idempotence of following in Fig.~\ref{fig:interpret/4dir/following}b.
Finally, we exhibit an axiom the model fails to adequately capture in Fig.~\ref{fig:interpret/4dir/left-following}:
\begin{equation}\label{eq:left-follows-axiom}
    \texttt{P}_2\texttt{ turns left.P}_1\texttt{ follows P}_2\texttt{.} = \texttt{P}_1\texttt{ follows P}_2\texttt{.P}_1\texttt{ turns left.P}_2\texttt{ turns left.}
\end{equation}
This axiom captures the idea that turns can be `copied through' \texttt{follows} when they act on the person being followed, however the model is unable to handle the quarter turn correctly.

\paragraph{Questions}
The question states are again very similar to the two directions model. We can express them in terms of the $\{\ket{\texttt{NW}}, \ket{\texttt{SE}}\}$ basis derived from the two-qubit \texttt{follows} projections (see App.~\ref{app:interpretability} for visualisation in Fig.~\ref{fig:interpret/questions}):

$$\ket{\texttt{same dir}} \approx \frac{1}{\sqrt{2}} (\ket{\texttt{NW}} \otimes \ket{\texttt{NW}} - \ket{\texttt{SE}} \otimes \ket{\texttt{SE}})$$
$$\ket{\texttt{opp dir}} \approx \frac{1}{\sqrt{2}} (\ket{\texttt{NW}} \otimes \ket{\texttt{SE}} - e^{-i\pi/8}\ket{\texttt{SE}} \otimes \ket{\texttt{NW}})$$

Once again, this shows the underlying two-directional nature of the model, in that it tests for correlation or anti-correlation along the $\{\ket{\texttt{NW}}, \ket{\texttt{SE}}\}$ basis.

\paragraph{Discussion}
The analysis above helps us to identify and explain biases that the model exhibits. The bias can be seen to arise from a bias in the dataset - there are very few instances where the final directions faced by the model are at right angles to each other, so the model can afford to ignore these instances at relatively little cost. We can hence identify the instances that are `difficult' for the model as those that depend on the correct resolution of a quarter turn at some point in the text. As these are a relative minority across the dataset, the model appears to perform better than it should. 

It is in fact expected that the model should struggle to perform well, as it is effectively trying to encode 2 bits of information (which of the four directions each actor is currently facing) into a single qubit: indeed the equivalent deterministic model for the four directional dataset (described in App.~\ref{app:interpretability} Fig.~\ref{fig:interpretability/clifford-4dir}) requires two qubits.

The fact that we were able to visualise the model in a modular way, and thus understand \textit{why} it appeared to work, allowed us to identify a bias in our dataset. This was possible in part due to the relative simplicity of the model, as well as its compositional nature, which allowed us to consider only one and two qubit components of the model in isolation, and from this infer its action on arbitrary texts. In the two-directions case, this broadly supported the empirical evidence that the model had correctly learnt to generalise, whilst also exposing a failure case to be wary of. In contrast, for four directions we were able to identify that the model had \textit{not} learnt to generalise in the way we had hoped, despite the otherwise promising empirical evidence.

\subsection{Interventions}
\begin{figure}[ht]
    \centering
    \includegraphics[width=\textwidth]{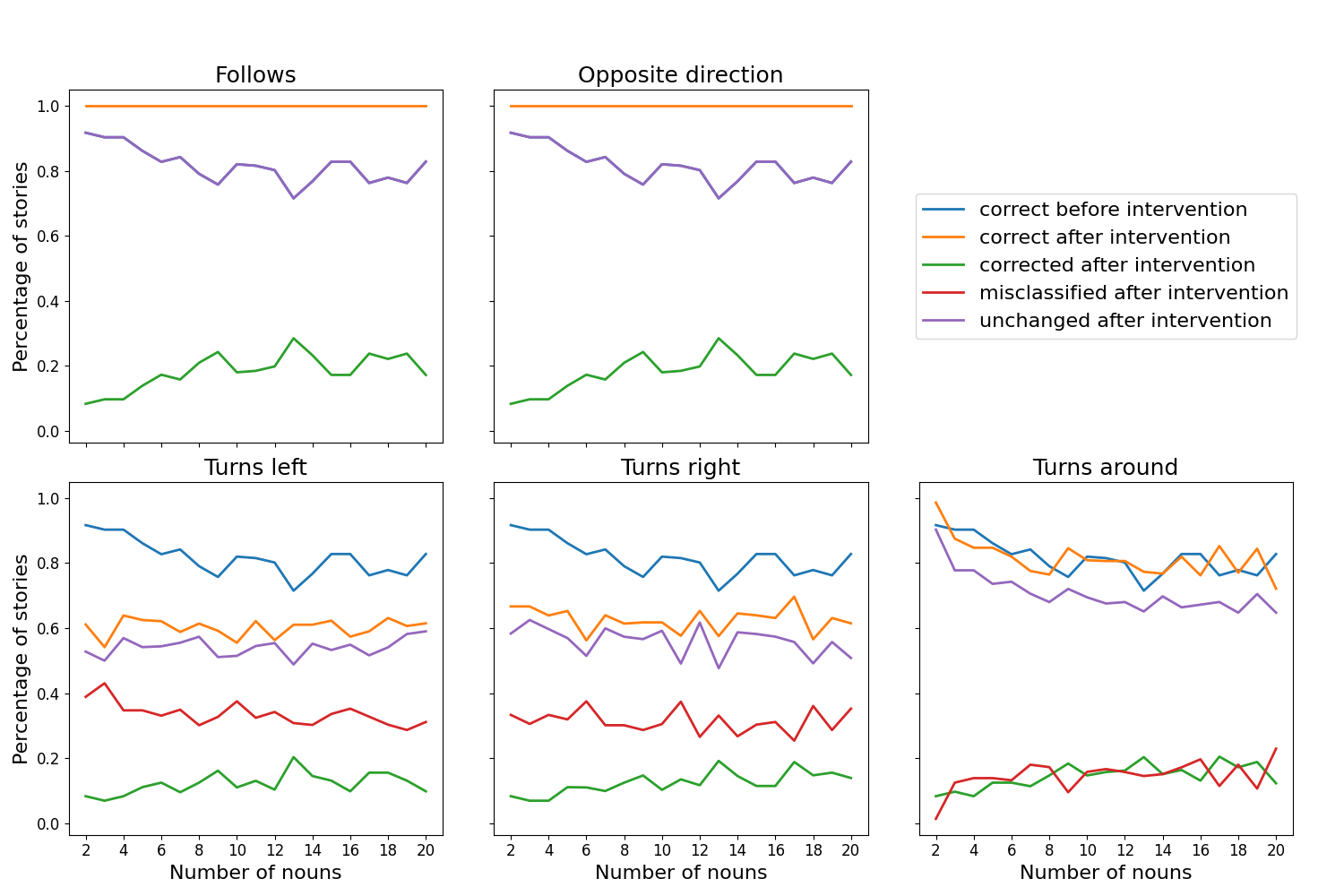}
    \caption{Effect of performing interventions at the end of stories. The lines for `correct before intervention' and `correct after intervention' indicate the total percentages of stories that were accurately classified without and with the intervention. The `corrected after intervention' line shows the percentage of stories incorrectly classified without the intervention but corrected with the intervention. Conversely, the `misclassified after intervention' line describes stories classified incorrectly after the intervention that were previously classified correctly.
    The `unchanged after intervention' line shows the percentage of stories that maintained their classification accuracy regardless of the intervention, although this might still imply a change in the story’s label. For instance, if in the original story, \texttt{Bob is not going in the same direction as Alice}, and the model accurately identified this, but after the intervention, \texttt{Bob is following Alice} and the model still classifies it accurately (now as \texttt{goes in the same direction}), this story would contribute to the `unchanged after intervention' percentage because its classification remained accurate despite the label change. 
     }
    \label{fig:interventions}
\end{figure}

Another way to better understand our model's robustness and inner workings is to carry out interventions on the original stories and investigate how these affect the model's output. Therefore, we generated an alternative test dataset from a subset of the different density four-directional datasets. This subset contains the full simple and deeper datasets with up to 20 nouns, less dense with up to 14 nouns, dense with up to 12 nouns, and superdense with up to 10 nouns. We chose this reduced subset for efficient simulation. At the end of each story, we added an action from the set of actions $\{ \texttt{turns right}$, $\texttt{turns left}$, $\texttt{turns around}$, $\texttt{follows}$, $\texttt{goes in the opposite direction of} \}$ to one or both actors involved in the question to see how it would affect the model's output. The single-actor actions were always applied to the second actor in the question, while the two-actor actions were applied to both actors in the occurring order. For instance, if we consider a story in which \texttt{Bob goes in the same direction as Alice}, which is classified correctly by our model, we add the sentence \texttt{Alice turns left}, so now they are not facing the same direction anymore and check whether our model classifies this new data point correctly.
The results are depicted in Fig.~\ref{fig:interventions}. 

\paragraph{Discussion}
Interventions with \texttt{turns left} and \texttt{turns right} reduce the overall accuracy, which can be seen because the `correct after intervention' line is lower than the `correct before intervention' line and the `misclassified after intervention' line is higher than the `corrected after intervention' line. 
For interventions with \texttt{turns around}, the overall accuracy stays about the same. The `correct before intervention' and `correct after intervention' lines are quite similar because about the same amount of stories are `corrected after intervention' as `misclassified after intervention'. Interventions with \texttt{follows} and \texttt{goes in the opposite direction of} increase the accuracy to $100\%$.

This aligns with the insights gained from examining the Bloch spheres in Sec.~\ref{sec:comp_interpret_4_dir}. The model faces difficulties in accurately classifying stories when the actors are positioned at 90 degrees relative to each other, likely a reflection of the bias within the dataset of having more stories with actors facing the same or opposite directions than being positioned at 90 degrees to each other. See Sec.~\ref{app:acc-bias} for further details. Adding \texttt{turns left} and \texttt{turns right} interventions to stories, where characters previously faced opposite directions, results in them now being positioned at 90-degree angles to one another. Conversely, stories where characters originally faced each other at 90 degrees are adjusted to face the same or opposite directions. Given that the majority of our original dataset comprises stories with actors facing the same or opposite directions, the majority of stories in our intervention dataset will have actors facing each other at 90 degrees. Thus, the `misclassified after intervention' line is higher than the `corrected after intervention' line.

Interventions with \texttt{turns around} do not affect the angle at which people are facing each other. Thus, the overall accuracy stays about the same. From Fig.~\ref{fig:interpret/2dir/directions-states}b iii) we see that the action \texttt{turns around} introduces some error. Depending on the story's course up to the point of the intervention this noise might be the cause for the `corrected after intervention' and `misclassified after intervention' stories.

The questions do not consider the absolute directions people are going but rather check the correlation of the two actors involved. The actions \texttt{follows} and \texttt{goes in the opposite direction of} set this correlation correctly as they project the state of the second actor onto the same or opposite state of the first actor regardless of the exact position of the first actor's state on the bloch sphere. Therefore interventions with \texttt{follows} and \texttt{goes in the opposite direction of} correct the labels of all stories and increase the accuracy to $100 \%$.
This also aligns with the finding that the model generally performs better on instances that require fewer inference steps. A more detailed analysis of the model's accuracy depending on the number of inference steps can be found in App.~\ref{app:acc-bias} and Fig.~\ref{fig:interpretability/acc-inf_steps}. 

\section{Discussion and future work}
\label{sec:conclusion}

In this work, we presented the first experimental implementation of the QDisCoCirc model, as it was described in \cite{tkb}, for an NLP task.
Specifically, we focused on the task of question answering for the two toy datasets we constructed.
The compositional nature of the model enables generalisation to larger instances than what the model was trained on, a property that mainstream neural architectures struggle with, or achieve only with significant tuning and curriculum design \cite{zhou2024transformers}.
Our setup provides a promising route for scalable quantum machine learning. It avoids trainability issues due to barren plateaus by training components classically and using them to compose larger instances. These instances are then  evaluated on a quantum computer only for inference.

Furthermore, we have constructed the model such that its semantic encoding follows the compositional structure of the input text, thus enabling interpretability. Given a trained model, we can then understand how the model performs a task by inspecting the quantum states and word embeddings.
In particular, when the model showed excellent performance, we were able to identify the mechanisms by which it solved the task, and verified that these matched what we would have intuitively expected.
When the model performs sub-optimally due to a smaller-than-ideal number of dimensions, it learns to compromise and take advantage of biases in the data to perform better than random guessing; again, the compositional structure allows us to understand this mechanism.

In future work, we aim to train a QDisCoCirc model on larger-scale real-world data, making use of the text-to-diagram parser of Ref. \cite{liu2023pipeline},
and explore other tasks to be framed in our compositional framework.
In general, semantic rewrites would not be hard-coded during training; however, the degree to which we expect semantic rewrites hold can be tested with a trained model.
Further, it is interesting to explore ways of pre-training quantum word embeddings such that they can be used solve a task only by composing them, or such that they can be used as good initial guesses for in-task fine-tuning.
In addition, the DisCoCirc framework invites many different variants of quantum semantics other than the generic ones we used here, which are based on expressive ansaetze. These lead to hard-to-simulate instances, while there are \emph{easy-to-simulate} quantum circuit families that would enable large-scale experimentation. Examples include Clifford circuits, which are however hard to optimise due to their combinatorial nature, and free fermions (i.e. matchgates) which come equipped with continuous parameters and therefore are easier to optimise~\cite{Matos_2023}.
A further direction to investigate would be to broaden the way in which we ask questions, for example, to allow the model to implement a quantum random access code~\cite{farkas2024simple} by interpreting the question outputs as bitsrings rather than probability distributions.
Note also that, while we only considered productivity here, other aspects of compositionality could be explored (see App.~\ref{app:compositionality}).
Last, but not least, one may also consider a neural-network DisCoCirc model, which would use a direct sum rather than the tensor product for the parallel composition of systems and would capture a different type of correlations natively.
Nevertheless, with any of these variants of DisCoCirc models, the by-construction compositional nature of our setup is the key differentiator with respect to usual language-modelling setups.

\paragraph{Acknowledgements:} We thank Richie Yeung, Boldizsar Poor, Benjamin Rodatz, Jonathon Liu, and Razin Shaikh for support with DisCoPy and discussions on the formulation of text as process diagrams. We thank Nikhil Khatri for help with with training the classical baselines. 
We also thank the TKET and Hardware teams at Quantinuum for their support in compiling, submitting, and monitoring jobs to the H1-1 machine.
Finally, we thank Harry Buhrman for discussions on information theory, and Ilyas Khan for supporting this long-term project.

\clearpage

\addcontentsline{toc}{section}{References}

{\small
\bibliographystyle{unsrt}
\bibliography{refs}
}

\clearpage

\appendix

\addcontentsline{toc}{section}{Appendices}
\section{Semantic functor}
\label{app:word-func}

\begin{figure}[ht]
    \centering
    \includegraphics[width=0.33\linewidth]{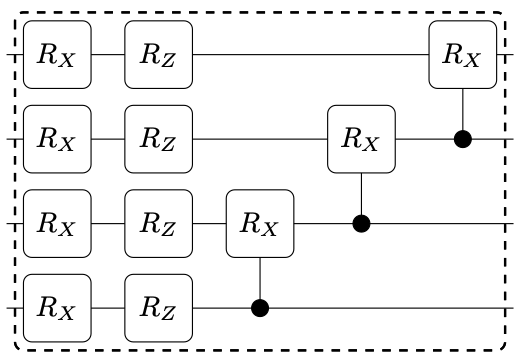}
    \caption{One layer of the ansatz used in this work, namely Circuit 4 from Ref. ~\cite{Sim_2019}. Specifically, here we show its $4$-qubit version. This parameterised quantum circuit is used to implement all unitaries throughout this work.}
    \label{fig:circuit4}
\end{figure}

\begin{figure}[ht]
    \centering
    $$\begin{tikzpicture}
	\begin{pgfonlayer}{nodelayer}
		\node [style=none] (0) at (-12, -3.75) {};
		\node [style=none] (1) at (-13.25, -3.75) {};
		\node [style=none] (2) at (-10.75, -3.75) {};
		\node [style=none] (3) at (-12, -3.25) {\texttt{person}};
		\node [style=none] (4) at (-12, -4.5) {};
		\node [style=none] (5) at (-13.25, -2.75) {};
		\node [style=none] (6) at (-10.75, -2.75) {};
		\node [style=none] (33) at (-6, -4.75) {};
		\node [style=none] (34) at (-6, -4.25) {};
		\node [style=none] (35) at (-6, -2.25) {$0$};
		\node [style=none] (73) at (0, -12.25) {};
		\node [style=none] (74) at (0, -7) {};
		\node [style=none] (117) at (16.75, -7) {};
		\node [style=none] (118) at (-13.75, -7) {};
		\node [style=none] (213) at (-9.25, -3.75) {$\leftrightarrow$};
		\node [style=none] (214) at (5.5, -3) {$\leftrightarrow$};
		\node [style=none] (216) at (-6.5, -9.5) {$\leftrightarrow$};
		\node [style=none] (219) at (-7.75, -4.25) {};
		\node [style=none] (220) at (-4.25, -4.25) {};
		\node [style=none] (221) at (-7.75, -3.25) {};
		\node [style=none] (222) at (-4.25, -3.25) {};
		\node [style=none] (223) at (-6, -3.75) {$U(\theta_\texttt{person})$};
		\node [style=none] (224) at (-6, -2.75) {};
		\node [style=none] (225) at (-6, -3.25) {};
		\node [style=none] (226) at (-6.75, -2.75) {};
		\node [style=none] (227) at (-5.25, -2.75) {};
		\node [style=none] (228) at (-6, -1.75) {};
		\node [style=none] (230) at (8.25, -3.5) {};
		\node [style=none] (232) at (7, -3.5) {};
		\node [style=none] (233) at (9.5, -3.5) {};
		\node [style=none] (234) at (7, -2.5) {};
		\node [style=none] (235) at (9.5, -2.5) {};
		\node [style=none] (236) at (8.25, -3) {$U(\theta_w)$};
		\node [style=none] (237) at (8.25, -1.75) {};
		\node [style=none] (238) at (8.25, -2.5) {};
		\node [style=none] (277) at (-1.25, -8.25) {};
		\node [style=none] (278) at (-1.25, -9) {};
		\node [style=none] (280) at (-3.75, -10) {};
		\node [style=none] (281) at (-0.75, -10) {};
		\node [style=none] (282) at (-2.25, -9.5) {$U(\theta_w)$};
		\node [style=none] (283) at (-3.75, -9) {};
		\node [style=none] (284) at (-0.75, -9) {};
		\node [style=none] (318) at (-1.3, -10.75) {};
		\node [style=none] (319) at (-1.3, -10) {};
		\node [style=none] (320) at (-1.75, -10.75) {};
		\node [style=none] (321) at (-0.85, -10.75) {};
		\node [style=none] (322) at (-1.625, -10.875) {};
		\node [style=none] (323) at (-0.95, -10.875) {};
		\node [style=none] (324) at (-1.5, -11) {};
		\node [style=none] (325) at (-1.1, -11) {};
		\node [style=none] (326) at (-1.375, -11.1) {};
		\node [style=none] (327) at (-1.225, -11.1) {};
		\node [style=none] (362) at (-3.25, -9) {};
		\node [style=none] (363) at (-3.25, -8.25) {};
		\node [style=none] (364) at (-3.25, -9) {};
		\node [style=none] (365) at (-3.25, -10) {};
		\node [style=none] (366) at (-3.25, -10.75) {};
		\node [style=none] (373) at (-11.5, -9) {};
		\node [style=none] (374) at (-12.5, -9) {};
		\node [style=none] (375) at (-9.5, -9) {};
		\node [style=none] (377) at (-11.5, -8.25) {};
		\node [style=none] (378) at (-12.5, -10) {};
		\node [style=none] (379) at (-9.5, -10) {};
		\node [style=none] (380) at (-10.5, -9) {};
		\node [style=none] (381) at (-10.5, -8.25) {};
		\node [style=none] (382) at (-10.5, -10) {};
		\node [style=none] (383) at (-10.5, -10.75) {};
		\node [style=none] (384) at (-11.5, -10) {};
		\node [style=none] (385) at (-11.5, -10.75) {};
		\node [style=none] (386) at (8.25, -4.25) {};
		\node [style=none] (498) at (0.5, -2.5) {};
		\node [style=none] (499) at (3.5, -2.5) {};
		\node [style=none] (500) at (2, -3) {$w$};
		\node [style=none] (502) at (0.5, -3.5) {};
		\node [style=none] (503) at (3.5, -3.5) {};
		\node [style=none] (504) at (2, -2.5) {};
		\node [style=none] (505) at (2, -1.75) {};
		\node [style=none] (506) at (2, -3.5) {};
		\node [style=none] (507) at (2, -4.25) {};
		\node [style=none] (566) at (7, -9.5) {$\leftrightarrow$};
		\node [style=none] (567) at (10.5, -8.25) {};
		\node [style=none] (568) at (10.5, -9) {};
		\node [style=none] (569) at (8.75, -10) {};
		\node [style=none] (570) at (11.25, -10) {};
		\node [style=none] (571) at (10, -9.5) {$U(\theta_w)$};
		\node [style=none] (572) at (8.75, -9) {};
		\node [style=none] (573) at (11.25, -9) {};
		\node [style=none] (574) at (10.45, -10.75) {};
		\node [style=none] (575) at (10.45, -10) {};
		\node [style=none] (584) at (9.5, -9) {};
		\node [style=none] (585) at (9.5, -8.25) {};
		\node [style=none] (586) at (9.5, -9) {};
		\node [style=none] (587) at (9.5, -10) {};
		\node [style=none] (588) at (9.5, -10.75) {};
		\node [style=none] (589) at (2.5, -9) {};
		\node [style=none] (590) at (1, -9) {};
		\node [style=none] (591) at (5, -9) {};
		\node [style=none] (592) at (3, -9.5) {$w$};
		\node [style=none] (593) at (2.5, -8.25) {};
		\node [style=none] (594) at (1, -10) {};
		\node [style=none] (595) at (5, -10) {};
		\node [style=none] (596) at (3.5, -9) {};
		\node [style=none] (597) at (3.5, -8.25) {};
		\node [style=none] (598) at (3.5, -10) {};
		\node [style=none] (599) at (3.5, -10.75) {};
		\node [style=none] (600) at (2.5, -10) {};
		\node [style=none] (601) at (2.5, -10.75) {};
		\node [style=none] (616) at (-3.25, -7) {};
		\node [style=none] (617) at (-3.25, -0.5) {};
		\node [style=none] (618) at (-2.25, -8.25) {};
		\node [style=none] (619) at (-2.25, -9) {};
		\node [style=none] (620) at (-2.25, -10) {};
		\node [style=none] (621) at (-2.25, -10.75) {};
		\node [style=none] (622) at (7, -5.75) {w $\in$ $\{$ \texttt{left},  \texttt{around}, \texttt{north}, \texttt{east}, \texttt{south}, \texttt{west} $\}$};
		\node [style=none] (623) at (-11, -9.5) {$\texttt{follows}$};
		\node [style=none] (625) at (6.5, -12) {w $\in$ $\{$ \texttt{same~dir}, \texttt{not~same~dir} $\}$};
		\node [style=none] (626) at (-1.25, -7.75) {$0$};
		\node [style=none] (627) at (-1.25, -8.25) {};
		\node [style=none] (629) at (-2, -8.25) {};
		\node [style=none] (630) at (-0.5, -8.25) {};
		\node [style=none] (631) at (-1.25, -7.25) {};
	\end{pgfonlayer}
	\begin{pgfonlayer}{edgelayer}
		\draw (2.center) to (1.center);
		\draw (0.center) to (4.center);
		\draw (6.center) to (5.center);
		\draw (6.center) to (2.center);
		\draw (5.center) to (1.center);
		\draw (34.center) to (33.center);
		\draw [style=box edge] (74.center) to (73.center);
		\draw [style=box edge] (117.center) to (118.center);
		\draw (220.center) to (219.center);
		\draw (222.center) to (221.center);
		\draw (222.center) to (220.center);
		\draw (221.center) to (219.center);
		\draw (224.center) to (225.center);
		\draw (228.center) to (226.center);
		\draw (226.center) to (227.center);
		\draw (227.center) to (228.center);
		\draw (233.center) to (232.center);
		\draw (235.center) to (234.center);
		\draw (235.center) to (233.center);
		\draw (234.center) to (232.center);
		\draw (237.center) to (238.center);
		\draw (277.center) to (278.center);
		\draw (281.center) to (280.center);
		\draw (284.center) to (283.center);
		\draw (284.center) to (281.center);
		\draw (283.center) to (280.center);
		\draw (318.center) to (319.center);
		\draw (322.center) to (323.center);
		\draw (321.center) to (320.center);
		\draw (324.center) to (325.center);
		\draw (327.center) to (326.center);
		\draw (363.center) to (364.center);
		\draw (365.center) to (366.center);
		\draw (375.center) to (374.center);
		\draw (373.center) to (377.center);
		\draw (379.center) to (378.center);
		\draw (379.center) to (375.center);
		\draw (378.center) to (374.center);
		\draw (381.center) to (380.center);
		\draw (382.center) to (383.center);
		\draw (384.center) to (385.center);
		\draw (230.center) to (386.center);
		\draw (499.center) to (498.center);
		\draw (503.center) to (502.center);
		\draw (503.center) to (499.center);
		\draw (502.center) to (498.center);
		\draw (505.center) to (504.center);
		\draw (506.center) to (507.center);
		\draw (567.center) to (568.center);
		\draw (570.center) to (569.center);
		\draw (573.center) to (572.center);
		\draw (573.center) to (570.center);
		\draw (572.center) to (569.center);
		\draw (574.center) to (575.center);
		\draw (585.center) to (586.center);
		\draw (587.center) to (588.center);
		\draw (591.center) to (590.center);
		\draw (589.center) to (593.center);
		\draw (595.center) to (594.center);
		\draw (595.center) to (591.center);
		\draw (594.center) to (590.center);
		\draw (597.center) to (596.center);
		\draw (598.center) to (599.center);
		\draw (600.center) to (601.center);
		\draw [style=box edge] (617.center) to (616.center);
		\draw (618.center) to (619.center);
		\draw (620.center) to (621.center);
		\draw (631.center) to (629.center);
		\draw (629.center) to (630.center);
		\draw (630.center) to (631.center);
	\end{pgfonlayer}
\end{tikzpicture}$$
    \caption{
        Semantic functor applied to the specific vocabulary used in this work. Each wire is assigned one qubit. All nouns are expressed as \texttt{person}. The word boxes (on the left) for \texttt{turns~left}, \texttt{turns~around}, \texttt{walks~north}, \texttt{walks~east}, \texttt{walks~south}, and \texttt{walks~west}, are assigned pure unitaries (on the right). The Box for \texttt{follows} gets an ancilla qubit which is then discarded. 
        For each word $w$, the associated unitary $U(\theta_w)$ is controlled by a set of parameters that is unique to that word. \texttt{turns~right} and \texttt{goes in the opposite direction of} are substituted according to the semantic rewrites in Fig~\ref{fig:axioms}.
    }
    \label{fig:word-func}
\end{figure}

The semantic functor, as outlined in Sec.~\ref{sec:qdiscocirc}, is defined in terms of some hyperparameters,
which we specify here.
Each wire is assigned one qubit.
Every unitary $U(\theta)$ is implemented using three layers of Circuit 4, which we show in Fig.~\ref{fig:circuit4}.
This parameterised quantum circuit is taken from Ref. \cite{Sim_2019},
which studied a diverse set of ansaetze with regard to their entangling capabilities and their expressivity.
Specifically for initialisations of single qubit states we used the \textit{Euler paramterisation} which is defined as $\ket{\psi} = R_x(\theta_3) R_z(\theta_2) R_x(\theta_1) \ket{0}$.

In Fig.~\ref{fig:word-func} we show in more detail the semantic functor as we have defined it for the specific vocabulary and question-answering task described in Sec.~\ref{sec:question-answering} and Sec.~\ref{sec:datasets}, including specifying which verbs are assigned pure unitaries and which are assigned channels with the addition of ancillae that are then discarded. The channels allow for meaning updating, in the spirit of Ref.~\cite{coecke2020meaning}. In theory, any box in the text circuit could be assigned a channel while ensuring that only one of the text circuit and question circuit is mixed, which is a requirement for question answering as formulated here to be valid (\cite{tkb}, Section 3). This allows information to be discarded, which is beneficial, but since it requires ancilla qubits and more two-qubit gates, we want to avoid this wherever possible. As we expect that only the \texttt{follows} box would need to discard information, we assign all other boxes as unitaries. Note that all actor names, such as \texttt{Alice} or \texttt{Bob}, in the stories, are replaced by the word \texttt{person} and share the same set of parameters $\theta_\texttt{person}$ among all circuits that prepare quantum noun-states.
The specific actors are identified by their position in the circuit, i.e. their corresponding wire, by the fact that the tensor product is non-commutative.

\section{Datasets}
\label{app:datasets}

Table~\ref{tab:dataset_sizes} below shows the number of data entries per dataset referenced in the paper.
\begin{table}[ht]
    \centering
    \begin{tabular}{| c | c | c | c | c | c | c | }
    \hline
    \rule[-0.5em]{0pt}{1.5em} \bf{dataset} & \multicolumn{3}{c|}{\bf{two-directional} } &  \multicolumn{3}{c|}{\bf{four-directional}} \\
    \rule[-0.5em]{0pt}{1.5em} (\# entries) & \textit{Train, Valid A} & \textit{Valid Comp} & \textit{Test} & \textit{Train, Valid A} & \textit{Valid Comp} & \textit{Test} \\
    \hline
    \rule[-0.5em]{0pt}{1.5em} \bf{simple} & 492 & 864 & 480 & 492 & 864 & 480 \\ 
    \hline
    \rule[-0.5em]{0pt}{1.5em} \bf{deeper} & 0 & 750 & 500 & 150 & 600 & 500\\  
    \hline
    \rule[-0.5em]{0pt}{1.5em} \bf{less-dense} & 0 & 750 & 500 & 150 & 600 & 500 \\
    \hline
    \rule[-0.5em]{0pt}{1.5em} \bf{dense} & 0 & 750 & 500 & 150 & 600 & 500 \\
    \hline
    \rule[-0.5em]{0pt}{1.5em} \bf{superdense} & 0 & 750 & 500 & 150 & 600 & 500 \\
    \hline
\end{tabular}
    \caption{Number of data entries per dataset. For the training on the four-directional dataset all densities were used so the \textit{Train} dataset size is 874 and \textit{Valid A} is 218 which is 0.8/0.2 of the total respectively.  Note that not all data entries for the datasets with 21-30 nouns (\textit{Test}) were used due to device constraints. For details of which data entries were used from these datasets see Sec.~\ref{sec:train-val-test-split} and Sec.~\ref{sec:H1-imp}.}
    \label{tab:dataset_sizes}
\end{table}

\begin{figure}[ht]
    \centering
    \begin{tabular}{c@{\hspace{2em}}c}
        \hspace{-0.75em}\includegraphics[scale=0.245]{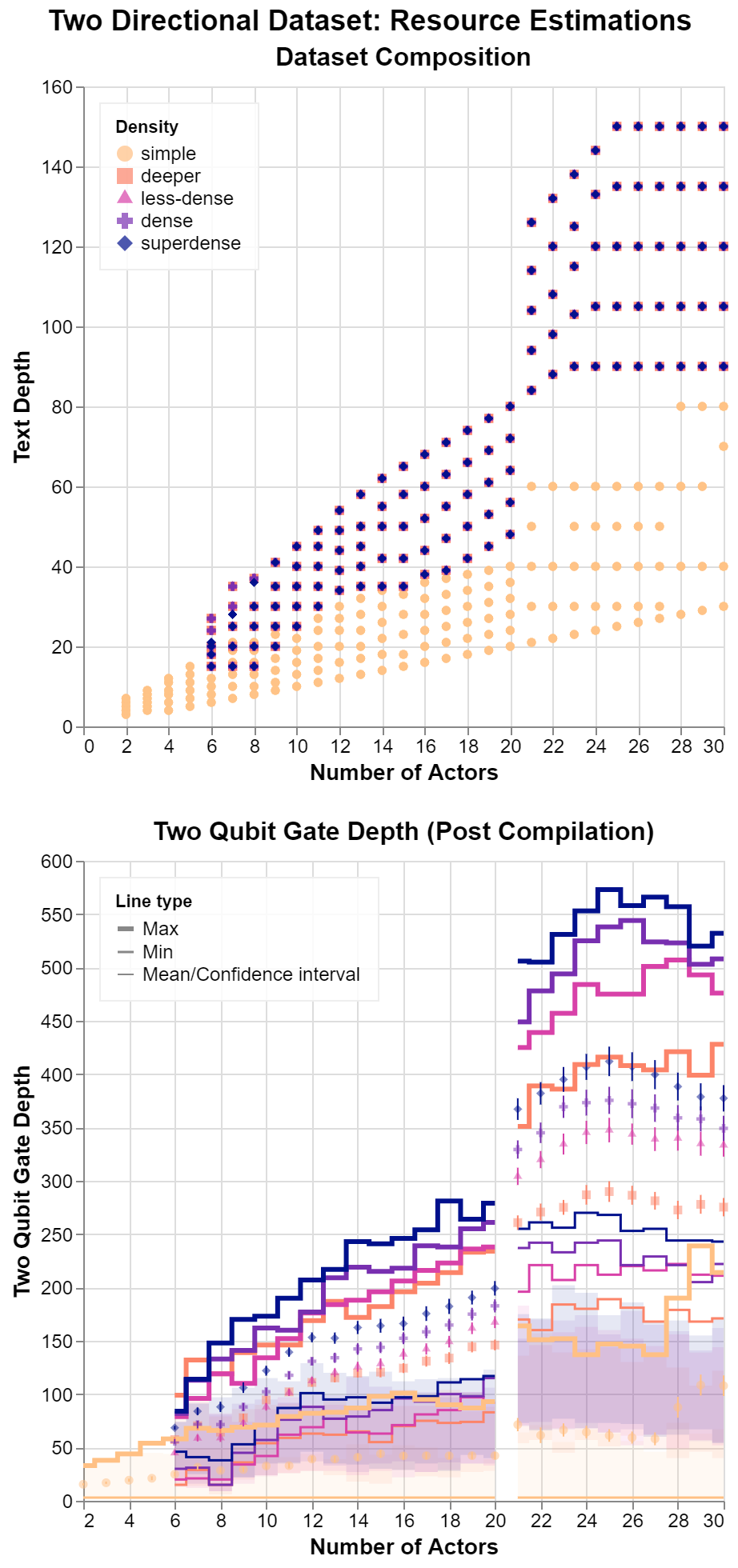} 
        & \includegraphics[scale=0.245]{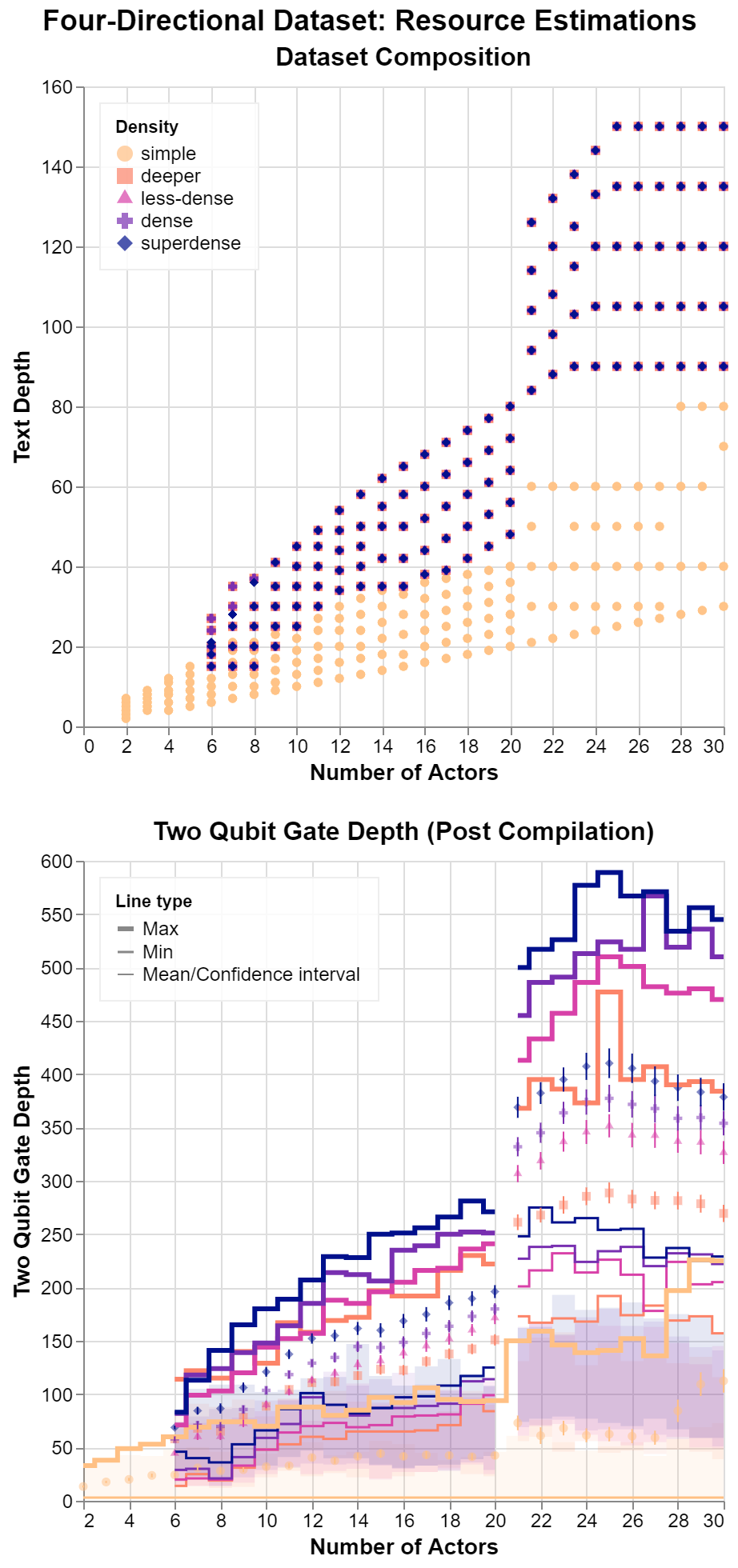} \\
        (a) & (b)
    \end{tabular}
    
    \caption{
    Further dataset characterisation: a. Two-directional dataset. b. Four-directional dataset. We exhibit the correlation between number of actors (text width) and text depth built into the data. Note that for densities other than simple, the dataset composition is almost exactly the same, and are overlapped on the display.
    The distribution of two-qubit gate depth is also shown, post-compilation, with qubit reuse according to the number of actors. For each dataset, we plot the maximum and minimum gate depth as lines, marking the mean value with a point. The vertical lines represent the 95\% confidence interval for the post-compilation mean, while the shaded area shows the extent of the pre-compilation gate depths for comparison.
    }
    \label{fig:datasets/shapes}
    \label{fig:datasets/2q-depth}
\end{figure}

\begin{figure}[ht]
    \centering
    \includegraphics[width=\textwidth]{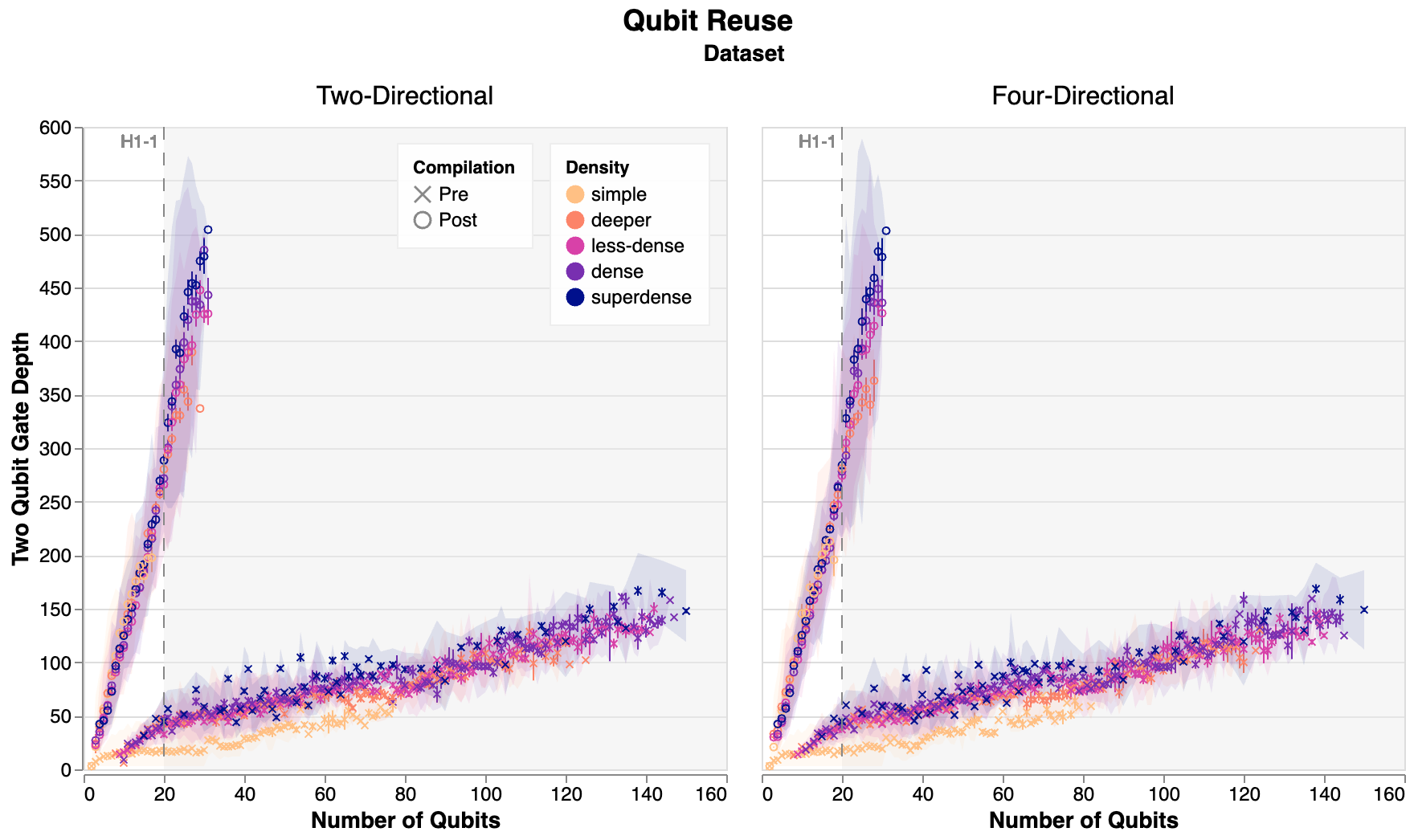}
    \caption{
    The effect of using qubit reuse when compiling the datasets for Quantinuum's H1-1 ion-trap device, which has 20 qubits. We plot the average, with 95\% confidence interval error bars, and shade the region covered by each dataset type. 
    }
    \label{fig:datasets/qubit-reuse}
\end{figure}

\begin{figure}[ht]
    \centering
    \includegraphics[width=\textwidth]{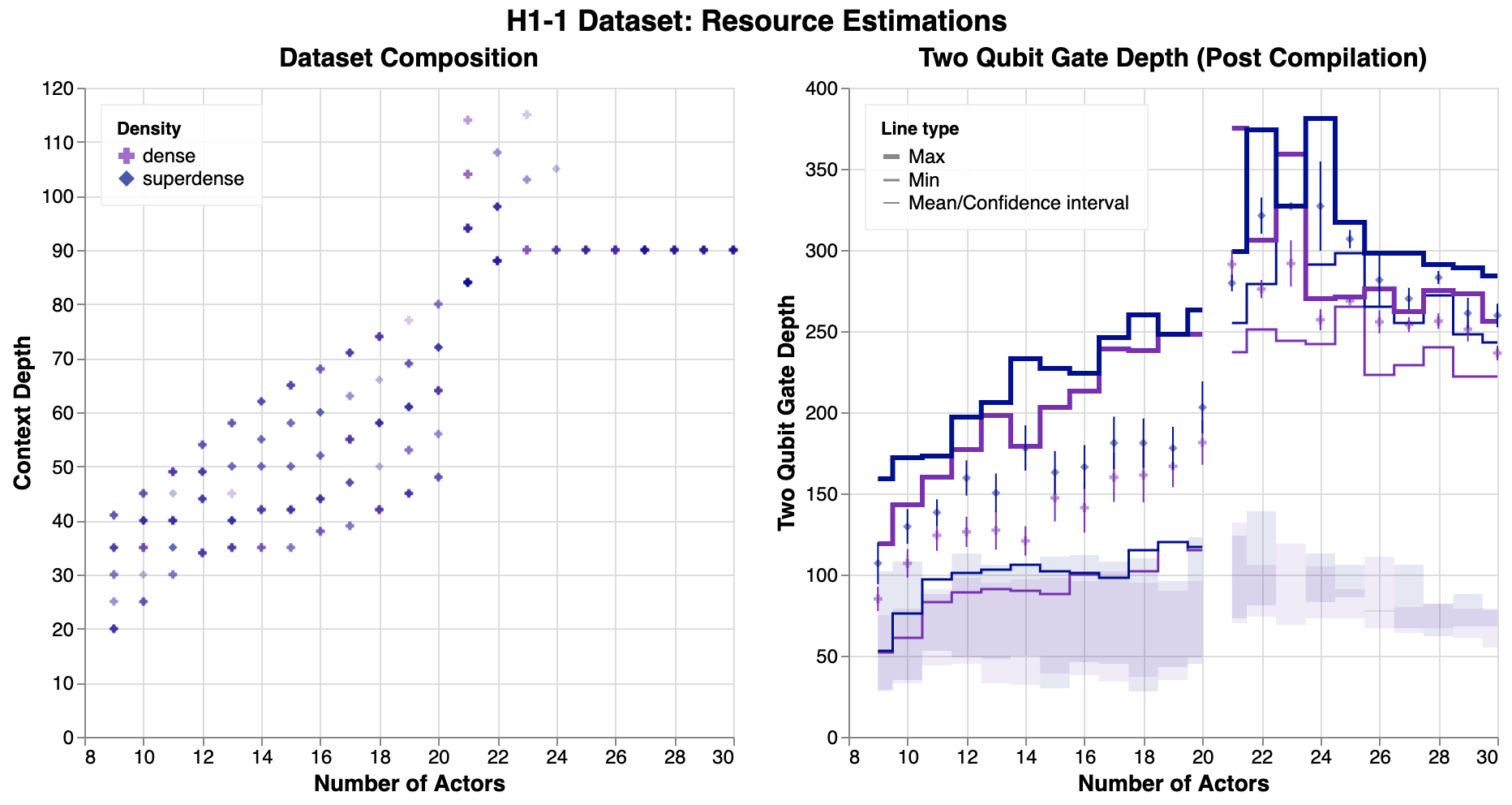}
    \caption{
    Dataset characterisation for the subset of the two-directional dataset sent to H1-1. The opacity of the points on the number of actors-text depth plot reflects the number of datapoints selected. The pre-compilation gate depths are shown as shaded regions, while the post compilation depths are displayed as lines, with mean and confidence intervals plotted as points and an error bar respectively.}
    \label{fig:datasets/2q-depth-H1}
\end{figure}
\begin{figure}[ht]
    \centering
    \includegraphics[width=\textwidth / 2]{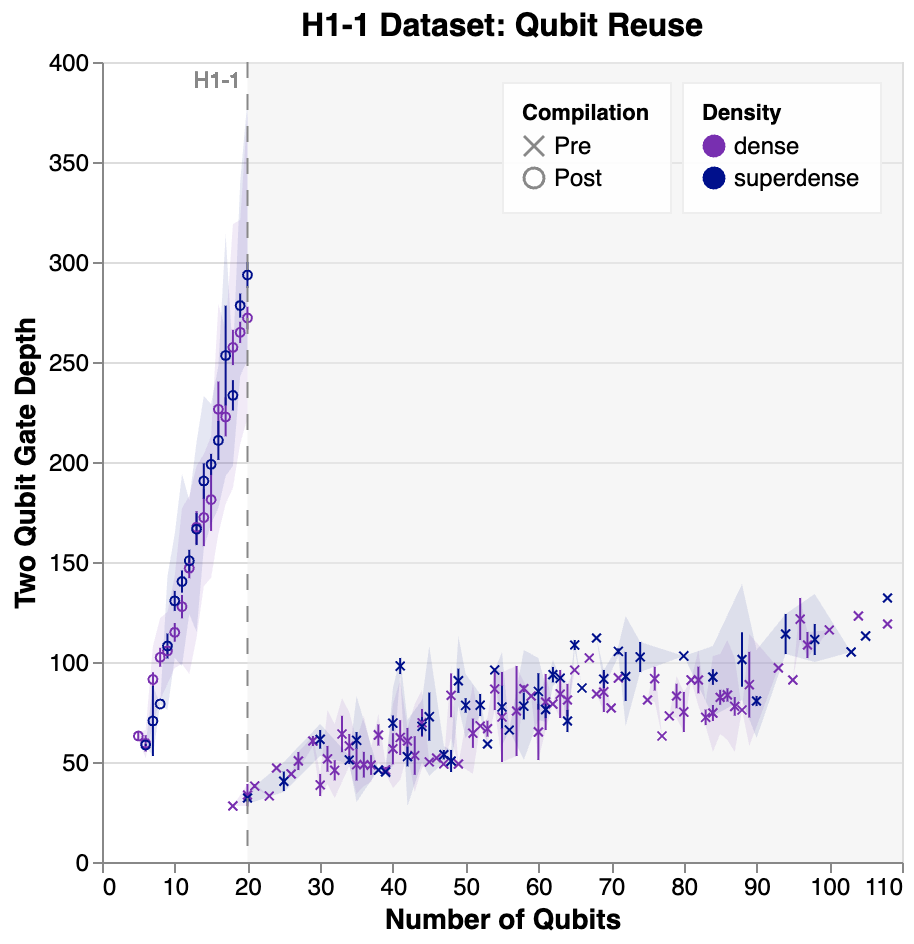}
    \caption{
    Effect of qubit reuse for the subset of the two-directional dataset sent to H1-1. We plot the average as a point, with 95\% confidence interval error bars, and shade the region covered by each dataset type.
    }
    \label{fig:datasets/qubit-reuse-H1}
\end{figure}

\begin{figure}[ht]
    \centering
    \includegraphics[width=\textwidth]{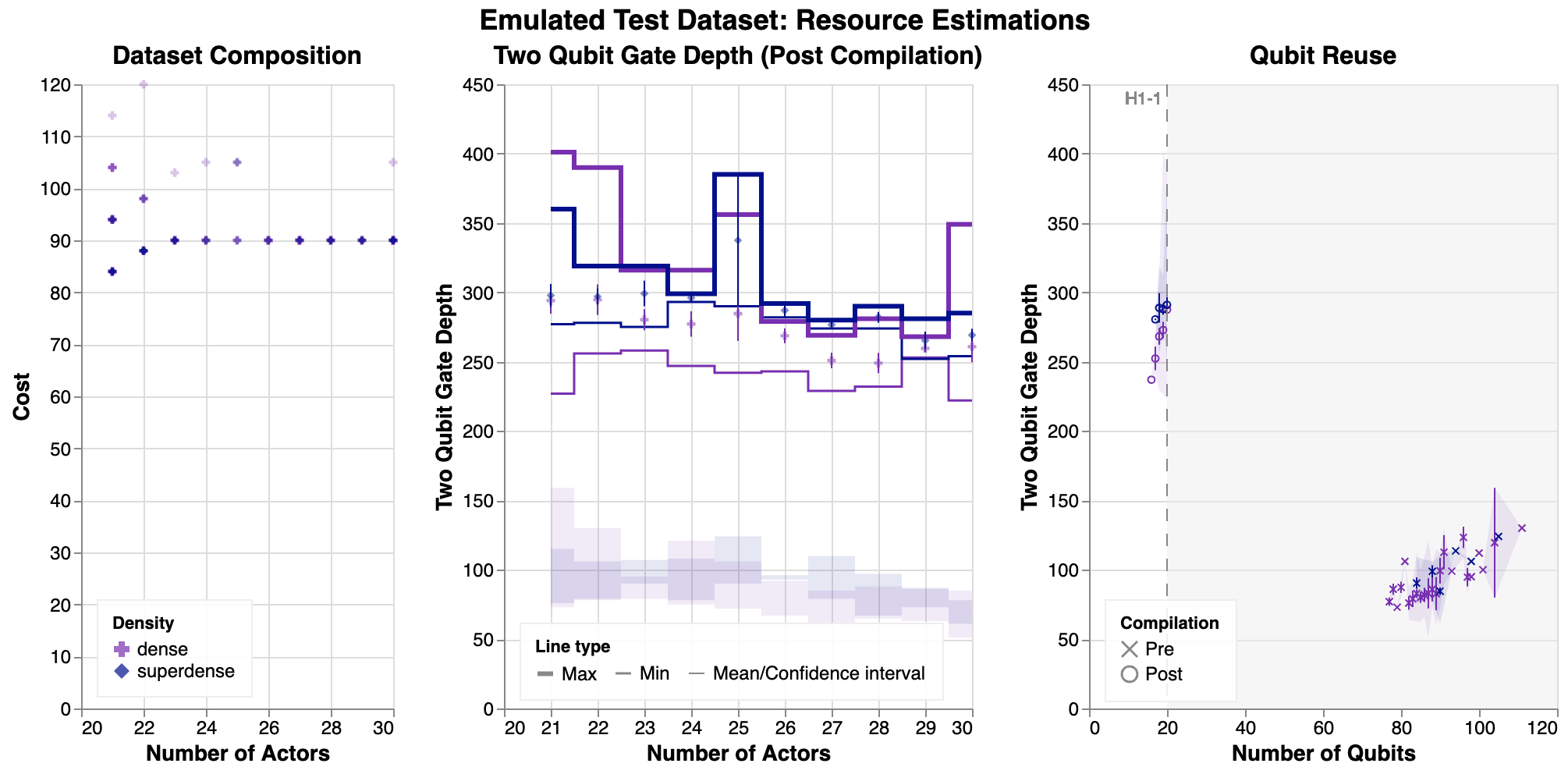}
    \caption{
    Dataset characterisation for the subset of the four-directional \textit{Test (qujax)} dataset. The opacity of the points on the dataset composition plot reflects the number of datapoints selected. The pre-compilation gate depths are shown as shaded regions, while the post compilation depths are displayed as lines, with mean and confidence intervals plotted as points and an error bar respectively. Finally, the effect of qubit reuse is also shown. For each number of qubits, the average two-qubit gate depth is shown pre- and post- compilation, with the 95\% confidence interval shaded.
    }
    \label{fig:datasets/4dir-emulated-estimations}
\end{figure}

\subsection{Story densities}

The density of a story is the number of two-actor interactions within the story divided by the number of sentences in the story. The density per dataset is shown in Table~\ref{tab:story_densities} below.

\begin{table}[ht]
    \centering
    \begin{tabular}{|c | c | c | c | c |}
    \hline
    \rule[-0.5em]{0pt}{1.5em} \bf{dataset} & \multicolumn{2}{c|}{\bf{two-directional} } &  \multicolumn{2}{c|}{\bf{four-directional}} \\
    \rule[-0.5em]{0pt}{1.5em} (\% density) & \textit{Train, Valid A, Valid Comp} & \textit{Test} & \textit{Train, Valid A, Valid Comp} & \textit{Test} \\
    
    \hline
    \rule[-0.5em]{0pt}{1.5em} \bf{simple} & 26.5 & 24.8 & 26.6 & 25.0 \\ 
    \hline
    \rule[-0.5em]{0pt}{1.5em} \bf{deeper} & 48.2 & 54.3 & 48.0 & 54.3\\  
    \hline
    \rule[-0.5em]{0pt}{1.5em} \bf{less-dense} & 50.0 & 66.5 & 50.4 & 66.6 \\
    \hline
    \rule[-0.5em]{0pt}{1.5em} \bf{dense} & 58.2 & 71.6 & 58.1 & 71.6 \\
    \hline
    \rule[-0.5em]{0pt}{1.5em}  \bf{superdense} & 68.7 & 77.5 & 68.7 & 77.5 \\
    \hline
\end{tabular}
    \caption{Average densities of the two-directional and four-directional datasets.}
    \label{tab:story_densities}
\end{table}

\section{Training}
\label{app:training}

\subsection{Hyperparameter tuning}

The hyperparameters we considered and their final tuned values are summarised in Table~\ref{tab:hyperparams} below. The optimisation of the tuning was for the best \textit{Valid A} accuracy, and the tuning was carried out using \texttt{Ax}~\cite{bakshy2018, AxTuningRepo}. Each tuning run was run for 15 epochs. From these runs we selected the best hyperparameters and ran some further training runs for 200 epochs. To select our final model, we first picked 2 candidate models from the tuning runs with high \textit{Valid A} accuracy. We then evaluated both on the \textit{Valid Comp} dataset, and selected the model with the best generalisation performance.

\begin{table}[ht]
    \centering
    \begin{tabular}{| c | c | c | }
    \hline
    \rule[-0.5em]{0pt}{1.5em} \bf{hyperparameter} & \bf{range considered}  & \bf{final value} \\
    \hline
    \rule[-0.5em]{0pt}{1.5em} learning rate & 0.0001 - 0.1 & 0.02840955 \\
    \hline
    \rule[-0.5em]{0pt}{1.5em} batch size & choice: 2$^0$, 2$^1$ - 2$^8$  & 256 \\
    \hline
    \rule[-0.5em]{0pt}{1.5em} init param seed & range:  0 - (2$^{32}$-1) & 1151618203 \\
    \hline
    \end{tabular}
    \caption{Training hyperparameters used and the ranges we considered when tuning on the four-directional dataset. The learning rate and init param seed were `range parameters' chosen from any value in the range specified, and the batch size was a `choice parameter' chosen from a list of the values of 2 to the power of 0 through 8.}
    \label{tab:hyperparams}
\end{table}

For the two-directional dataset, we did not tune the hyperparameters. Table \ref{tab:hyperparams-2dir} summarises the values used.

\begin{table}[ht]
    \centering
    \begin{tabular}{| c | c | }
    \hline
    \rule[-0.5em]{0pt}{1.5em} \bf{hyperparameter} & \bf{final value} \\
    \hline
    \rule[-0.5em]{0pt}{1.5em} learning rate & 0.005 \\
    \hline
    \rule[-0.5em]{0pt}{1.5em} batch size & 1 \\
    \hline
    \end{tabular}
    \caption{Training hyperparameters used for the two-directional dataset. The initial parameters were recorded, but not explicitly seeded.}
    \label{tab:hyperparams-2dir}
\end{table}

\subsection{Cross-validation}
\begin{figure}[ht]
    \centering
    \includegraphics[height=\textheight * 8 / 10]{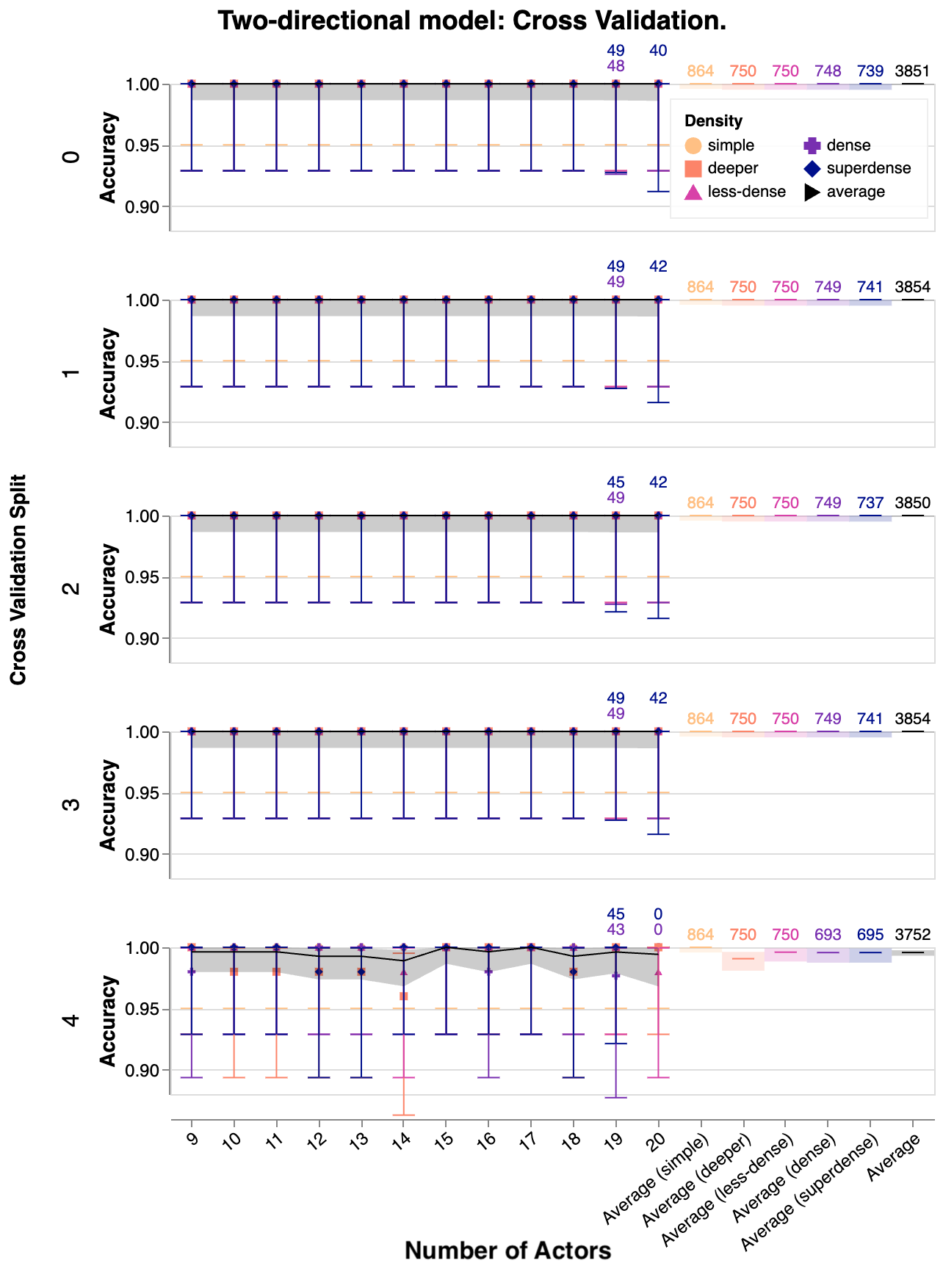}
    \caption{Cross-validated compositional generalisation for the two directions dataset on the \textit{Valid Comp} split, shown per cross-validation split. All but one model managed to generalise perfectly, matching the performance of our final model. Note that some of the larger datapoints failed to evaluate. The total number of instances for each number of actors was 72 for simple and 50 for all other densities. We report the number evaluated for the averages, and when not all available datapoints were evaluated. The error bars show the 95\% confidence interval for the average accuracy of each split.}
    \label{fig:compogen-crossval-2dir}
\end{figure}

We perform 5-fold cross-validation using the hyperparameters chosen for the two-directional dataset, with 5 iterations per fold. Each model was trained for 10 epochs. For each fold, we pick the model with highest \textit{Valid A} accuracy and evaluate it on the \textit{Valid Comp} dataset (see Sec.~\ref{sec:train-val-test-split}). The results are summarised in Fig.~\ref{fig:compogen-crossval-2dir}. From this we see that the generalisation capacity of the model is broadly independent of the initialisation, however further selection beyond \textit{Valid A} is required to ensure that the final model generalises \textit{compositionally}: in our case, although all models achieved 100\% accuracy on the \textit{Train} and \textit{Valid A} datasets, we see that the last model failed to generalise as well on the \textit{Valid Comp} dataset.

\section{Tests of compositionality}
\label{app:compositionality}

In this work, we focused on one test of compositionality, namely \emph{productivity},
as defined in Ref.\cite{hupkes2020compositionality}:
the ability of a model that is trained on small examples to use the learned rules to generalize to larger instances.
Ref.\cite{hupkes2020compositionality} introduces four other measures of compositionality. They can also be adapted to the DisCoCirc framework.
For all of them, we can provide quantitative measures in terms of differences in test and train accuracies, in a way that can be easily compared with other baseline models.

\emph{Systematicity} is akin to the vanilla generalisation performance that most supervised learning setups measure by performing inference on a held-out test set of the same type of data as those that the model was trained on.
In DisCoCirc, one would measure the effect of swaping word-boxes with other ones of the same shape (ie part of speech), and measure the effect this intervention has on the test accuracy.
In other words, we test on instances created by the reshuffling of components that the model has already been trained on, where the reshuffling respects the grammatical structure.

Another property of compositionality is \emph{substitutivity}, which tests to what effect equivalent instances that have different representation lead to the same prediction. In language, this is essentially analogous to paraphrase detection, and in DisCoCirc this can be quantified by measuring the effect of the application of axioms, such as the semantic rewrites of Fig.~\ref{fig:axioms}, to the data has on the model's prediction. Although we are replacing diagram segments as in the case of test systematicity, in the case of substitutivity, we do not expect a change in the final labels.

The property of \emph{overgeneralisation} is an aspect of compositionality that reflects the degree to which the model is attempting to apply rules it has learned to situations where it should not. Even though this property shows a manner in which the model fails at a task, this quantifies that the model indeed does not just memorise the data but rather actually learns some underlying rules.
To test this, we can compare the performance of models trained on datasets with increasing amounts of noise on a noiseless test dataset. Assuming that the noise added is random, and corrupts some of the labels in the training data, we can identify two different trends - overgeneralising and overfitting.
Provided the noise is not so significant the dataset becomes effectively random, we can identify a model that overgeneralises as one who's test accuracy is higher than seen in training, as it learns the underlying compositional rules by ignoring the noise in the training set. Conversely, a model that overfits will achieve high accuracy in training by memorising samples, but will have worse accuracy on the test set as these memorised examples will no longer be relevant.

Finally, \emph{localism} measures how local versus global the compositional structure is. DisCoCirc models are inherently local, however we can also test whether the semantics compose locally or globally using feature ablation, with the help of entanglement measures between the quantum systems (carried by the wires).

\section{Compositional generalisation}
\label{app:compogen}

\begin{figure}[ht]
    \centering
    \includegraphics[width=\textwidth]{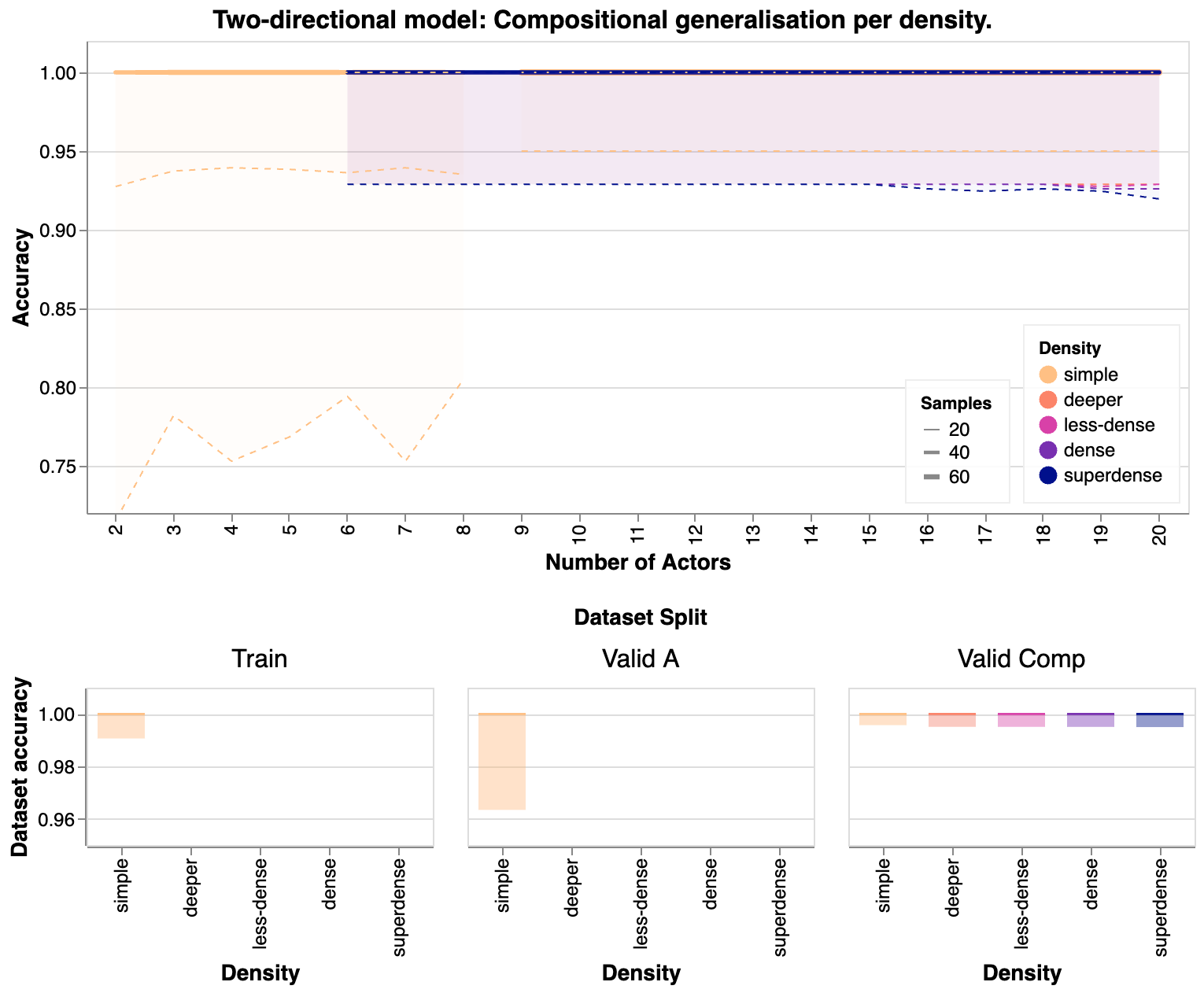}
    \caption{Compositional generalisation of the two-directional datasets per density, for the simulated datapoints only. The training set (simple up to 8 nouns) is shown separately. The model achieved 100\% accuracy on all datasets. The 95\% confidence interval is shown as a shaded region; the variation in width reflects the number of datapoints that were evaluated. Note that for the deeper to superdense datasets, the error bars are mostly overlapping.}
    \label{fig:compogen-2-dir-densities}
\end{figure}

\begin{figure}[ht]
    \centering
    \includegraphics[width=\textwidth]{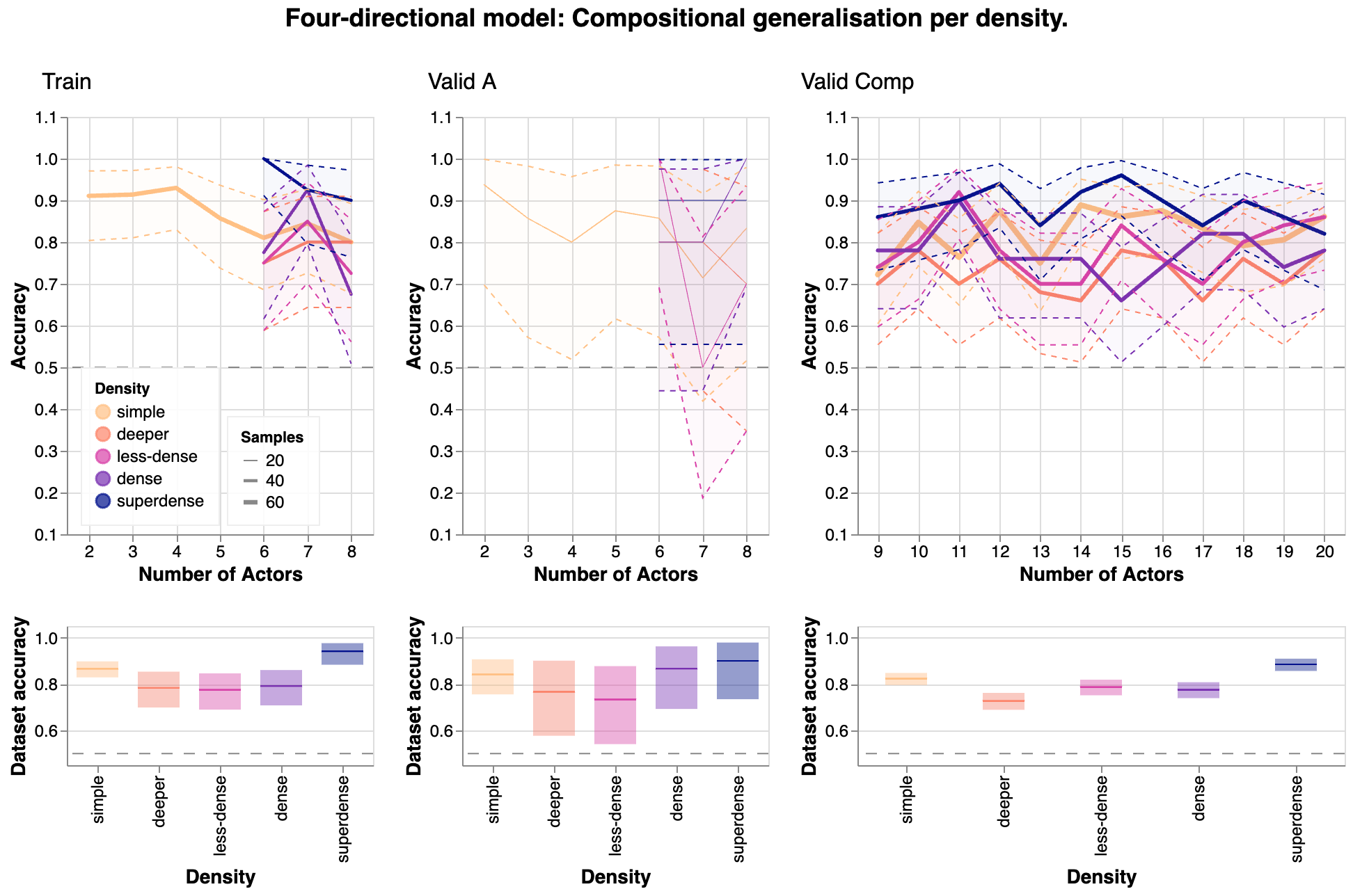}
    \caption{Compositional generalisation of the four-directional datasets per density. The model was trained on all datapoints up to 8 nouns. The shaded regions represent the 95\% confidence intervals for the mean.}
    \label{fig:compogen-4-dir-densities}
\end{figure}

Here we give a more detailed breakdown of the compositional generalisation curves, looking at each density separately. Fig~\ref{fig:compogen-2-dir-densities} shows the two-directional model, whilst Fig~\ref{fig:compogen-4-dir-densities} displays the four-directional model.

\subsection{H1-1 and \textit{Test (qujax)} sub-datasets}
\begin{figure}[ht]
    \centering
    \includegraphics[width=\textwidth]{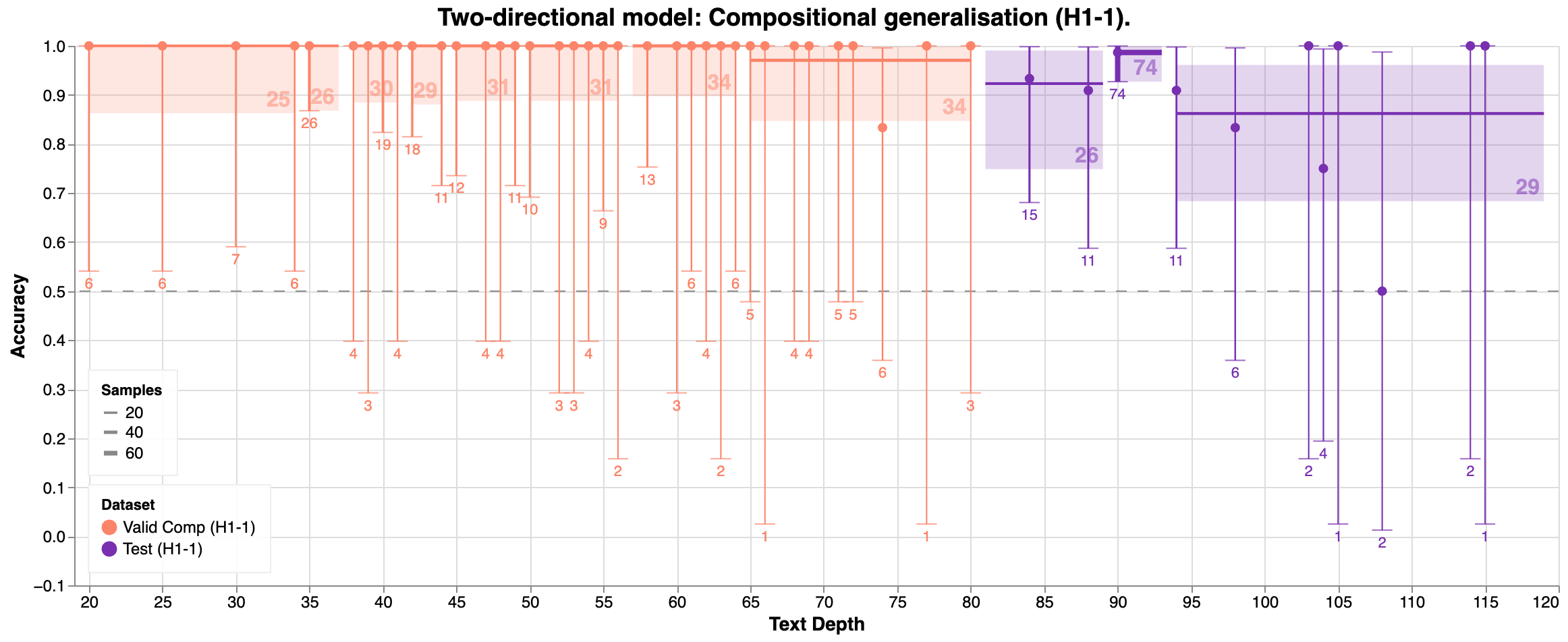}
    \caption{Compositional generalisation for the H1-1 samples per sentences in the text. The error bars shown represent the 95\% confidence interval for each depth, calculated using the exact Clopper-Pearson method \cite{clopper_pearson_1934}, which is sensitive to the number of samples. Since the number of samples per depth is very uneven, we also plot confidence intervals for binned data, targeting around 30 samples per bin. The average per bin is drawn as a line covering the binned region, with a shaded confidence interval. The number of datapoints per bin is noted inside each binned region.}
    \label{fig:compogen-H1-depth}
\end{figure}

\begin{figure}[ht]
    \centering
    \includegraphics[width=\textwidth]{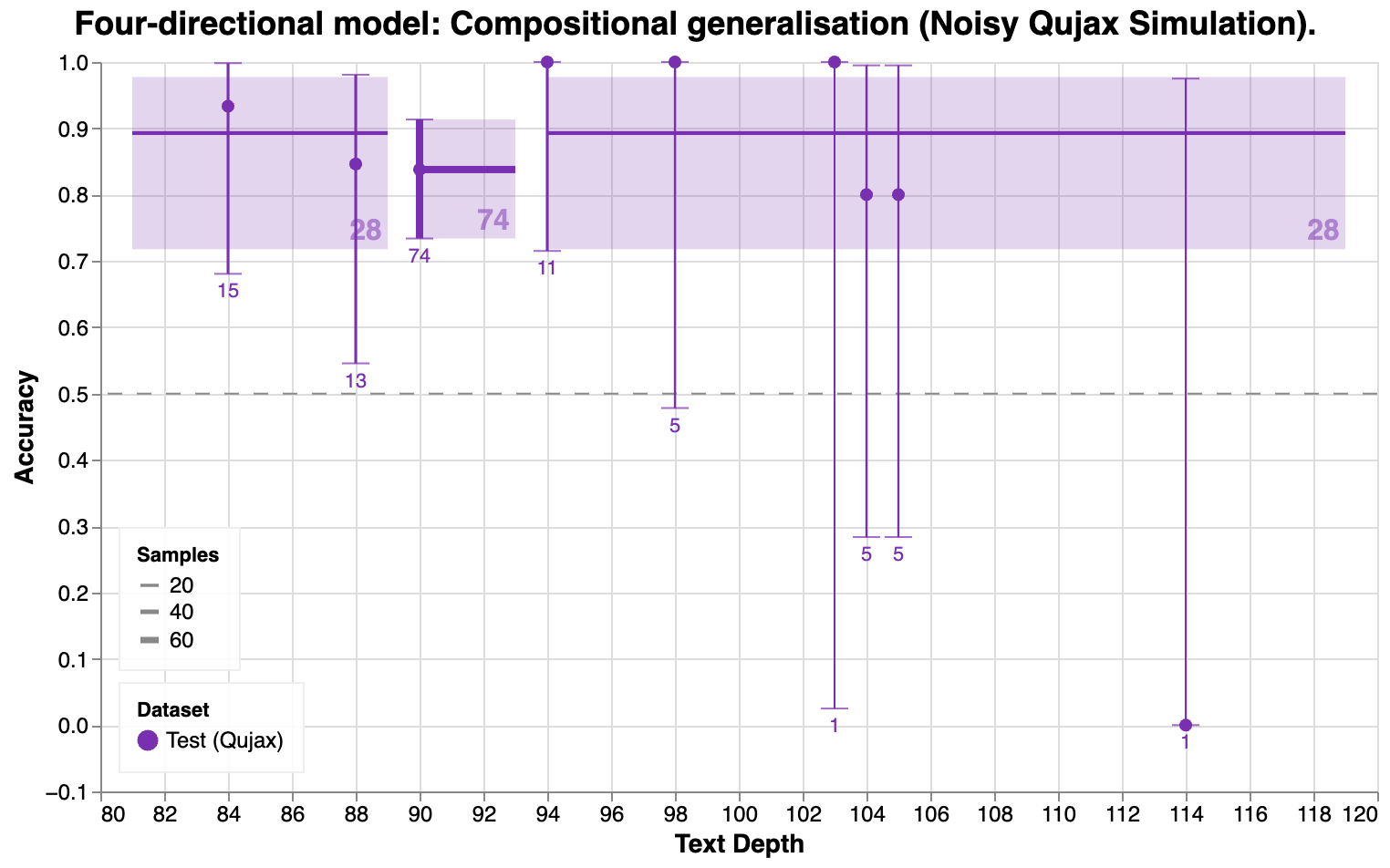}
    \caption{Compositional generalisation for the four-directional \textit{Test (qujax)} samples, per number of sentences in the story. The error bars shown represent the 95\% confidence interval for each depth, calculated using the exact Clopper-Pearson method. As with Fig.~\ref{fig:compogen-H1-depth}, we also plot the mean and confidence intervals for binned data, using the same bins as for the \textit{Test (H1-1)} dataset. The average per bin is drawn as a line covering the binned region, with a shaded confidence interval. The number of datapoints per bin is noted inside each binned region.}
    \label{fig:compogen-4dir-test-depth}
\end{figure}

Both the four- and two-directional datasets correlate the number of actors with text depth, however both the \textit{Test (H1-1)} and \textit{Test (qujax)} subsets that were actually evaluated were sampled from those that could be compiled down to 20 qubits. Fig.~\ref{fig:datasets/2q-depth-H1} visualises the subset of datapoints that were sent to H1-1, and Fig.~\ref{fig:datasets/4dir-emulated-estimations} visualises the samples from the four-directional dataset. We see that above 20 actors, there is instead an anti-correlation between number of actors and text depth, though all instances had text depths above that seen in any of the \textit{Valid Comp} instances. We visualise compositional generalisation in terms of the text depth directly in Fig.~\ref{fig:compogen-H1-depth} and Fig.~\ref{fig:compogen-4dir-test-depth} for the samples sent to H1-1, and the four-directional \textit{Test (qujax)} set. Of particular note is that we had few samples for each depth, so the confidence intervals for the mean are very wide, however in the cases where more samples were present, the mean remains high, and above random guessing.

\section{Compositional interpretability}
\label{app:interpretability}

\subsection{Clifford models}

\begin{figure}[ht]
    \centering
    $$\input{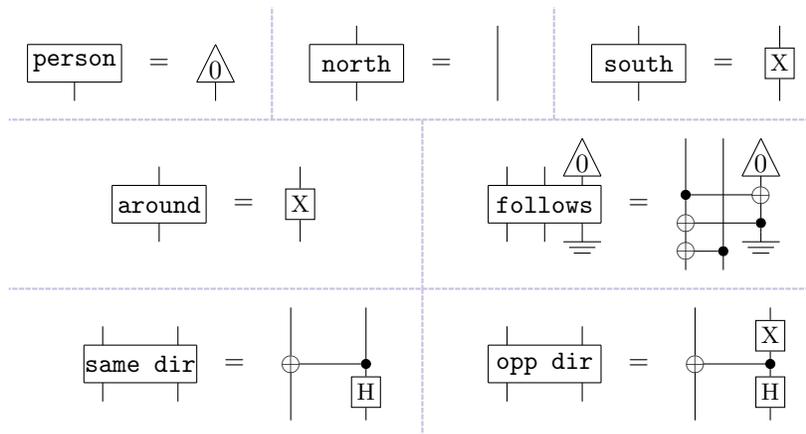}$$
    \caption{
    An implementation of a deterministic model for the two-directional dataset. 
    The model associates $\ket{\texttt{North}}$ with $\ket{0}$ and $\ket{\texttt{South}}$ with $\ket{1}$. Turning around is implemented as a bit-flip. \texttt{Follows} discards the first qubit, then copies the second qubit along the computational basis, such that the state of the first qubit will match that of the second when projected onto the computational basis. Finally, the questions become bell states checking for correlation or anti-correlation along the computational basis.
    }
    \label{fig:interpretability/clifford-2dir}
\end{figure}

\begin{figure}[ht]
    \centering
    \makebox[\textwidth][c]{$$\input{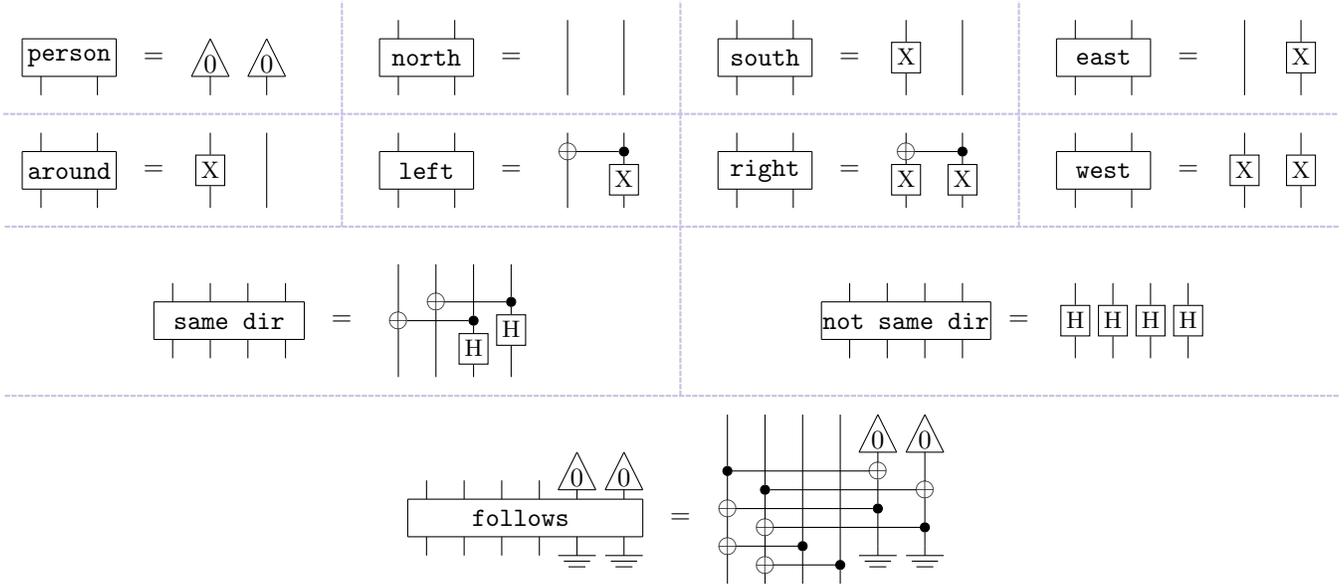}$$}
    \caption{
    An implementation of a deterministic model for the four-directional dataset. 
    Here, each of the cardinal directions is assigned a state of the computational basis: $\ket{\texttt{North}} = \ket{0,0}$, $\ket{\texttt{South}} = \ket{1,0}$, $\ket{\texttt{East}} = \ket{0,1}$, $\ket{\texttt{West}} = \ket{1,1}$. We can interpret these states as numbers, such that the rotations can be implemented as addition (modulo 4): \texttt{left} acts as $+1$, \texttt{around} as $+2$ and \texttt{right} as $-1 \equiv +3$. Like the two-directional model, \texttt{Follows} discards the wires corresponding to the first actor, then copies the state of the second noun along the computation basis. The positive question \texttt{same dir} acts as a bell state, so the overlap will be one when the states are the same and zero otherwise, whilst the overlap with \texttt{not same dir} will always be one half, and thus the maximum of the two checks if the states are the same or different.
    }
    \label{fig:interpretability/clifford-4dir}
\end{figure}

Given the simplicity of the task, we are able to construct a perfect, deterministic, compositional model that solves the task by hand. The two-directional model is described in Fig.~\ref{fig:interpretability/clifford-2dir}, while Fig.~\ref{fig:interpretability/clifford-4dir} describes the four-directional model. These models are constructed to perfectly implement the axioms required to solve the task. In the two-directional case, this provides a target for the trained model to strive towards, as well as a baseline for comparison.

\subsection{Extra Bloch spheres}

\begin{figure}[ht]
    \centering
     \includegraphics[scale=0.6]{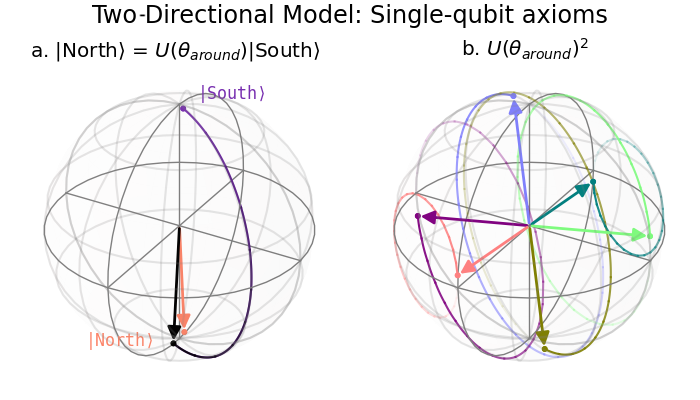}
     $$\begin{tikzpicture}
	\begin{pgfonlayer}{nodelayer}
		\node [style=none] (194) at (26.75, -2.75) {$=$};
		\node [style=none] (270) at (24.75, -4.5) {};
		\node [style=none] (271) at (24.75, -4) {};
		\node [style=none] (323) at (24.75, -3.5) {\texttt{north}};
		\node [style=none] (324) at (23.5, -3) {};
		\node [style=none] (325) at (26, -3) {};
		\node [style=none] (326) at (26, -4) {};
		\node [style=none] (327) at (23.5, -4) {};
		\node [style=none] (328) at (24.75, -2.5) {};
		\node [style=none] (329) at (24.75, -3) {};
		\node [style=none] (330) at (28.75, -3.75) {};
		\node [style=none] (331) at (28.75, -3.25) {};
		\node [style=none] (332) at (28.75, -2.75) {\texttt{south}};
		\node [style=none] (333) at (27.5, -2.25) {};
		\node [style=none] (334) at (30, -2.25) {};
		\node [style=none] (335) at (30, -3.25) {};
		\node [style=none] (336) at (27.5, -3.25) {};
		\node [style=none] (337) at (28.75, -1.75) {};
		\node [style=none] (338) at (28.75, -2.25) {};
		\node [style=none] (339) at (28.75, -5.25) {};
		\node [style=none] (340) at (28.75, -4.75) {};
		\node [style=none] (341) at (28.75, -4.25) {\texttt{around}};
		\node [style=none] (342) at (27.5, -3.75) {};
		\node [style=none] (343) at (30, -3.75) {};
		\node [style=none] (344) at (30, -4.75) {};
		\node [style=none] (345) at (27.5, -4.75) {};
		\node [style=none] (346) at (34.25, -3) {$=$};
		\node [style=none] (354) at (33.5, -2) {};
		\node [style=none] (355) at (33.5, -4) {};
		\node [style=none] (356) at (36.25, -3.25) {};
		\node [style=none] (357) at (36.25, -2.75) {};
		\node [style=none] (358) at (36.25, -2.25) {\texttt{around}};
		\node [style=none] (359) at (35, -1.75) {};
		\node [style=none] (360) at (37.5, -1.75) {};
		\node [style=none] (361) at (37.5, -2.75) {};
		\node [style=none] (362) at (35, -2.75) {};
		\node [style=none] (363) at (36.25, -1.25) {};
		\node [style=none] (364) at (36.25, -1.75) {};
		\node [style=none] (365) at (36.25, -4.75) {};
		\node [style=none] (366) at (36.25, -4.25) {};
		\node [style=none] (367) at (36.25, -3.75) {\texttt{around}};
		\node [style=none] (368) at (35, -3.25) {};
		\node [style=none] (369) at (37.5, -3.25) {};
		\node [style=none] (370) at (37.5, -4.25) {};
		\node [style=none] (371) at (35, -4.25) {};
		\node [style=none] (372) at (28.75, -1.25) {\texttt{person}};
		\node [style=none] (373) at (27.5, -0.75) {};
		\node [style=none] (374) at (30, -0.75) {};
		\node [style=none] (375) at (30, -1.75) {};
		\node [style=none] (376) at (27.5, -1.75) {};
		\node [style=none] (377) at (24.75, -2) {\texttt{person}};
		\node [style=none] (378) at (23.5, -1.5) {};
		\node [style=none] (379) at (26, -1.5) {};
		\node [style=none] (380) at (26, -2.5) {};
		\node [style=none] (381) at (23.5, -2.5) {};
	\end{pgfonlayer}
	\begin{pgfonlayer}{edgelayer}
		\draw (271.center) to (270.center);
		\draw (327.center) to (326.center);
		\draw (326.center) to (325.center);
		\draw (325.center) to (324.center);
		\draw (324.center) to (327.center);
		\draw (328.center) to (329.center);
		\draw (331.center) to (330.center);
		\draw (336.center) to (335.center);
		\draw (335.center) to (334.center);
		\draw (334.center) to (333.center);
		\draw (333.center) to (336.center);
		\draw (337.center) to (338.center);
		\draw (340.center) to (339.center);
		\draw (345.center) to (344.center);
		\draw (344.center) to (343.center);
		\draw (343.center) to (342.center);
		\draw (342.center) to (345.center);
		\draw (354.center) to (355.center);
		\draw (357.center) to (356.center);
		\draw (362.center) to (361.center);
		\draw (361.center) to (360.center);
		\draw (360.center) to (359.center);
		\draw (359.center) to (362.center);
		\draw (363.center) to (364.center);
		\draw (366.center) to (365.center);
		\draw (371.center) to (370.center);
		\draw (370.center) to (369.center);
		\draw (369.center) to (368.center);
		\draw (368.center) to (371.center);
		\draw (376.center) to (375.center);
		\draw (375.center) to (374.center);
		\draw (374.center) to (373.center);
		\draw (373.center) to (376.center);
		\draw (381.center) to (380.center);
		\draw (380.center) to (379.center);
		\draw (379.center) to (378.center);
		\draw (378.center) to (381.center);
	\end{pgfonlayer}
\end{tikzpicture}$$
     
    \caption{Single qubit axioms for the two-directional dataset. We verify that the two-directional model satisfies them approximately: a. The final states (highlighted as vectors) are very close to one another. Note that in this case the colours are to be taken as a visual aid only. b. We see that the reference points are almost mapped back to their original positions. The slight discrepancy is what leads to the peirodicity shown in Fig.~\ref{fig:interpret/2dir/directions-states}b.iii.
    }
    \label{fig:interpret/2dir-1q-axioms}
\end{figure}

We visualise the models in more detail.  Fig.~\ref{fig:interpret/2dir-1q-axioms} demonstrates the single qubit axioms holding approximately for the two directional dataset. The first axiom establishes the required relationship between the \texttt{North} and \texttt{South} initalisations, and the second establishes the expected spatial axiom for turning around twice.

\begin{figure}[ht]
    \centering
    \includegraphics[scale=0.6]{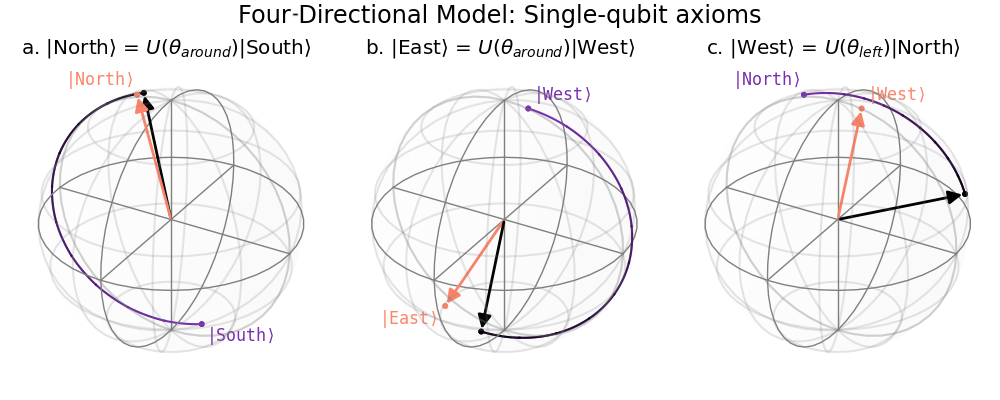}
    $$\begin{tikzpicture}
	\begin{pgfonlayer}{nodelayer}
		\node [style=none] (383) at (45, -3) {$=$};
		\node [style=none] (384) at (43, -4.75) {};
		\node [style=none] (385) at (43, -4.25) {};
		\node [style=none] (386) at (43, -3.75) {\texttt{east}};
		\node [style=none] (387) at (41.75, -3.25) {};
		\node [style=none] (388) at (44.25, -3.25) {};
		\node [style=none] (389) at (44.25, -4.25) {};
		\node [style=none] (390) at (41.75, -4.25) {};
		\node [style=none] (391) at (43, -2.75) {};
		\node [style=none] (392) at (43, -3.25) {};
		\node [style=none] (393) at (47, -4) {};
		\node [style=none] (394) at (47, -3.5) {};
		\node [style=none] (395) at (47, -3) {\texttt{west}};
		\node [style=none] (396) at (45.75, -2.5) {};
		\node [style=none] (397) at (48.25, -2.5) {};
		\node [style=none] (398) at (48.25, -3.5) {};
		\node [style=none] (399) at (45.75, -3.5) {};
		\node [style=none] (400) at (47, -2) {};
		\node [style=none] (401) at (47, -2.5) {};
		\node [style=none] (402) at (47, -5.5) {};
		\node [style=none] (403) at (47, -5) {};
		\node [style=none] (404) at (47, -4.5) {\texttt{around}};
		\node [style=none] (405) at (45.75, -4) {};
		\node [style=none] (406) at (48.25, -4) {};
		\node [style=none] (407) at (48.25, -5) {};
		\node [style=none] (408) at (45.75, -5) {};
		\node [style=none] (428) at (47, -1.5) {\texttt{person}};
		\node [style=none] (429) at (45.75, -1) {};
		\node [style=none] (430) at (48.25, -1) {};
		\node [style=none] (431) at (48.25, -2) {};
		\node [style=none] (432) at (45.75, -2) {};
		\node [style=none] (433) at (43, -2.25) {\texttt{person}};
		\node [style=none] (434) at (41.75, -1.75) {};
		\node [style=none] (435) at (44.25, -1.75) {};
		\node [style=none] (436) at (44.25, -2.75) {};
		\node [style=none] (437) at (41.75, -2.75) {};
		\node [style=none] (439) at (55, -3) {$=$};
		\node [style=none] (440) at (53, -4.75) {};
		\node [style=none] (441) at (53, -4.25) {};
		\node [style=none] (442) at (53, -3.75) {\texttt{west}};
		\node [style=none] (443) at (51.75, -3.25) {};
		\node [style=none] (444) at (54.25, -3.25) {};
		\node [style=none] (445) at (54.25, -4.25) {};
		\node [style=none] (446) at (51.75, -4.25) {};
		\node [style=none] (447) at (53, -2.75) {};
		\node [style=none] (448) at (53, -3.25) {};
		\node [style=none] (449) at (57, -4) {};
		\node [style=none] (450) at (57, -3.5) {};
		\node [style=none] (451) at (57, -3) {\texttt{north}};
		\node [style=none] (452) at (55.75, -2.5) {};
		\node [style=none] (453) at (58.25, -2.5) {};
		\node [style=none] (454) at (58.25, -3.5) {};
		\node [style=none] (455) at (55.75, -3.5) {};
		\node [style=none] (456) at (57, -2) {};
		\node [style=none] (457) at (57, -2.5) {};
		\node [style=none] (458) at (57, -5.5) {};
		\node [style=none] (459) at (57, -5) {};
		\node [style=none] (460) at (57, -4.5) {\texttt{left}};
		\node [style=none] (461) at (55.75, -4) {};
		\node [style=none] (462) at (58.25, -4) {};
		\node [style=none] (463) at (58.25, -5) {};
		\node [style=none] (464) at (55.75, -5) {};
		\node [style=none] (465) at (57, -1.5) {\texttt{person}};
		\node [style=none] (466) at (55.75, -1) {};
		\node [style=none] (467) at (58.25, -1) {};
		\node [style=none] (468) at (58.25, -2) {};
		\node [style=none] (469) at (55.75, -2) {};
		\node [style=none] (470) at (53, -2.25) {\texttt{person}};
		\node [style=none] (471) at (51.75, -1.75) {};
		\node [style=none] (472) at (54.25, -1.75) {};
		\node [style=none] (473) at (54.25, -2.75) {};
		\node [style=none] (474) at (51.75, -2.75) {};
		\node [style=none] (475) at (35, -3) {$=$};
		\node [style=none] (476) at (33, -4.75) {};
		\node [style=none] (477) at (33, -4.25) {};
		\node [style=none] (478) at (33, -3.75) {\texttt{north}};
		\node [style=none] (479) at (31.75, -3.25) {};
		\node [style=none] (480) at (34.25, -3.25) {};
		\node [style=none] (481) at (34.25, -4.25) {};
		\node [style=none] (482) at (31.75, -4.25) {};
		\node [style=none] (483) at (33, -2.75) {};
		\node [style=none] (484) at (33, -3.25) {};
		\node [style=none] (485) at (37, -4) {};
		\node [style=none] (486) at (37, -3.5) {};
		\node [style=none] (487) at (37, -3) {\texttt{south}};
		\node [style=none] (488) at (35.75, -2.5) {};
		\node [style=none] (489) at (38.25, -2.5) {};
		\node [style=none] (490) at (38.25, -3.5) {};
		\node [style=none] (491) at (35.75, -3.5) {};
		\node [style=none] (492) at (37, -2) {};
		\node [style=none] (493) at (37, -2.5) {};
		\node [style=none] (494) at (37, -5.5) {};
		\node [style=none] (495) at (37, -5) {};
		\node [style=none] (496) at (37, -4.5) {\texttt{around}};
		\node [style=none] (497) at (35.75, -4) {};
		\node [style=none] (498) at (38.25, -4) {};
		\node [style=none] (499) at (38.25, -5) {};
		\node [style=none] (500) at (35.75, -5) {};
		\node [style=none] (501) at (37, -1.5) {\texttt{person}};
		\node [style=none] (502) at (35.75, -1) {};
		\node [style=none] (503) at (38.25, -1) {};
		\node [style=none] (504) at (38.25, -2) {};
		\node [style=none] (505) at (35.75, -2) {};
		\node [style=none] (506) at (33, -2.25) {\texttt{person}};
		\node [style=none] (507) at (31.75, -1.75) {};
		\node [style=none] (508) at (34.25, -1.75) {};
		\node [style=none] (509) at (34.25, -2.75) {};
		\node [style=none] (510) at (31.75, -2.75) {};
	\end{pgfonlayer}
	\begin{pgfonlayer}{edgelayer}
		\draw (385.center) to (384.center);
		\draw (390.center) to (389.center);
		\draw (389.center) to (388.center);
		\draw (388.center) to (387.center);
		\draw (387.center) to (390.center);
		\draw (391.center) to (392.center);
		\draw (394.center) to (393.center);
		\draw (399.center) to (398.center);
		\draw (398.center) to (397.center);
		\draw (397.center) to (396.center);
		\draw (396.center) to (399.center);
		\draw (400.center) to (401.center);
		\draw (403.center) to (402.center);
		\draw (408.center) to (407.center);
		\draw (407.center) to (406.center);
		\draw (406.center) to (405.center);
		\draw (405.center) to (408.center);
		\draw (432.center) to (431.center);
		\draw (431.center) to (430.center);
		\draw (430.center) to (429.center);
		\draw (429.center) to (432.center);
		\draw (437.center) to (436.center);
		\draw (436.center) to (435.center);
		\draw (435.center) to (434.center);
		\draw (434.center) to (437.center);
		\draw (441.center) to (440.center);
		\draw (446.center) to (445.center);
		\draw (445.center) to (444.center);
		\draw (444.center) to (443.center);
		\draw (443.center) to (446.center);
		\draw (447.center) to (448.center);
		\draw (450.center) to (449.center);
		\draw (455.center) to (454.center);
		\draw (454.center) to (453.center);
		\draw (453.center) to (452.center);
		\draw (452.center) to (455.center);
		\draw (456.center) to (457.center);
		\draw (459.center) to (458.center);
		\draw (464.center) to (463.center);
		\draw (463.center) to (462.center);
		\draw (462.center) to (461.center);
		\draw (461.center) to (464.center);
		\draw (469.center) to (468.center);
		\draw (468.center) to (467.center);
		\draw (467.center) to (466.center);
		\draw (466.center) to (469.center);
		\draw (474.center) to (473.center);
		\draw (473.center) to (472.center);
		\draw (472.center) to (471.center);
		\draw (471.center) to (474.center);
		\draw (477.center) to (476.center);
		\draw (482.center) to (481.center);
		\draw (481.center) to (480.center);
		\draw (480.center) to (479.center);
		\draw (479.center) to (482.center);
		\draw (483.center) to (484.center);
		\draw (486.center) to (485.center);
		\draw (491.center) to (490.center);
		\draw (490.center) to (489.center);
		\draw (489.center) to (488.center);
		\draw (488.center) to (491.center);
		\draw (492.center) to (493.center);
		\draw (495.center) to (494.center);
		\draw (500.center) to (499.center);
		\draw (499.center) to (498.center);
		\draw (498.center) to (497.center);
		\draw (497.center) to (500.center);
		\draw (505.center) to (504.center);
		\draw (504.center) to (503.center);
		\draw (503.center) to (502.center);
		\draw (502.center) to (505.center);
		\draw (510.center) to (509.center);
		\draw (509.center) to (508.center);
		\draw (508.center) to (507.center);
		\draw (507.center) to (510.center);
	\end{pgfonlayer}
\end{tikzpicture}$$
    
    \caption{Single qubit axioms for the four-directional dataset. a,b. the axioms relating  $\ket{\texttt{North}}$ to  $\ket{\texttt{South}}$, and  $\ket{\texttt{East}}$ to  $\ket{\texttt{West}}$ hold approximately. c. The model fails to satisfy the final axiom relating $\ket{\texttt{North}}$ to $\ket{\texttt{west}}$. Note that the initial states are coloured to as a visual aid only.
    }
    \label{fig:interpret/4dir-1q-axioms}
\end{figure}

\begin{figure}[ht]
    \centering
    \includegraphics[width=\textwidth]{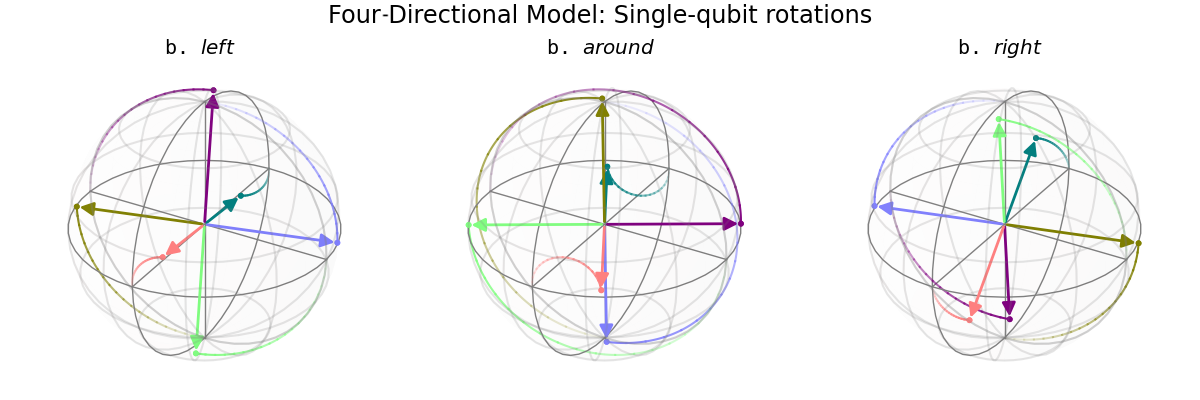}
    \caption{Four-directional single qubit rotations. \texttt{around} and \texttt{right} are defined in terms of \texttt{left}.
    }
    \label{fig:interpret/4dir/1q-rotations}
\end{figure}

An equivalent consideration of the single qubit axioms for the four-directional model fails for certain axioms, as demonstrated in \cref{fig:interpret/4dir-1q-axioms}, due to the directional pairings being incompatible with the quarter-turn \texttt{left}.
We also visualise the derived \texttt{right} and \texttt{around} rotations for the four-directional model in \cref{fig:interpret/4dir/1q-rotations}.


\begin{figure}[ht]
    \centering
    \begin{tabular}{c}
        \includegraphics[width=\textwidth]{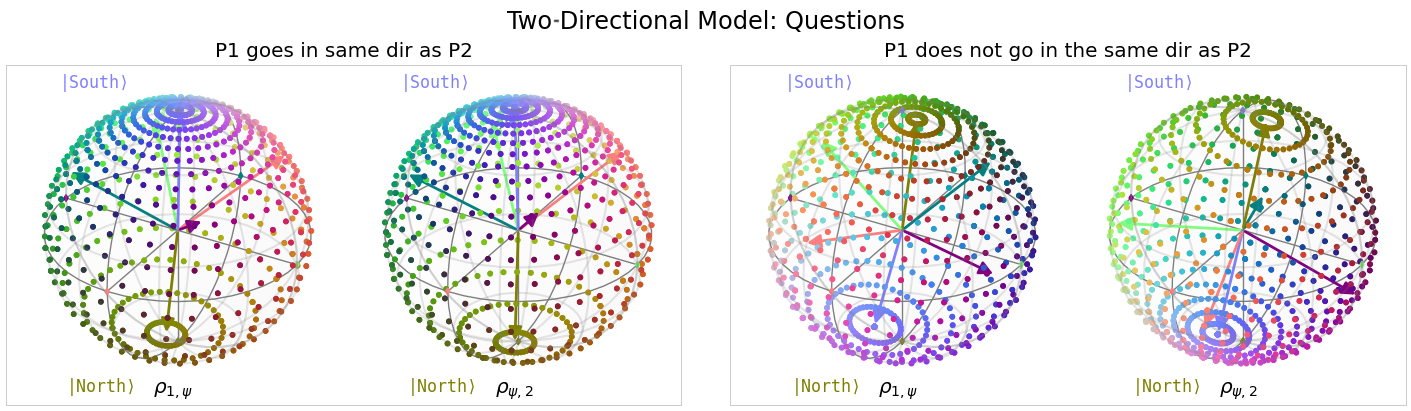} \\
        (a) \\
        \\
        \includegraphics[width=\textwidth]{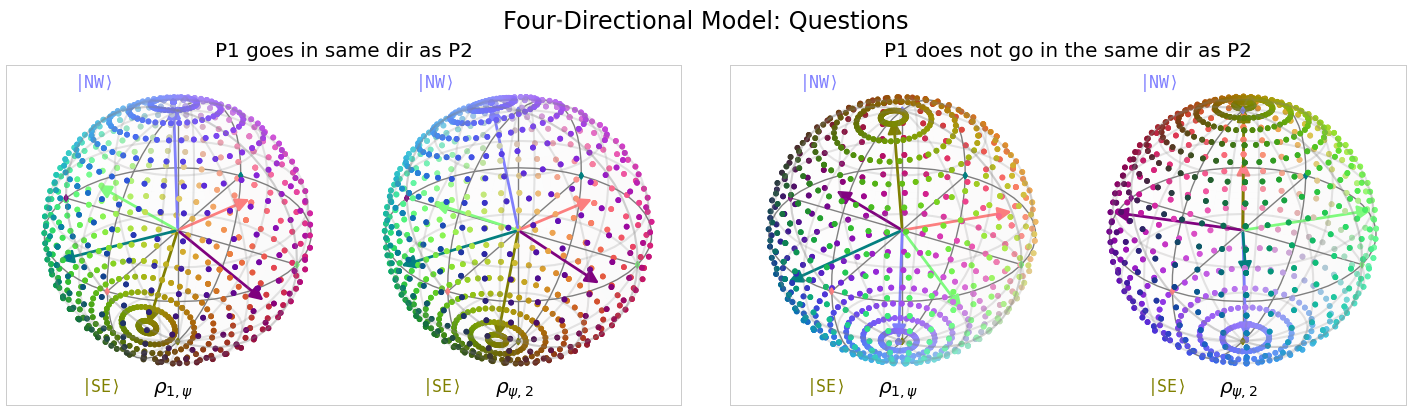} \\
        (b)
    \end{tabular}
    \caption{
    Visualising the questions for the two and four direction models. The main directions distinguished by the model are labelled and coloured according to the state identified. These are all maximally entangled pure states as both paired surfaces extend over their entire spheres. a. Two directional model i. \texttt{same dir}: we see here that $\rho_{1, \psi} \approx \rho_{\psi, 2}$, such that states are correlated along $\ket{\texttt{North, North}}$ and $\ket{\texttt{South, South}}$ with no phase shift. We also see that the possible states $\rho_{1, \psi}, \rho_{\psi, 2}$ are slightly concentrated towards the $\ket{0}$ state, which suggests the model can tolerate more error when the final state of the text \textit{should} be $\ket{\texttt{South, South}}$, but has instead deviated towards the equator. 
    ii. \texttt{not same dir}: Conversely, the negative question expresses an anti-correlation along $\ket{\texttt{North}}$ - $\ket{\texttt{South}}$, with a slight phase shift between the two surfaces. $\rho_{1, \psi}$ and $\rho_{\psi, 2}$ also exhibit a slight skew towards $\ket{\texttt{South}}$ and $\ket{\texttt{North}}$ respectively, which would allow the model to tolerate more error when the final text state is expected to be $\ket{\texttt{South, North}}$.
    b. Four directional model:  This model exhibits similar behaviour to the two-directional one, with similar, though less prominent, skews (recall that $\ket{\texttt{NW}} \approx \ket{0}$ for the four directional model), such that \texttt{same dir} favours $\ket{\texttt{SW, SW}}$ and \texttt{not same dir} favours $\ket{\texttt{NE, SW}}$.
    }
    \label{fig:interpret/questions}
\end{figure}

Visualising the question states, as in Fig.~\ref{fig:interpret/questions}, we see that in the case of both datasets, the model differentiates between correlation and anti-correlation primarily along the basis along which the directions are encoded. In the four directional case especially, the correlations along complementary bases become more relevant due to the presence of quarter-turns. Indeed, the model remains able to correctly classify a majority of instances where one actor undergoes a quarter-turn after a \texttt{follows} gate.

\begin{figure}[ht]
    \centering
    \includegraphics[width=\textwidth]{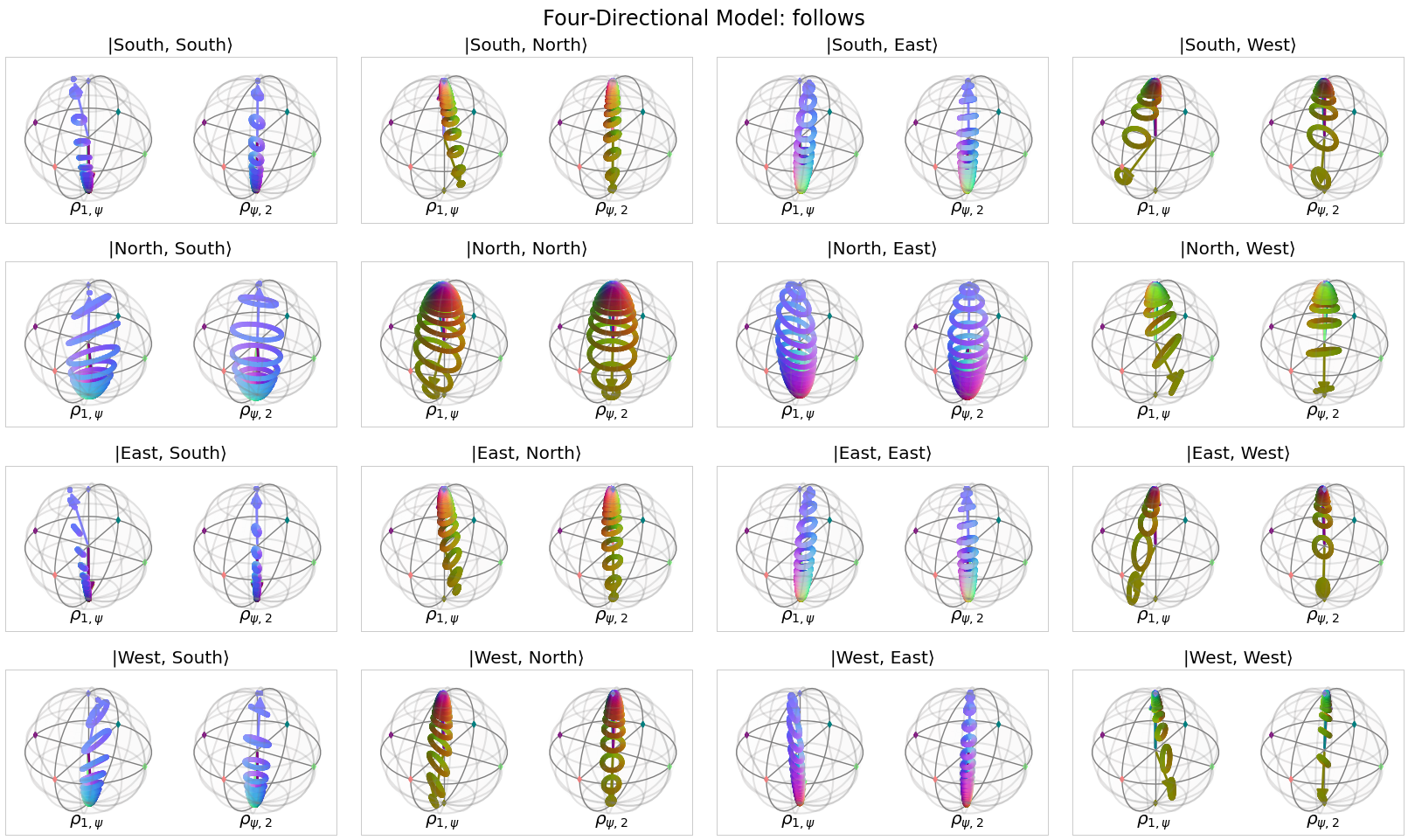}
    \caption{
    Full characterisation of the four direction following gate. Each paired Bloch sphere labelled $\ket{x, y}$ describes the state $U(\theta_{\texttt{follows}})\ket{x, y}$. We see the pairing of directions, as well as the independence of the final state on the initial state $\ket{x}$ of the first qubit.
    }
    \label{fig:interpret/4dir/following-full}
\end{figure}

Fig.~\ref{fig:interpret/4dir/following-full} shows the effect of applying \texttt{follows} to a pair of actors facing each of the possible initial cardinal directions. As with the two directional model, the action of \texttt{follows} is largely independent of the state of the subject qubit, however the grouping of \texttt{North} with \texttt{West} and \texttt{South} with \texttt{East} is evident.

\subsection{Extra accuracy/bias analysis}
\label{app:acc-bias}

\begin{figure}[ht]
    \centering
    \includegraphics[width=\textwidth * 2/ 3]{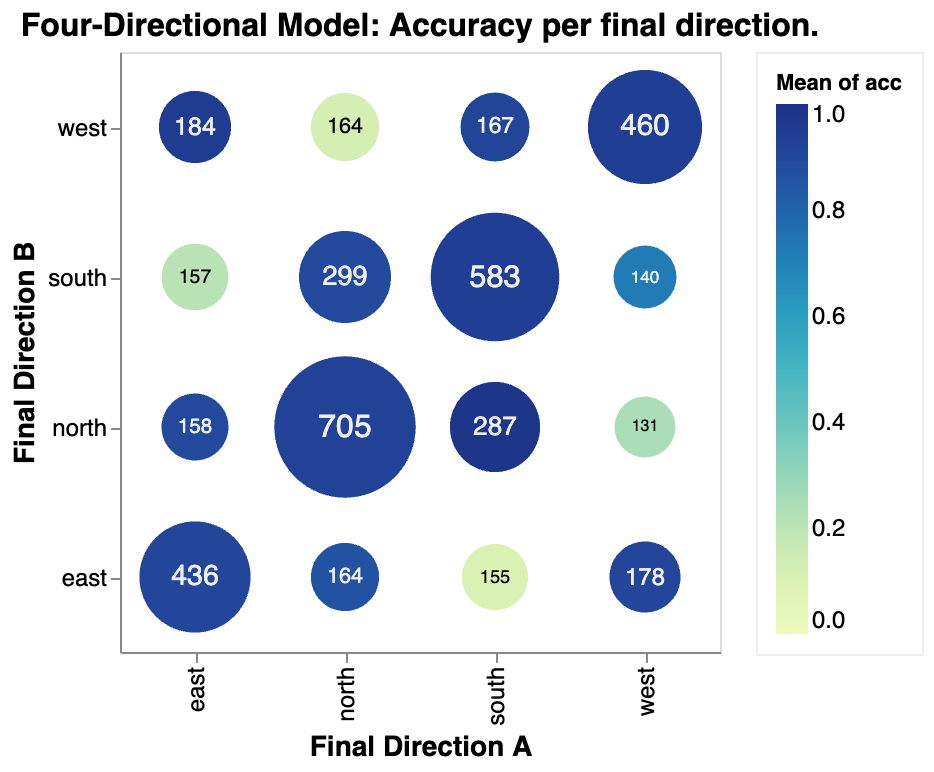}
    \caption{Comparing model accuracy against the final direction faced by each person. The accuracy is much worse when the final directions pair \texttt{North} with \texttt{West}, or \texttt{South} with \texttt{East}. This corresponds to the internal pairings the model makes.}
    \label{fig:interpretability/acc-final-dir}
\end{figure}

Fig.~\ref{fig:interpretability/acc-final-dir} shows the accuracy for different combinations of directions the actors involved in the final question face, contextualised by the number of such data points. We can see that the model correctly classifies stories in which the actors are going in the same or opposite direction. Stories where the actors are positioned at 90 degrees to each other are only classified correctly for the combinations \texttt{north-east} and \texttt{south-west}. These are the cases where the directions are located on opposite poles of the Bloch sphere in Fig.~\ref{fig:interpret/4dir/directions}~a. due to the clustering of directions. The figure also shows that there are only very few instances with people facing directions at 90 degrees to each other compared to the other combinations. The high amount of stories with people going in the same direction arises from the need to generate a balanced dataset in terms of labels. Other than that the proportion of different combinations of directions was not controlled in the data generation and the strong imbalance originates from favouring two-actor actions (which result in people facing the same or opposite directions) over single-actor actions (which can result in all possible combinations including 90 degrees angles between the actors) in order to increase story density. Therefore the denser the dataset, the fewer quarter-turns in the stories.
Thus, the performance of the four-directional model can be seen to arise by exploiting this unintentional bias in our dataset. 

\begin{figure}[ht]
    \centering
    \includegraphics[width=\textwidth ]{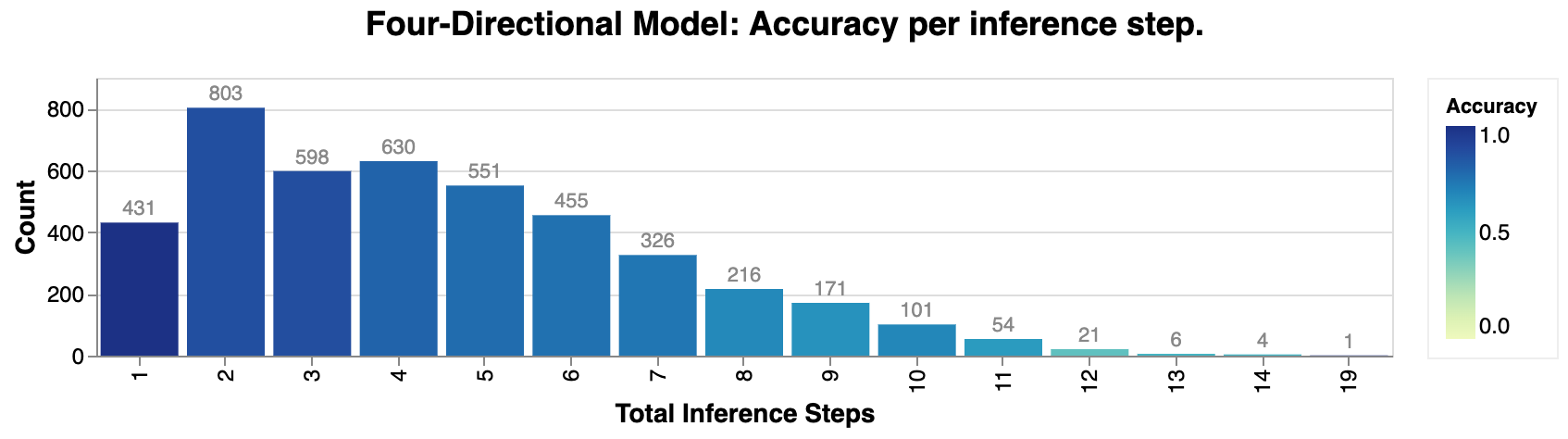}
    \caption{Model accuracy against the total inference steps required to solve the question. The height of the bars corresponds to the number of datapoints in that region, which is also labelled. The accuracy tends to decrease as the inference steps are increased.}
    \label{fig:interpretability/acc-inf_steps}
\end{figure}

Fig.~\ref{fig:interpretability/acc-inf_steps} shows the correlation between accuracy and the number of sentences required to solve the task. This metric intends to capture the task's expected `difficulty'. This particular metric was not controlled during data generation, and favouring connectivity in our generation method resulted in most stories requiring fewer inference steps. We see that the model struggles more with the `harder' instances. 
From Sec.~\ref{sec:comp_interpret_4_dir} we know that \texttt{turns right, turns left} and \texttt{turns around} introduce an error that accumulates and results in an accuracy drop for stories that require more inference steps.

\section{Tensor contraction resources}
\label{app:classical-resources}

\begin{figure}[ht]
    \centering
    \includegraphics[scale=0.65]{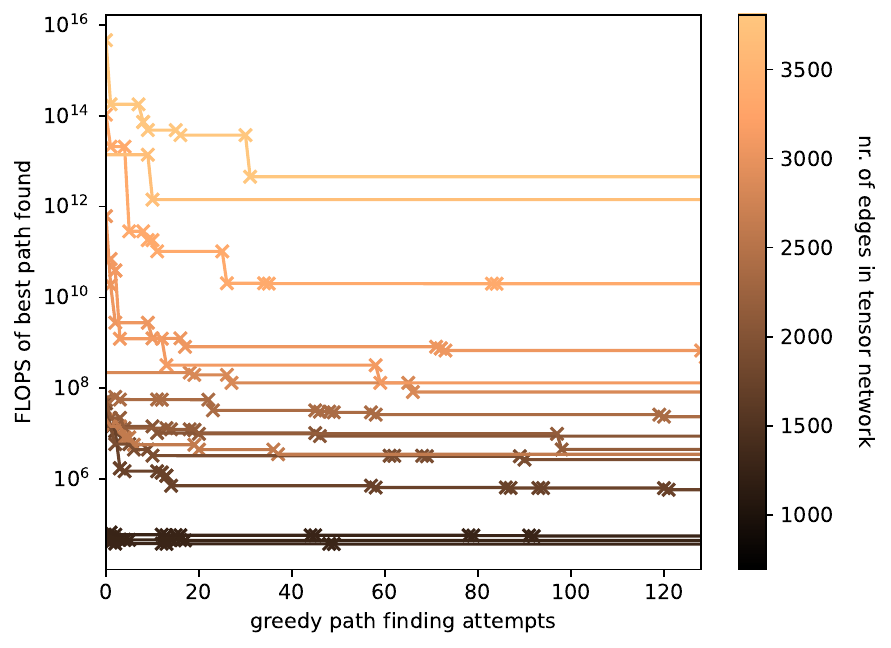}
    \caption{FLOPS required to perform a tensor contraction using the path that minimises this number plotted against the number of random greedy path optimiser runs performed so far. We see that drops in the required number of FLOPS become more sporadic as the number of runs increases, suggesting that there are diminishing returns in the time spent looking for the optimal path. Data is for $15$ points out of ``dense" class of the four-directions ``following" dataset, sampled uniformly according to the total nr. of edges in the tensor network, which is illustrated using the colour of the plot lines.
    }
    \label{fig:flops_vs_nr_of_greedy_runs}
\end{figure}

\begin{figure}[ht]
    \centering
    \includegraphics[scale=0.65]{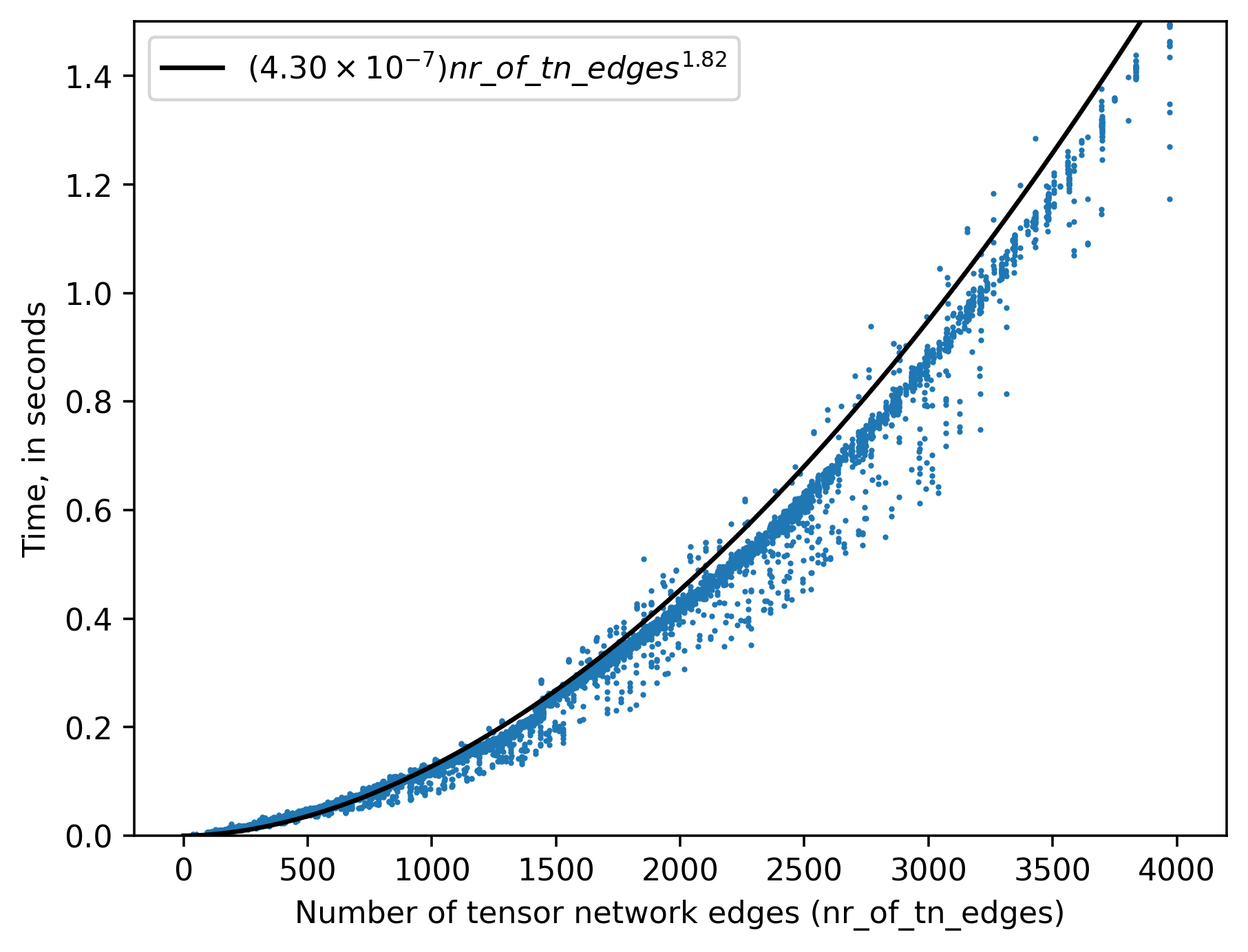}
    \caption{Time spent computing a single run of the random greedy path optimiser plotted against the number of edges in the tensor network. We see an approximately quadratic relationship between these quantities, for which we perform a fit with parameters given in the plot legend. Points correspond to the full four-directions ``following" dataset, where only one of the positive and negative answer circuits is picked for each point (these circuits have the same tensor structure). 
    }
    \label{fig:time_vs_nr_of_greedy_runs}
\end{figure}

Throughout this work, tensor contraction paths were found by running a stochastic greedy optimizer in the \texttt{opt\_einsum} Python package \cite{Smith2018}. This optimiser works by building the contraction step-by-step, at each stage choosing between the available contractions with a probability given by a Boltzmann distribution, i.e. with a probability proportional to
\begin{equation}
    \exp\left(-\frac{\epsilon}{kT}\right)
\end{equation}
where $\epsilon$ is determined by the number of FLOPS (floating-point operations per second) the contraction takes. This process repeats a set number of times $N$ and the path with the lowest total number of FLOPS is returned. In Fig.~\ref{fig:flops_vs_nr_of_greedy_runs}, we plot the number of FLOPS of the best contraction path found against the number of times the random greedy optimiser has been run. We see a large initial drop in the number of FLOPS required as the number of runs increases, and then sporadic drops in this number. For the experiments in this paper, we found that $128$ runs represented a good trade-off between time spent looking for an optimal path and improvement in the contraction path. In Fig.~\ref{fig:time_vs_nr_of_greedy_runs} we plot the time taken to compute one run of the random greedy optimiser as a function of edges present in the tensor network for the four-directional ``following" dataset.

\begin{figure}[ht]
    \centering
    \includegraphics[width=\textwidth]{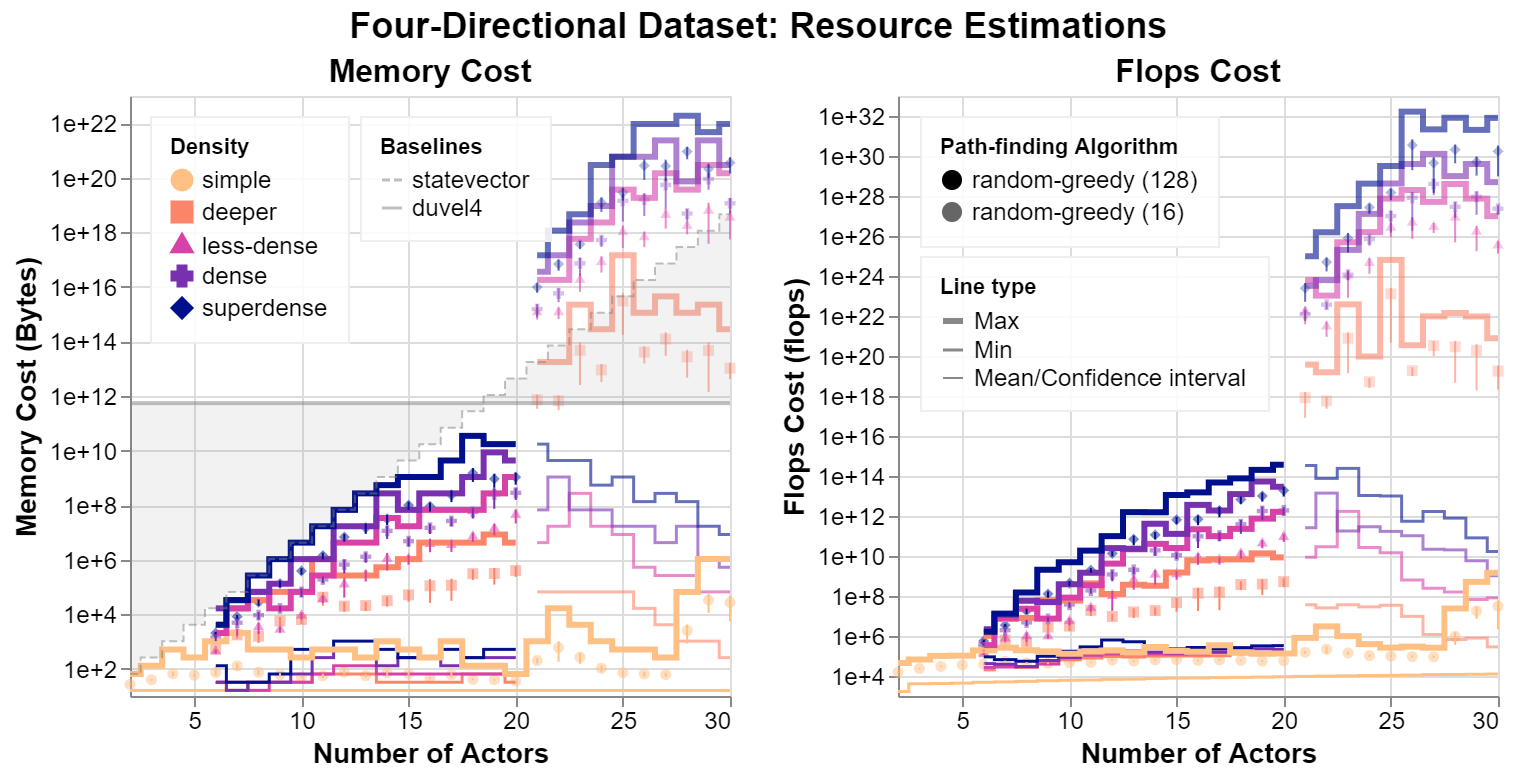}
    \caption{Resource estimates for the four-directional dataset, which are done as in Fig.~\ref{fig:resources/2-dir-summary} for the two-directional dataset. For each density, we plot the required memory and FLOPS for the exact evaluation of each data point as a full tensor contraction. In each case, the thickest line shows the most expensive contraction, while the thin line gives the lower bound. The mean for each number of actors width is plotted as a point, with a vertical line showing the confidence interval for the mean. Fewer iterations were used to compute the paths for data points with more than 20 actors, to keep the time spent reasonable (see the rest of App.~\ref{app:classical-resources} for further details). We show two baselines for the memory cost to contextualise this data. To train and evaluate the models we used Quantinuum's \textsl{duvel4} server, which has 512GB of memory. The second baseline reflects performing a statevector-style simulation of the circuit. We assume the statevector requires at most $n + 1$ qubits, that is: one qubit per actor, $n$ actors, and a single reused ancilla qubit. The circuit must be evaluated as a doubled state (or density matrix) due to the presence of discards. The total memory cost is then $2^{2(n + 1)}$.
    }
    \label{fig:resources/4-dir-summary}
\end{figure}

All simulations involving tensor contractions were run on our \textsl{duvel4} machine on 16-core hyper-threaded Intel\textsuperscript{\textregistered} Xeon\textsuperscript{\textregistered} Gold 5317 CPU and 512GB of RAM. While we experimented with using a GPU to perform the tensor contractions, we found no improvement over using a CPU for this purpose. This is because the tensor networks, having different shapes, can not be directly batched and executed in parallel on a GPU. While further processing could be employed to batch sub-components of the tensor network having the same shape, we leave this for future work. In Fig.~\ref{fig:resources/4-dir-summary}, we estimate the tensor contraction costs for the four-directional dataset for the best path found; this is also done in Fig.~\ref{fig:resources/2-dir-summary} of the main text for the two-directional dataset.

\section{Conversion of ansatz to H1-1 native gates} \label{sec:ansatz_to_hardware}

The ansatz employed (see Section~\ref{sec:qdiscocirc}) uses $CR_X$ $2$-qubit gates and $R_Z$ $1$-qubit gates. The $R_Z$ gate is native to the H1-1 archtecture~\cite{h11_data_sheet}, while the $CR_X$ gate can easily be converted to a $R_{ZZ}$ gate. Indeed, given $i\neq j$ indexing qubits of the device, it is the case that
\begin{align}
    CR_X(\theta, i,j) = H(j) CR_Z(\theta, i, j) H(j) = H(j) R_Z\left(-\frac{\theta}{2}, i\right) R_Z\left(-\frac{\theta}{2}, j\right) R_{ZZ}\left(\frac{\theta}{2}, i, j\right) H(j),
\end{align}
where $H(j)$ are Hadamard gates acting on the $j$th qubit and $CR_X(\theta, i,j)$ is a controlled $X$ rotation acting on the $j$th qubit controlled by the $i$th qubit (and analogously for $CR_Z$). In practice, this conversion between $2$-qubit gates is not explicitly performed and is instead delegated to the \texttt{TKET} compiler~\cite{tket}.

\section{Noise simulation details} \label{sec:noise_details}

The depolarising channel on two qubits $\{i, j\}$ acting on a mixed state $\sigma$ can be written as
\begin{align}
    D(\sigma) &= (1-p)\sigma + p \text{Tr}_{\{i,j\}}(\sigma) \otimes \frac{I_2}{4}, \label{eq:noise} \\
    &= \sum_{P \in \mathcal{P}^2} \alpha_P P \sigma P, \\
    \alpha_P &= \begin{cases}
    1 - \frac{15}{16}p, &\text{if} \; P = I \otimes I\\
    \frac{p}{16}, &\text{otherwise}
    \end{cases}\label{eq:noise_kraus}
\end{align}
where $\frac{I_2}{4}$ is the fully mixed state on the qubits, $\text{Tr}_{\{i,j\}}$ is the partial trace over the subsystem defined by the qubits, and $p$ is the probability of noise occurring.

We simulate noise using a sampling approach as follows: suppose we wish to apply a $2$-qubit gate $U$, but our system is affected by a depolarising noise channel. Suppose also that we are interested in computing the expectation value $\text{tr}(\sigma O)$ of some operator $O$. Assuming a pure initial state $\sigma_0 = \ket{\psi_0}\bra{\psi_0}$, this expectation can be expressed as
\begin{align}
\alpha_{I \otimes I} \langle U\ket{\psi_0} \rangle_O  + \sum_{P \in \mathcal{P}^2 \setminus I \otimes I} \alpha_P \langle P \ket{\psi_0} \rangle_O
\end{align}
The above equality means that we can sample expectation values of several pure state simulations according to the probabilites $\alpha_P$ to arrive at the expectation value of the operator $O$ under the noise model, which would otherwise require mixed state simulation. Note that, in practice, we will have multiple noisy gates in our circuit. To apply this technique, every time we encounter a noisy gate, we sample and apply a gate according to the probabilities $\alpha_P$. This technique is often called \emph{quantum trajectories} or \emph{Monte Carlo (wavefunction) simulation}. It avoids the memory required to simulate mixed states using density matrices (which is double that of pure state simulation using statevectors) in exchange for an approximate result that depends on the number of samples taken (but which is unbiased, i.e. will converge to the exact value as the number of samples increases). We use qujax~\cite{qujax}, a Python library which leverages Google's JAX \cite{jax2018github} platform, to just-in-time compile the statevector simulations using the XLA~\cite{xla} framework. This greatly speeds up simulation, and allows us to batch circuit execution, which can be done over both the number of samples and the noise strengths used. These simulations were run on our \textsl{duvel1} machine, on a NVIDIA A100 80GB GPU.

\section{Classical baselines}
\label{app:baselines}
We trained a transformer and an LSTM neural network to explore their generalisation capabilities across the two-directional and four-directional datasets. It is important to note that there is no fair direct comparison between the quantum and classical performances, due to inherent differences in their frameworks. 

An implementation choice we made in training the QDisCoCirc model that could not be replicated in the classical baselines is replacing all names with \texttt{person} (see App.~\ref{app:word-func}). Instead, we opted for a uniform representation of all 30 names occurring in the \textit{Train, Valid Comp and Test} datasets. We randomly replaced the names in the \textit{Train} dataset with names from this distribution and used this alternative \textit{Train} dataset for training the classical baselines to make sure the model learns representations for all names occurring.
As a result, the two-directional vocabulary comprises approximately 62\% names and the four-directional 58\%, while the QDisCoCirc model has to learn only the word \texttt{person}. 

Another crucial difference is the absence of hardcoded semantic rewrites, as opposed to the QDisCoCirc pipeline, see Fig.~\ref{fig:axioms}. However, we maintained the same \textit{Train}/\textit{Valid A} and \textit{Valid Comp} splits as shown in Table~\ref{tab:data_splits} of the main text. The two-directional models were trained only on the simple dataset, to mirror the training methodology from Sec.~\ref{sec:train-val-test-split}

Furthermore, we evaluated GPT-4  ~\cite{openai2023gpt4} and tracked its performance across all densities of both datasets. What follows is a description of the training methodology and a detailed evaluation of the performance of the classical models.

\subsection{Transformer}
We trained a custom made transformer neural network to explore its generalisation capabilities across the two-directional and four-directional datasets. We used \texttt{PyTorch}~\cite{Paszke_PyTorch_An_Imperative_2019} to track the parameters and the Adam optimizer ~\cite{adam_kingma_2015} to train them. The hyperparameters we considered and their final tuned values are summarised in Table~\ref{tab:TransfHyperparams} below. We tuned the hyperparameters using \texttt{Ax}~\cite{bakshy2018, AxTuningRepo}, and aimed for the best \textit{Valid A} accuracy.

\begin{table}[ht]
    \centering
    \begin{tabular}{| c | c | c | c |}
    \hline
    \rule[-0.5em]{0pt}{1.5em} \bf{hyperparameter} & \bf{range/choice values} & \multicolumn{2}{c|}{\bf{final value}} \\
    \cline{3-4}
    \rule[-0.5em]{0pt}{1.5em} & & \bf{two-directional} & \bf{four-directional} \\
    \hline
    \rule[-0.5em]{0pt}{1.5em} learning rate & 0.0001 - 0.0003 & 0.00014 & 0.00017 \\
    \hline
    \rule[-0.5em]{0pt}{1.5em} batch size & 32, 64, 128 & 32 & 32 \\
    \hline
    \rule[-0.5em]{0pt}{1.5em} dropout & 0.3 - 0.75 & 0.413 & 0.565 \\
    \hline
    \rule[-0.5em]{0pt}{1.5em} hidden dimensions & 512 - 768 & 569 & 565 \\
    \hline
    \rule[-0.5em]{0pt}{1.5em} model dimensions & 256, 512, 768 & 256 & 512 \\
    \hline
    \rule[-0.5em]{0pt}{1.5em} number of layers & 1 - 4 & 4 & 3 \\
    \hline
    \rule[-0.5em]{0pt}{1.5em} number of heads & 2, 4, 8 & 2 & 4 \\
    \hline
    \rule[-0.5em]{0pt}{1.5em} seed & 0 - 1000 & 446 & 215 \\
    \hline
    \end{tabular}
    \caption{Training hyperparameters used and the ranges we considered when tuning the transformer on the two-directional and four-directional datasets.}
    \label{tab:TransfHyperparams}
\end{table}

\subsubsection{Transformer Architecture}

Our implementation is based on the transformer architecture introduced in ~\cite{vaswani2023attention} and adapted to our needs as described below.

The input tokens are processed through an embedding layer, the dimensionality of which varies between the two-directional and four-directional models, as detailed in Table~\ref{tab:TransfHyperparams}. Positional encoding ensures that the sequential nature of the data is retained. The core of the model is the transformer encoder, consisting of several layers. The model dimensionality, attention heads, hidden layers, and dropout rate differ across the two-directional and four-directional models, as indicated in Table~\ref{tab:TransfHyperparams}. The final linear classification layer maps the encoder's high-dimensional output to the target classes.

\begin{figure}[ht!]
    \centering
    \begin{subfigure}[b]{0.9\textwidth}
        \centering
        \includegraphics[width=\textwidth]{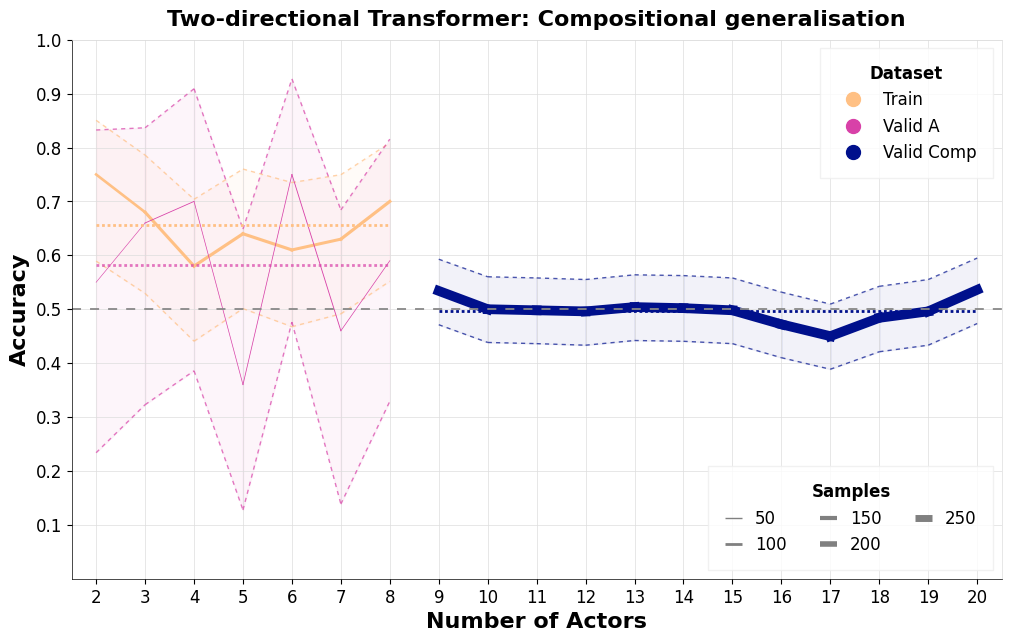}
        \label{fig:2dir_transf_compgen}
        (a)
    \end{subfigure}
    \hfill
    \begin{subfigure}[b]{0.9\textwidth}
        \centering
        \includegraphics[width=\textwidth]{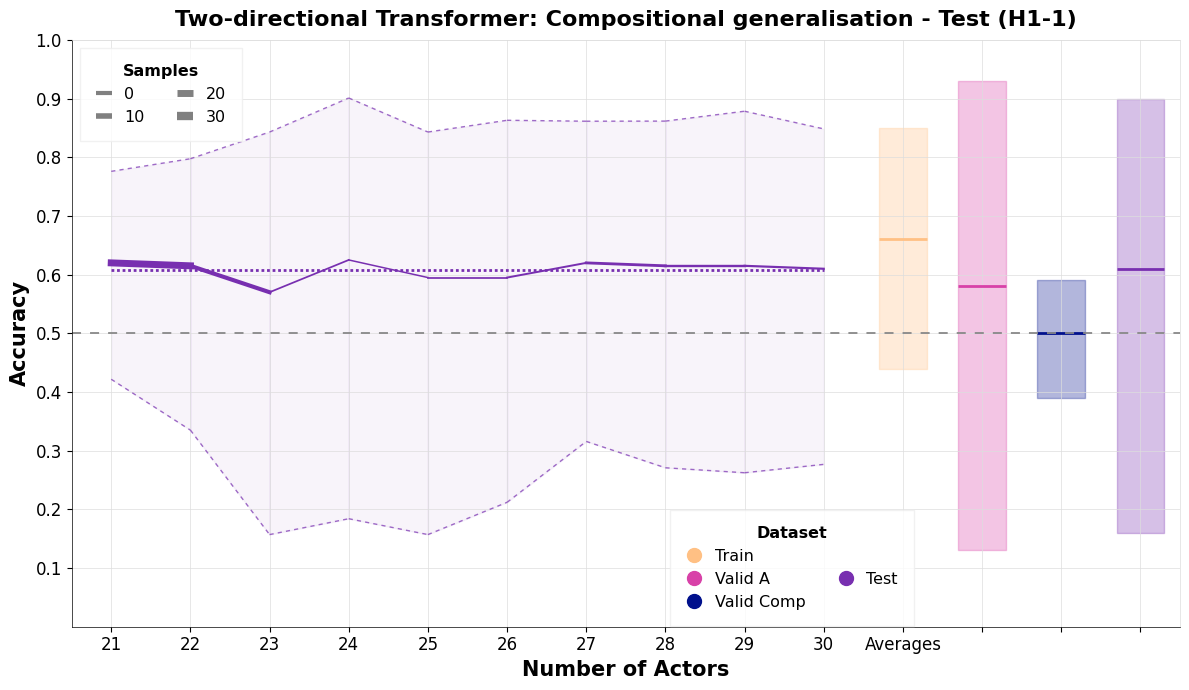}
        \label{fig:2dir_transf_test}
        (b)
    \end{subfigure}
    \caption{Compositional generalisation of the two-directional transformer across different datasets. Horizontal lines indicate average accuracy, and the dashed grey line shows the baseline accuracy. The shaded areas depict the 95\% Clopper-Pearson confidence interval for the mean. Line thickness varies with dataset size (see Table~\ref{tab:dataset_sizes}). (a) Performance on \textit{Train}, \textit{Valid A}, and \textit{Valid Comp} datasets. The model was trained only on the Simple dataset, mirroring the training methodology from Sec.~\ref{sec:train-val-test-split} (b) Performance on the \textit{Test (H1-1)} dataset. }
    \label{fig:2dir_transf_combined}
\end{figure}

\begin{figure}[ht!]
    \centering
    \includegraphics[width=\textwidth]{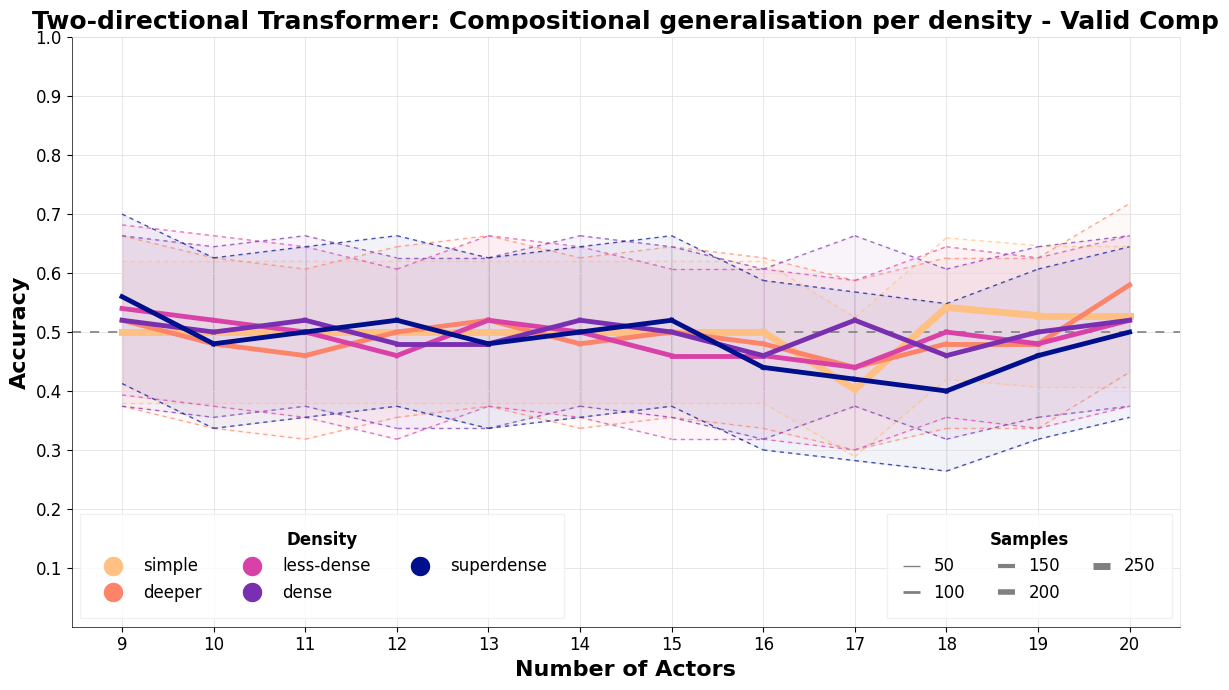}
    \caption{Compositional generalisation of the two-directional transformer, breakdown by density. The plot shows the \textit{Valid Comp} results exclusively, since the \textit{Train} and \textit{Valid A} datasets contain only the Simple density and their results can be found in Fig.~\ref{fig:2dir_transf_combined}(a). }
    \label{fig:2dir_transf_density}
\end{figure}

\begin{figure}[ht!]
    \centering
    \begin{subfigure}[b]{.9\textwidth}
        \centering
        \includegraphics[width=\textwidth]{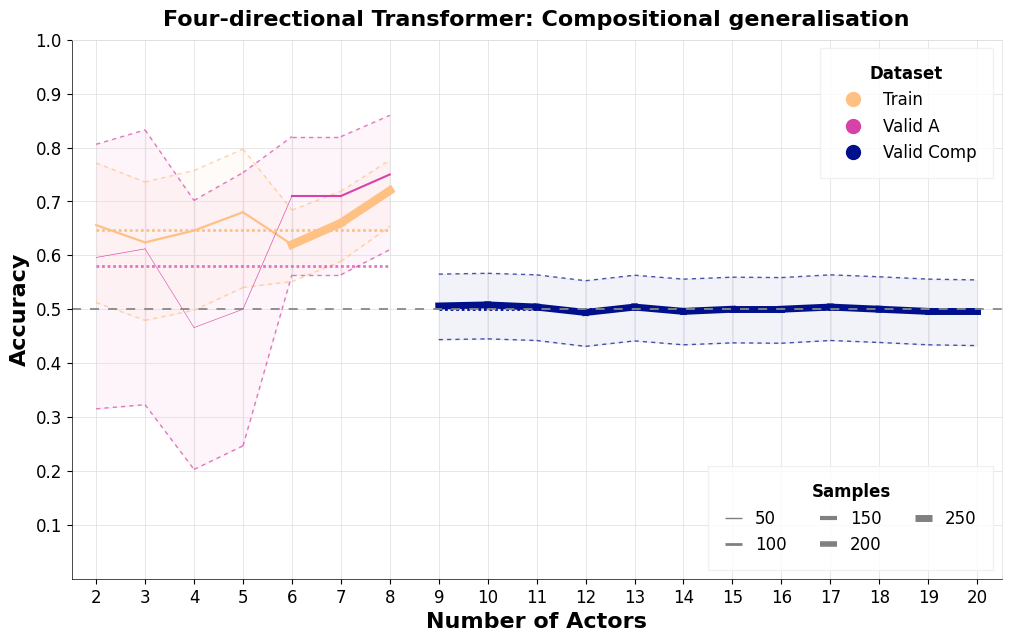}
        \label{fig:4dir_transf_compgen}
        \\
        (a)
    \end{subfigure}
    \hfill
    \begin{subfigure}[b]{\textwidth}
        \centering
        \includegraphics[width=.9\textwidth]{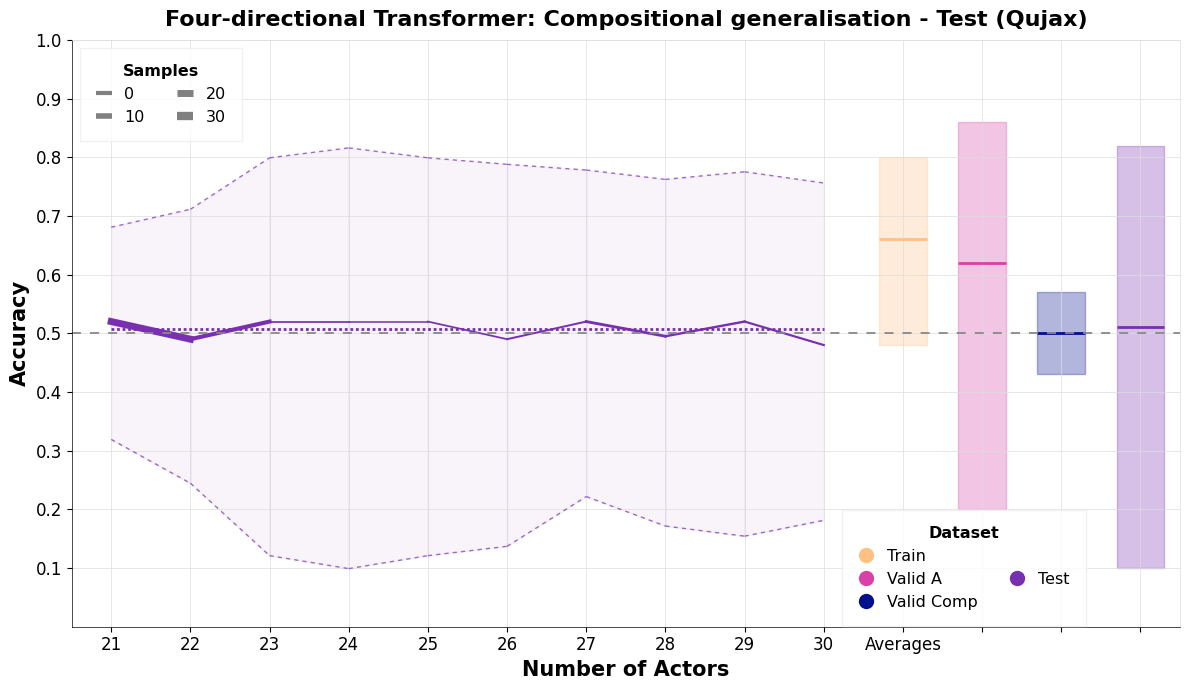}
        \label{fig:4dir_transf_test}
        \\
        (b)
    \end{subfigure}
    \caption{Compositional generalisation of the four-directional transformer across different datasets. Horizontal lines indicate average accuracy, and the dashed grey line shows the baseline accuracy. The shaded areas depict the 95\% Clopper-Pearson confidence interval for the mean. Line thickness varies with dataset size (see Table~\ref{tab:dataset_sizes}). (a) Performance on \textit{Train}, \textit{Valid A}, and \textit{Valid Comp} datasets. (b) Performance on the \textit{Test (qujax)} dataset. %
    }
    \label{fig:4dir_transf_combined}
\end{figure}

\begin{figure}[ht!]
    \centering
    \includegraphics[width=\textwidth]{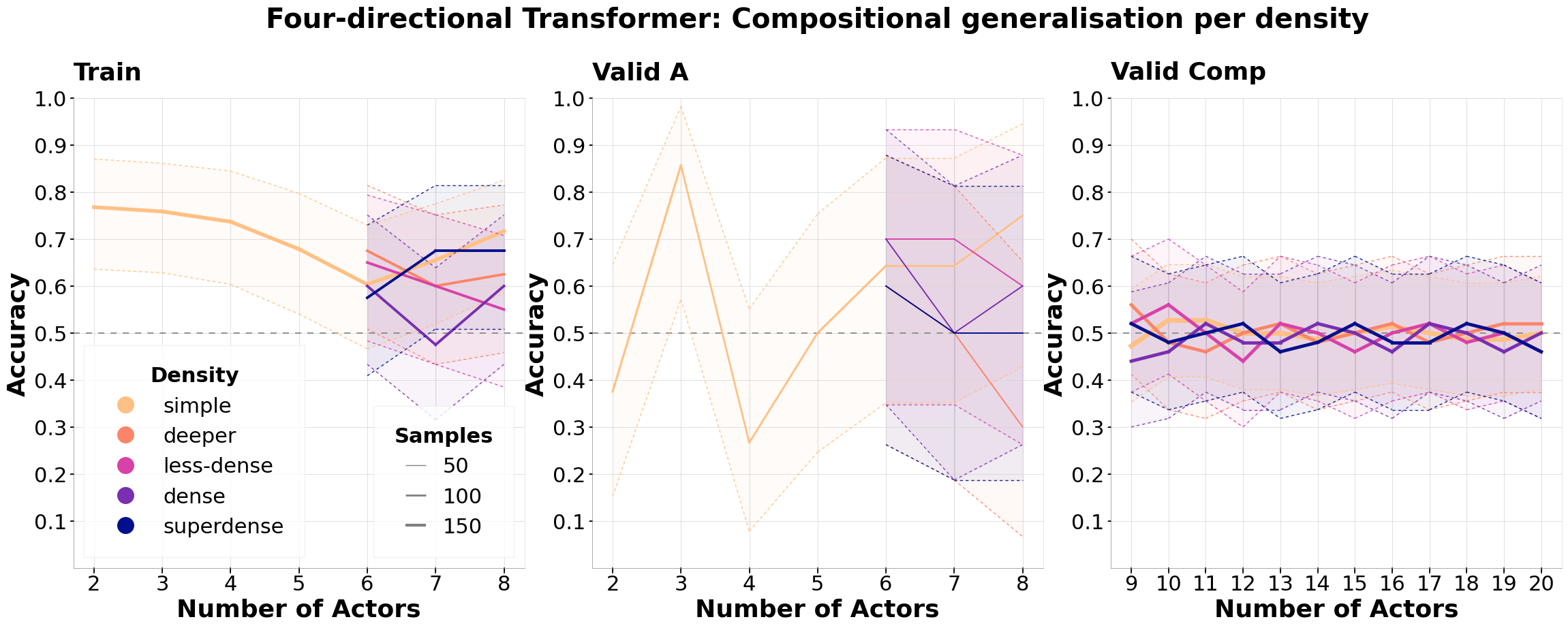}
    \caption{Compositional generalisation of the four-directional transformer, breakdown by density. } 
    \label{fig:4dir_transf_density}
\end{figure}

\subsubsection{Transformer Results: Two-Directional}

The \textit{Train} accuracy of the two-directional transformer model (Fig.~\ref{fig:2dir_transf_combined}(a)) fluctuates around 66\% across different text widths and never drops below 58\%. The confidence intervals drop below 50\% for text widths 4, 6 and 7.

The \textit{Valid A} accuracy follows a similar trend (Fig.~\ref{fig:2dir_transf_combined}(a)) surpassing the \textit{Train} accuracy on stories with 4 and 6
, but it is on average 8\% lower. Partially due to the smaller number of data points in the \textit{Valid A} dataset, the confidence interval drops well below 50\%.
When evaluated on stories from the \textit{Valid Comp} dataset with a higher number of actors in the stories,
the model’s performance is about 50\% across all text widths (Fig.~\ref{fig:2dir_transf_combined}(a)). We remind the reader that `text width' refers to the number of actors in a story. In our task, this performance means random guessing, suggesting that the model is not able to generalise productively on text widths 9 to 20. 

Figure~\ref{fig:2dir_transf_density} shows the \textit{Valid Comp} accuracy split by the different story densities. It shows that the model's accuracy is consistently random across all story densities for varying text widths and no particular story density outperforms or pulls down the overall accuracy.

When evaluating the transformer on the \textit{Test (H1-1)}
dataset (Fig.~\ref{fig:2dir_transf_combined}(b)) with a text width up to 30, we observe that performance hovers around 61\%, with a slight drop at 23 nouns. The confidence intervals fall significantly below 50\%, reaching as low as 15\%, partly due to the limited number of data points. In this scenario, the transformer appears to generalize well, performing better than on the \textit{Valid Comp} dataset, and even the \textit{Valid A} dataset (Fig.~\ref{fig:2dir_transf_combined}(b)).

To delve deeper into this performance, we tested the model on the entire `dense' and `superdense' \textit{Test} dataset with text width 21 to 30, rather than the selected subset of data points (which were the ones that compiled, with qubit reuse, down to 20 qubits or less due to hardware constraints) for the equivalent QDisCoCirc experiment. This performance was similar to that on the \textit{Valid Comp} dataset, with accuracy hovering around 52\%.

\subsubsection{Transformer Results: Four-Directional}

The \textit{Train} accuracy of the four-directional transformer model (Fig.~\ref{fig:4dir_transf_combined}(a)) fluctuates around 66\%. The \textit{Valid A} accuracy deviates less from the \textit{Train} accuracy for this model, by only about 4\% on average. While the confidence intervals of the training accuracies only drop below 50\% for text width 3, the confidence interval for \textit{Valid A} is below 50\% for most text widths, with the exception of 6, 7 and 8 nouns, largely due to the small number of datapoints. While there is an increase of the \textit{Valid A} accuracy on 6, 7 and 8 nouns, 
the four-directional transformer does not generalise to larger text widths from the \textit{Valid Comp} dataset and shows an accuracy of 50\% (Fig. ~\ref{fig:4dir_transf_combined}(a)).

When evaluating the transformer on the \textit{Test (qujax)} dataset (Fig.~\ref{fig:4dir_transf_combined}(b)), we see that the performance hovers around 51\%, with very slight fluctuations. The confidence intervals dip well below 50\%, reaching 10\%, due to the small number of datapoints. In the four-directional case, the model is not able to generalise.

Figure~\ref{fig:4dir_transf_density} shows the \textit{Valid Comp} accuracy split by the different story densities. It shows that the model's accuracy is consistently random across all story densities for varying text widths and no particular story density outperforms or pulls down the overall accuracy.

\subsubsection{Discussion and future work}
\label{app:transf-discussion}

We compare the trained transformer and QDisCoCirc models. The two-directional QDisCoCirc model achieves 100\% accuracy across the \textit{Train}, \textit{Valid A}, and \textit{Valid Comp} datasets, while the four-directional version maintains accuracy above 80\%. In contrast, both the two- and four-directional transformers consistently score below 80\%, with performance on the \textit{Valid Comp} dataset fluctuating around random guessing levels. 

The sharp drop in accuracy from 8 to 9 nouns -from \textit{Valid A} to \textit{Valid Comp}- indicates that the transformers were unable to generalize to texts with different distributions of text widths and text depths. In contrast, the QDisCoCirc model consistently outperforms the transformers, excelling at compositional generalization. This holds true for the four-directional \textit{Test (qujax)} dataset and even for the two-directional equivalent, when we consider the whole of the \textit{Test} set and not just the restricted set chosen due to hardware constraints, as further detailed in the analysis of the results.

When comparing the two models, it is essential to acknowledge the differences in the training methodology of the two models mentioned at the beginning of this section.  One reason for the model's poor performance might be that the characters' names comprise a large portion of the vocabulary. This causes the model to focus too much on specific character names and their behaviors, rather than understanding the underlying patterns of their interactions. In addition, our datasets are rather small compared to typical datasets used for training transformers. Thus, the limited \textit{Train} data (Table~\ref{tab:data_splits}, main text) is another main reason for the transformer's poor performance. 

For future work, we propose a more sophisticated and equitable experimental setup for the transformer models, while aligning closely with the QDisCoCirc training setup. This will involve modifications to the dataset: each data point corresponding to a story will be used to generate multiple samples. Each of these will maintain the original story structure but vary the names, aiming for a uniform representation across the augmented dataset. To compute the label for each original data point, all of the derived samples will be evaluated, and the final label will be calculated as the average across sub-stories. This intends to emulate the replacement of specific names with \texttt{person} in the QDisCoCirc setup.

Additionally, we could fix the pair of characters involved in the question (e.g., by fixing the question to always be `Does Bob go in the same direction as Alice?'), in all stories and their sub-stories. This second modification will allow the model to know which characters are relevant from the start of the text, simplifying the task and potentially allowing the model to learn a more compositional representation.

\subsection{LSTM}

We trained an LSTM neural network to explore its generalisation capabilities across the two-directional and four-directional datasets. We used \texttt{Tensorflow}~\cite{tensorflow2015-whitepaper} to track the parameters and
the Adam optimizer ~\cite{adam_kingma_2015} to train them.

The hyperparameters we considered and their final tuned values are summarised in Table~\ref{tab:LSTMhyperparams} below. We tuned for the best \textit{Valid A} accuracy using \texttt{Ax}~\cite{bakshy2018, AxTuningRepo}.

\begin{table}[ht]
    \centering
    \begin{tabular}{| c | c | c | c |}
    \hline
    \rule[-0.5em]{0pt}{1.5em} \bf{hyperparameter} & \bf{range/choice values} & \multicolumn{2}{c|}{\bf{final value}} \\
    \cline{3-4}
    \rule[-0.5em]{0pt}{1.5em} & & \bf{two-directional} & \bf{four-directional} \\
    \hline
    \rule[-0.5em]{0pt}{1.5em} learning rate & 0.00001 - 0.01 & 0.001 & 0.00008 \\
    \hline
    \rule[-0.5em]{0pt}{1.5em} batch size & 32, 64, 128 & 128 & 64 \\
    \hline
    \rule[-0.5em]{0pt}{1.5em} dropout & 0.35 - 0.65 & 0.39 & 0.38 \\
    \hline
    \rule[-0.5em]{0pt}{1.5em} l1 regularisation & 1e-6 - 1e-4 & 5e-5 & 8e-5 \\
    \hline
    \rule[-0.5em]{0pt}{1.5em} l2 regularisation & 1e-6 - 1e-4 & 3e-5 & 8e-5 \\
    \hline
    \rule[-0.5em]{0pt}{1.5em} hidden layers & 20 - 150 & 74 & 134 \\
    \hline
    \end{tabular}
    \caption{Training hyperparameters used and the ranges we considered when tuning the LSTM on the two-directional and four-directional datasets.}
    \label{tab:LSTMhyperparams}
\end{table}

\subsubsection{LSTM Results: Two-Directional}

\begin{figure}[ht!]
    \centering
    \begin{subfigure}[b]{\textwidth}
        \centering
        \includegraphics[width=.9\textwidth]{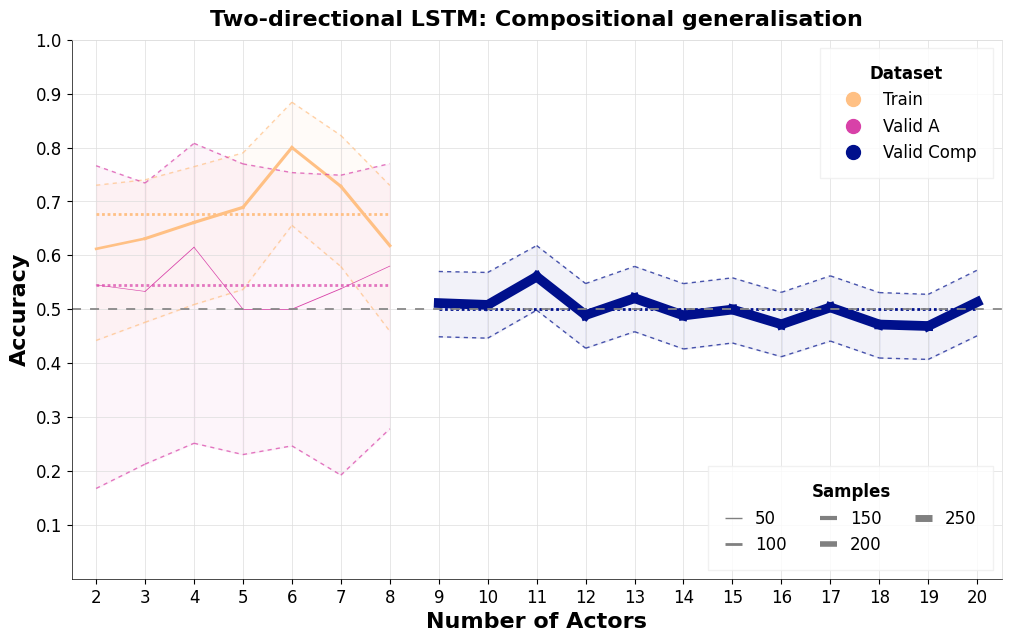}
        \label{fig:2dir_lstm_compgen}
        \\
        (a)
    \end{subfigure}
    \hfill
    \begin{subfigure}[b]{\textwidth}
        \centering
        \includegraphics[width=.9\textwidth]{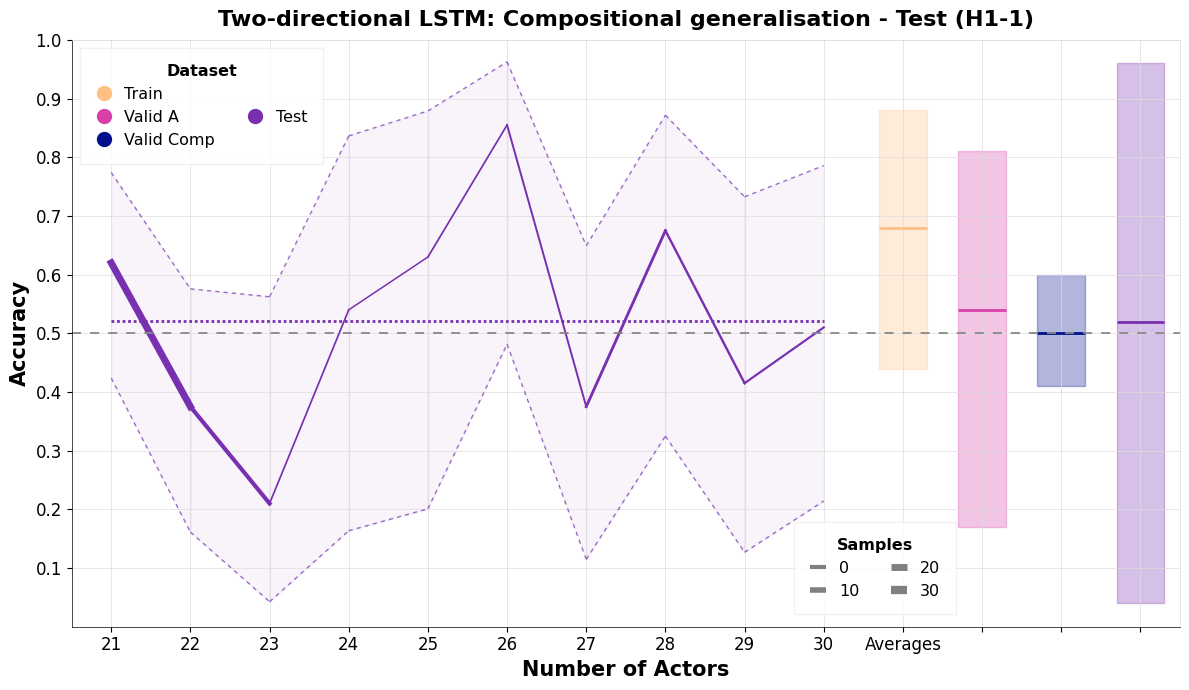}
        \label{fig:2dir_lstm_test}
        \\
        (b)
    \end{subfigure}
    \caption{Compositional generalisation of the two-directional LSTM across different datasets. The model was trained only on the Simple dataset, mirroring the training methodology from Sec.~\ref{sec:train-val-test-split}. Horizontal lines indicate average accuracy, and the dashed grey line shows the baseline accuracy. The shaded areas depict the 95\% Clopper-Pearson confidence interval for the mean. Line thickness varies with dataset size (see Table~\ref{tab:dataset_sizes}). (a) Performance on \textit{Train}, \textit{Valid A}, and \textit{Valid Comp} datasets. (b) Performance on the \textit{Test (H1-1)} dataset.}
    \label{fig:2dir_lstm_combined}
\end{figure}

\begin{figure}[ht!]
    \centering
    \includegraphics[width=\textwidth]{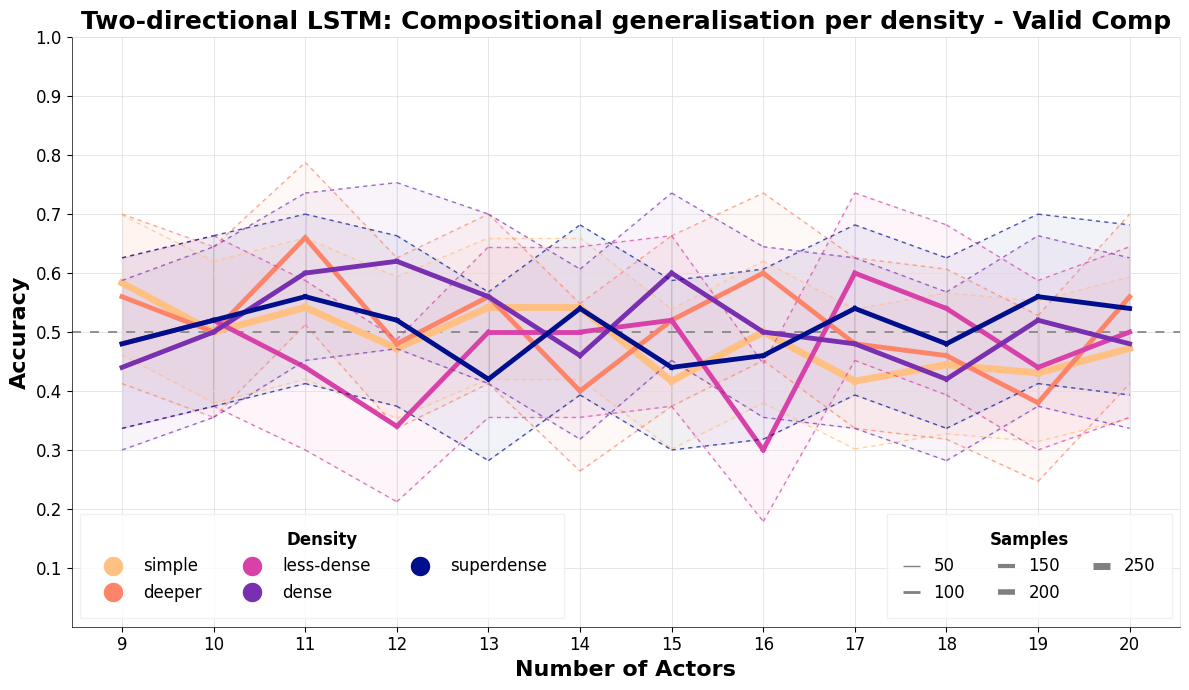}
    \caption{Compositional generalisation of the two-directional LSTM, breakdown by density. The plot shows the \textit{Valid Comp} results exclusively, since the \textit{Train} and \textit{Valid A} datasets contain only the Simple density and their results can be found in Fig.~\ref{fig:2dir_lstm_combined}(a). 
    }
    \label{fig:2dir_lstm_density}
\end{figure}

The \textit{Train} accuracy of the two-directional LSTM model (Fig.~\ref{fig:2dir_lstm_combined}(a)) fluctuates around 68\% across different text widths. The average validation accuracy \textit{Valid A} is lower at 54\%. The \textit{Valid Comp} accuracy fluctuates around 50\% for all story widths, which means random guessing for the presented task. It should be noted again that the particularly wide confidence intervals, especially for the \textit{Valid A} curve, are partly due to the smaller number of data points available. 

When evaluating the model on the \textit{Test (H1-1)} dataset (Fig.~\ref{fig:2dir_lstm_combined}(b)), we see that the performance hovers around 51\%, with prominent fluctuations across text widths.
The confidence intervals dip well below 50\%, reaching 10\%, due to the small number of datapoints. Considering that the accuracy is around random guessing, we conclude that the model is not able to generalise.

Fig.~\ref{fig:2dir_lstm_density} shows the \textit{Valid Comp} accuracy split by the different story densities. It shows that the model's accuracy is consistently random across all story densities for varying text widths and no particular story density outperforms or drags down the overall accuracy.

\subsubsection{LSTM Results: Four-Directional}

\begin{figure}[ht!]
    \centering
    \begin{subfigure}[b]{\textwidth}
        \centering
        \includegraphics[width=.9\textwidth]{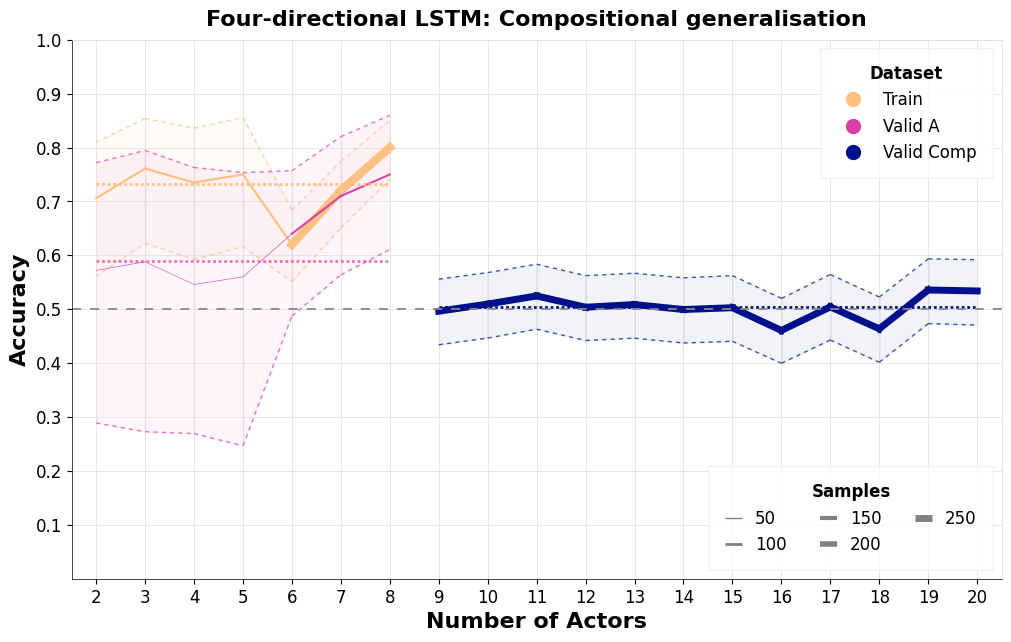}
        \label{fig:4dir_lstm_compgen}
        \\
        (a)
    \end{subfigure}
    \hfill
    \begin{subfigure}[b]{\textwidth}
        \centering
        \includegraphics[width=.9\textwidth]{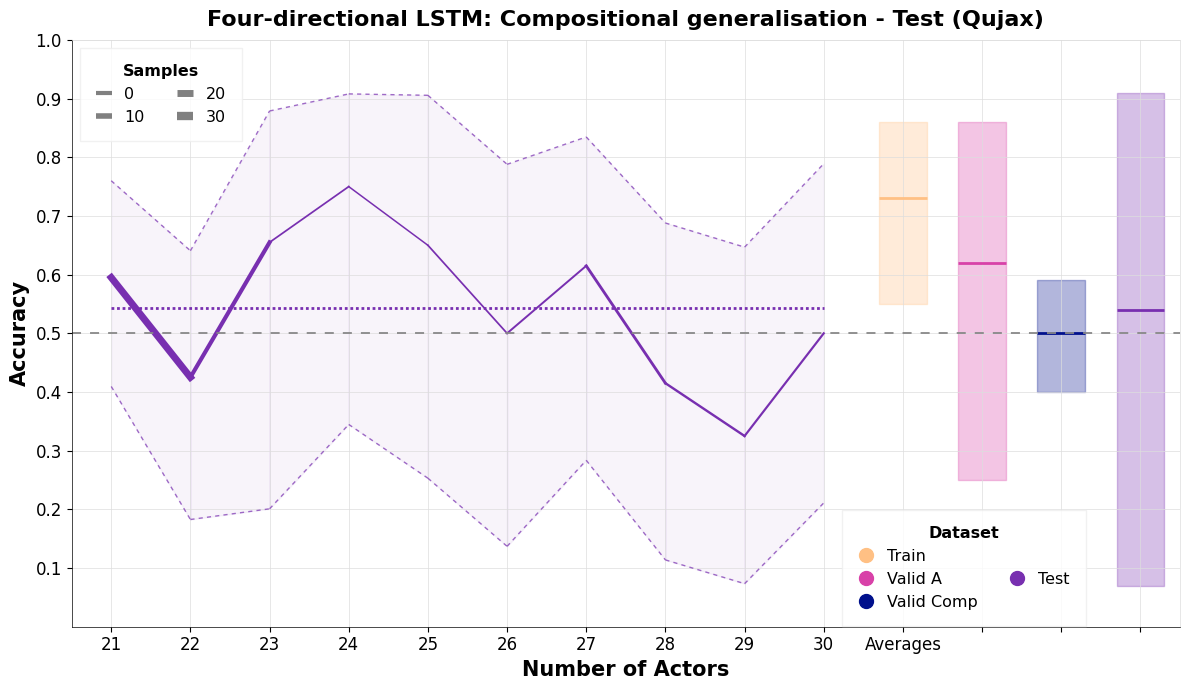}
        \label{fig:4dir_lstm_test}
        \\
        (b)
    \end{subfigure}
    \caption{Compositional generalisation of the four-directional LSTM across different datasets. Horizontal lines indicate average accuracy, and the dashed grey line shows the baseline accuracy. The shaded areas depict the 95\% Clopper-Pearson confidence interval for the mean. Line thickness varies with dataset size (see Table~\ref{tab:dataset_sizes}). (a) Performance on \textit{Train}, \textit{Valid A}, and \textit{Valid Comp} datasets. (b) Performance on the \textit{Test (qujax)} dataset.}
    \label{fig:4dir_lstm_combined}
\end{figure}

\begin{figure}[ht!]
    \centering
    \includegraphics[width=\textwidth]{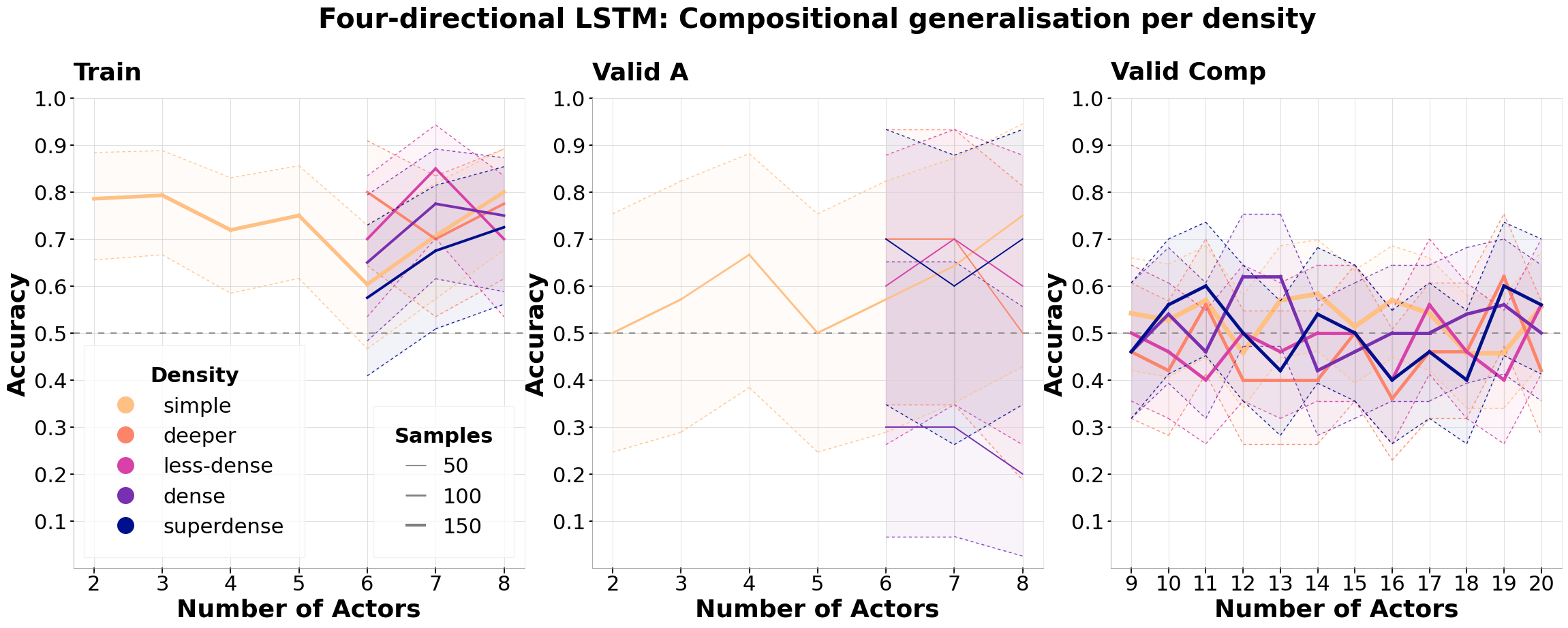}
    \caption{Compositional generalisation of the four-directional LSTM, breakdown by density.} 
    \label{fig:4dir_lstm_density}
\end{figure}

The \textit{Train} accuracy of the four-directional LSTM model (Fig.~\ref{fig:4dir_lstm_combined}(a)) fluctuates around 73\% and is relatively stable across text widths. The \textit{Valid A} accuracy is significantly lower than the \textit{Train} accuracy, at 62\% on average, suggesting substantial overfitting.

When evaluating the model on the \textit{Test (qujax)} dataset (Fig.~\ref{fig:4dir_lstm_combined}(b)), we see that the performance hovers around 54\%, again with prominent fluctuations present. The confidence intervals dip well below 50\%, reaching 7\%, due to the small number of datapoints. Considering that the accuracy on the \textit{Valid Comp} and \textit{Test (qujax)} datasets is around random guessing, we conclude that the model is not able to generalise.

Just like in the two-directional case, Fig.~\ref{fig:4dir_lstm_density} shows the \textit{Valid Comp} accuracy split by the different story densities. It shows that the model's accuracy is consistently random across all story densities for varying text widths and no particular story density outperforms or drags down the overall accuracy.

\subsubsection{Discussion}

In all cases the LSTM was trained on, the difference between the \textit{Train} and \textit{Valid A} accuracies was at least 10\%, indicating overfitting. The LSTM consistently performs worse than QDisCoCirc and is unable to generalise to larger text width instances on either the \textit{Valid Comp} or \textit{Test} datasets. The LSTMs' bad generalisation performance mirrors this architecture's broader performance problems, as seen in tasks like the bAbI datasets~\cite{Su2015SolvingTP}.

Despite the inherent limitations of the LSTM, data augmentation as suggested in the transformer discussion \ref{app:transf-discussion} might enhance the model's performance.

\subsection{GPT-4}

\begin{figure}[ht]
    \centering
    \includegraphics[width=\textwidth]{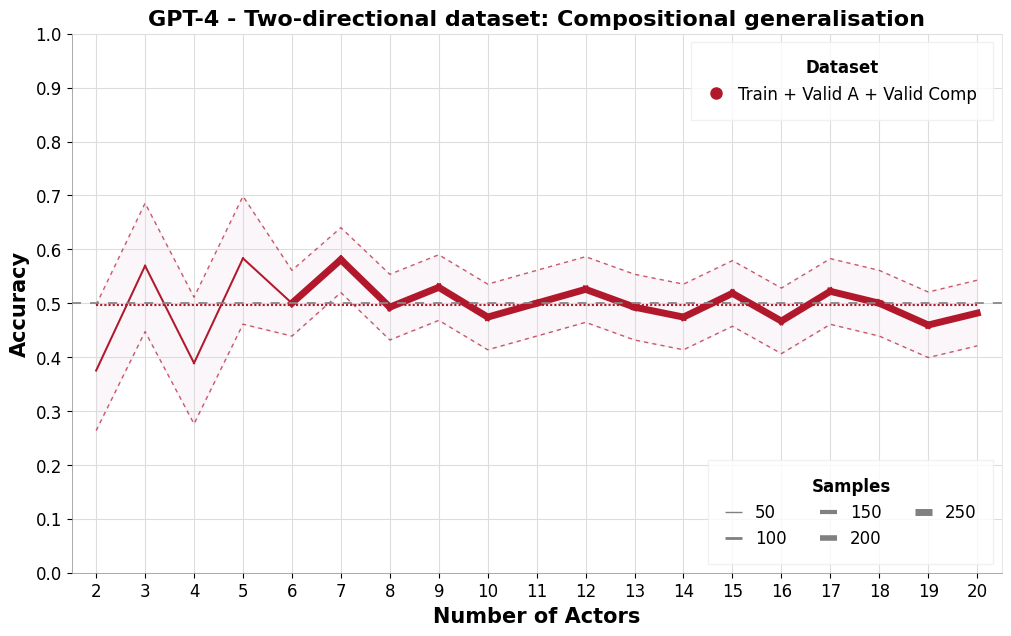}
    \caption{GPT-4 test accuracy for the two-directional dataset.}
    \label{fig:gpt_per_width_2d}
\end{figure}

\begin{figure}[ht]
    \centering
    \includegraphics[width=\textwidth]{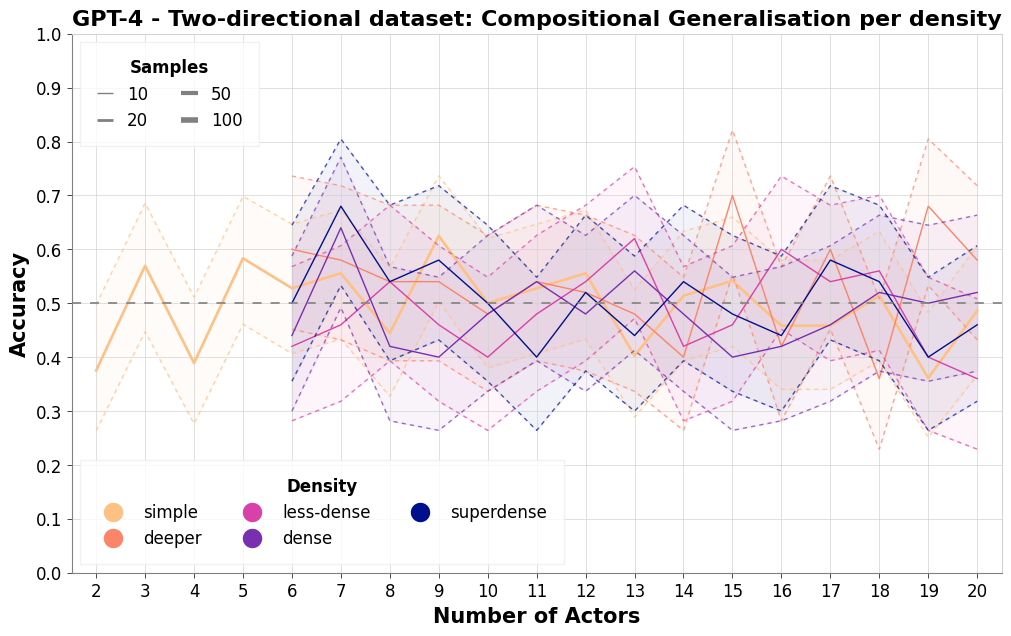}
    \caption{GPT-4 test accuracy for the two-directional dataset - breakdown by density.}
    \label{fig:gpt_per_width_2d_dens}
\end{figure}

\begin{figure}[ht]
    \centering
    \includegraphics[width=\textwidth]{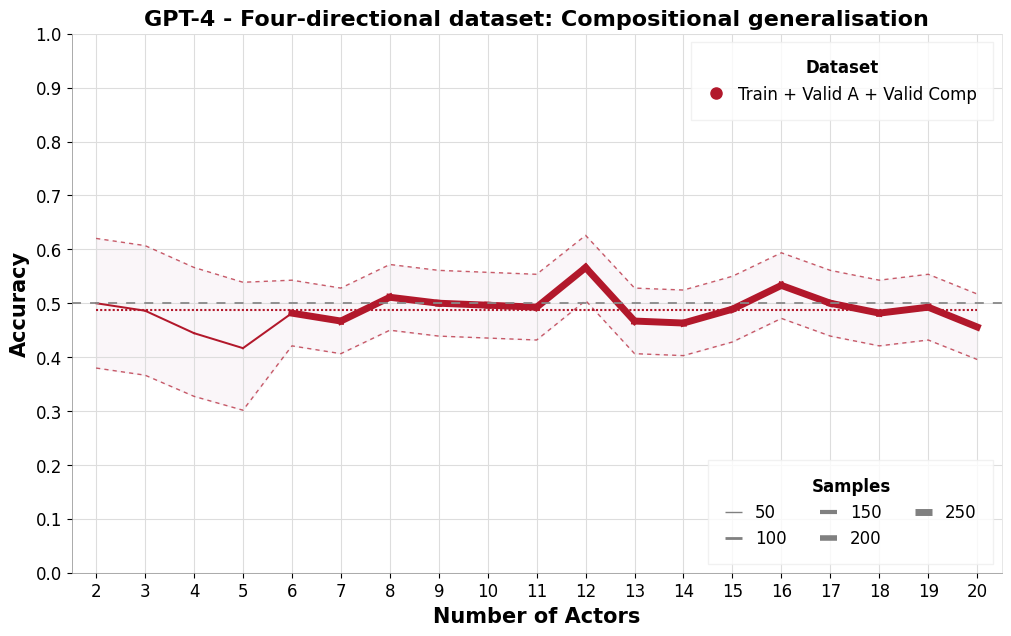}
    \caption{GPT-4 test accuracy for the four-directional dataset.}
    \label{fig:gpt_per_width}
\end{figure}

\begin{figure}[ht]
    \centering
    \includegraphics[width=\textwidth]{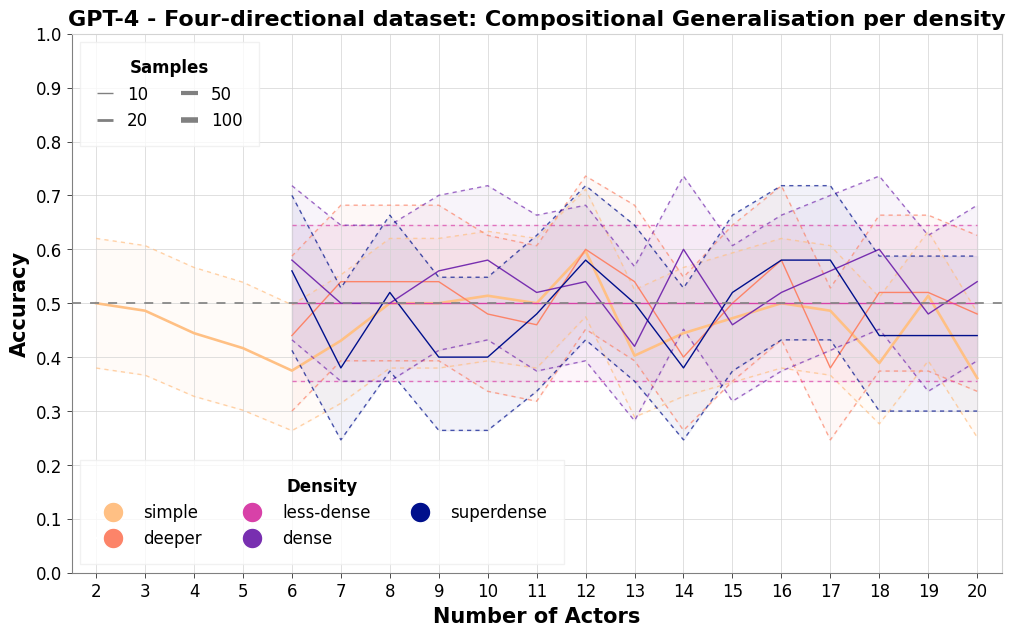}
    \caption{GPT-4 test accuracy for the four-directional dataset - breakdown by density.}
    \label{fig:gpt_per_width_dens}
\end{figure}

Finally, we assessed GPT-4 using both the two-directional and four-directional datasets, on 25/04/2024. We tested GPT-4 on the \textit{Train}, \textit{Valid A} and \textit{Valid Comp} datasets combined, consisting of 3756 entries in the two-directional and 4368 entries in the four-directional dataset. The stories were submitted in batches of 50. We compared GPT-4's responses to the correct labels and analyzed the outcomes, as illustrated in Figs.~\ref{fig:gpt_per_width_2d},~\ref{fig:gpt_per_width}. Moreover, we conducted a detailed analysis of performance based on story density, as shown in Figs.~\ref{fig:gpt_per_width_2d_dens} ,~\ref{fig:gpt_per_width_dens}.

We evaluated GPT-4's performance on the two-directional and four-directional datasets across various story densities and widths. 

The model's performance on both two- and four-directional datasets is on average close, or equal to 50\%, across all text widths, meaning random guessing in the given task (Figs.~\ref{fig:gpt_per_width_2d},~\ref{fig:gpt_per_width}). 
We restricted our evaluation on stories of up to 20 nouns, since the performance is consistently random, and we did not expect it to improve on larger text widths. A detailed breakdown of the performance by story density reveals a similar behaviour (Figs.~\ref{fig:gpt_per_width_2d_dens},~\ref{fig:gpt_per_width_dens}). The predictions for the four-directional less-dense dataset were imbalanced and GPT-4 predicted negative labels for all stories. We did not observe such imbalances for any of the other two- and four-directional story density datasets.

\end{document}